\documentclass[graybox]{svmult}

\usepackage{mathptmx}       % selects Times Roman as basic font
\usepackage{helvet}         % selects Helvetica as sans-serif font
\usepackage{courier}        % selects Courier as typewriter font
\usepackage{type1cm}        % activate if the above 3 fonts are
\usepackage{makeidx}         % allows index generation
\usepackage{graphicx}        % standard LaTeX graphics tool
\usepackage{multicol}        % used for the two-column index
\usepackage[bottom]{footmisc}% places footnotes at page bottom

\makeindex             % used for the subject index
\begin{document}

\title*{Nonlinear Lattice Waves in Random Potentials}
\author{Sergej Flach}
\institute{Sergej Flach \at New Zealand Institute for Advanced Study, Centre for Theoretical Chemistry and Physics, Massey University, 0745 Auckland New Zealand. \email{s.flach@massey.ac.nz}
%\and Name of Second Author \at Name, Address of Institute \email{name@email.address}
}
\maketitle

\abstract{Localization of waves by disorder is a fundamental physical problem encompassing a diverse spectrum of theoretical, experimental and numerical studies in the context of metal-insulator transition, quantum Hall effect, light propagation in photonic crystals, and dynamics of ultra-cold atoms in optical arrays. Large intensity light can induce nonlinear response, ultracold atomic gases can be tuned into
an interacting regime, which leads again to nonlinear wave equations on a mean field level.
The interplay between disorder and nonlinearity, their localizing and delocalizing effects is currently an intriguing and challenging issue in the field. We will discuss 
recent advances in the dynamics of nonlinear lattice waves in random potentials. In the absence of nonlinear terms in the wave equations, Anderson localization is leading to a halt of wave packet spreading.
 Nonlinearity couples localized eigenstates and, potentially, enables spreading and destruction of Anderson localization due to nonintegrability, chaos and decoherence.
The spreading process is characterized by universal subdiffusive laws due to nonlinear diffusion. We review extensive computational studies for one- and two-dimensional
systems with tunable nonlinearity power. We also briefly discuss extensions to other cases where the linear wave equation features
localization: Aubry-Andre localization with quasiperiodic potentials, Wannier-Stark localization with dc fields, and dynamical localization in momentum space with kicked rotors.
}

\section{Introduction}
\label{sec1}

In this chapter we will discuss the mechanisms of wave packet spreading in nonlinear disordered lattice systems.
More specifically, we will consider cases when the corresponding linear wave equations 
show Anderson localization, and the localization length is bounded from above by a finite value.

There are several reasons to analyze such situations. Wave propagation in spatially disordered
media has been of practical interest since the early times of studies of conductivity in solids. In particular,
it became of much practical interest for the conductance properties of electrons in semiconductor devices more
than half a century ago. It was probably these issues which motivated P. W. Anderson to perform his
groundbreaking lattice wave studies on what is now called Anderson localization \cite{PWA58}. With evolving modern technology, wave propagation
became of importance also in photonic and acoustic devices in structured materials \cite{Exp,Exp2}. Finally, recent advances in the control of
ultracold atoms in optical potentials made it possible to observe Anderson localization there as well \cite{BECEXP}.

In many if not all cases wave-wave interactions can be of importance, or can even be controlled experimentally.
Short range interactions hold for s-wave scattering of atoms. When many quantum particles interact,
mean field approximations often lead to effective nonlinear wave equations. Electron-electron interactions in solids and mesoscopic devices are also interesting candidates
with the twist of a new statistics of fermions. 
As a result, nonlinear wave equations
in disordered media become of practical importance. High intensity light beams propagating through structured optical devices
induce a nonlinear response of the medium and subsequent nonlinear contributions to the light wave equations.
While electronic excitations often suffer from dephasing due to interactions with other degrees of freedom (e.g. phonons),
the level of phase coherence can be controlled in a much stronger way for ultracold atomic gases and light.

There is moreover a fundamental mathematical interest in the understanding, how Anderson localization is modified 
in the presence of nonlinear terms in the wave equations. All of the above motivates the choice of corresponding
linear wave equations with finite upper bounds on the localization length. Then, the linear equations
admit no transport. Analyzing transport properties of nonlinear disordered wave equations allows to 
observe and characterize the influence of wave-wave interactions on Anderson localization in a straightforward
way. Finite upper bounds on the localization length for the corresponding linear wave equations are obtained for few band problems, which
are essentially emulating waves on lattices. Finite upper bounds on the localization length also allow to exclude overlap of initial states with eigenstates of
the linear equation which have a localization length larger than the considered system size, or which even have a diverging localization length.

No matter how tempting the general research theme  of this chapter is, one has to break it down to a list of more specific questions to be addressed. Let us attempt to file such a list, without pretending completeness:
\begin{itemize}
\item
I.  In his pioneering work Anderson addressed the 
fate of an initially localized wave packet during the subsequent evolution within the Schr\"odinger equation for a single particle \cite{PWA58}.
In particular, he showed that the return probability stays finite for infinite times, which essentially proves localization for the whole wave packet for all times.
What is the outcome for the case with nonlinear terms? 
\item
II. Anderson localization is equivalent to the statement that all eigenstates of the 
corresponding time-independent Schr\"odinger equation are spatially localized. Can these stationary states be continued into the nonlinear wave equation?
What are the properties of such stationary states in the nonlinear wave equation?
\item
III. The linear wave equation which enjoys Anderson localization yields zero conductivity, i.e. the system is an insulator, which is particularly true even for
finite densities of an infinitely extended wave state. Will the conductivity stay zero for nonlinear wave equations, or become finite? 
\item
IV. If qualitatively new physics is found in the presence of nonlinear terms in any of the above cases, how does it reconnect back to the linear equation which enjoys Anderson localization ?
\item
V. Quantizing the field equations leads to many-body interactions. What is the outcome for the above cases?
\item
VI. Wave localization for linear wave equations can be obtained also with quasiperiodic potentials (or in general correlated random potentials), even for potentials with nonzero
dc bias, but also for kicked systems (dynamical localization in momentum space). What is the outcome for all above cases (where applicable) then?
\end{itemize}
We will mainly focus on the first item. 
A number of studies was devoted to that subject  (see e.g. 
\cite{Mol98,PS08,Shep08,kkfa08,fks08,skkf09} as entree appetizers). 
It goes beyond the capabilities of this chapter to give a full account on all publications for the listed items. Some of them will be briefly discussed.
Nevertheless we list also a number of key publications as entree appetizers for the items II-VI: II \cite{gksa00,aisf07}, III \cite{dmb11,sfmvinl11}, IV \cite{mjgksa10,sa11,mvitvlsf11,dmb12}, 
V \cite{dmbilabla06,dokrksf11,mvitvlsf14}, VI \cite{dls94,ggjdbsf11,dokrksf09,mltvletal12}.

The chapter is structured in the following way. In section \ref{sec2} we introduce the models , and briefly discuss Anderson localization in section \ref{sec3}. In section \ref{sec4} 
we then proceed with adding nonlinear terms
to the wave equations. Using a secular normal form approach we demonstrate that a number of approximate treatments of the nonlinear terms keep localization intact,
in contrast to a large number of numerical observations. We identify omitted resonances, their occurence probabilities, and formulate expected dynamical regimes on that basis.
Section \ref{sec4} is closed with a technical discussion of different ways to characterize the evolution of wave packets. A number of numerical results on wave packet spreading are discussed in section \ref{sec5}.
Section \ref{sec6} is devoted to the formulation of an effective noise theory which is capable of describing the numerical observations. The various additional predictions of
the effective noise theory, along with their numerical tests, are presented in section \ref{sec7}. A short discussion of wave packet dynamics in related models with correlated potentials 
is given in section \ref{sec8}. Section \ref{sec9} closes this chapter with a discussion the probalistici restoring of Anderson localization for weak nonlinearity, and issues open for future
research.

\section{Lattice wave equations}
\label{sec2}

For the sake of simplicity we will first discuss one-dimensional lattice models, and subsequently generalize.
We will use the Hamiltonian of the disordered discrete nonlinear Schr\"odinger equation
(DNLS)
\begin{equation}
\mathcal{H}_{D}= \sum_{l} \epsilon_{l} 
|\psi_{l}|^2+\frac{\beta}{2} |\psi_{l}|^{4}
- (\psi_{l+1}\psi_l^*  +\psi_{l+1}^* \psi_l)
\label{RDNLS}
\end{equation}
with complex variables $\psi_{l}$, lattice site indices $l$ and nonlinearity
strength $\beta \geq 0$.   The uncorrelated random on-site energies $\epsilon_{l}$
are distributed with the probability density distribution (PDF) $\mathcal{P}_{\epsilon}(|x| \leq W/2) = 1/W$ and $\mathcal{P}_{\epsilon}(|x| > W/2) =0$.
where $W$ denotes the disorder
strength.  The equations of motion are generated by $\dot{\psi}_{l} = \partial
\mathcal{H}_{D}/ \partial (i \psi^{\star}_{l})$:
\begin{equation}
i\dot{\psi_{l}}= \epsilon_{l} \psi_{l}
+\beta |\psi_{l}|^{2}\psi_{l}
-\psi_{l+1} - \psi_{l-1}\;.
\label{RDNLS-EOM}
\end{equation}
Eqs.~(\ref{RDNLS-EOM}) conserve the energy (\ref{RDNLS}) and the norm $S
= \sum_{l}|\psi_l|^2$.  Note that varying the norm of an
initial wave packet is strictly equivalent to varying $\beta$.
Note also that the transformation $\psi_l \rightarrow (-1)^l \psi^*_l$, $\beta \rightarrow -\beta$, $\epsilon_l \rightarrow -\epsilon_l$ leaves the
equations of motion invariant. Therefore the sign of the nonlinear coefficient $\beta$ can be fixed without loss of generality.
Eqs.~(\ref{RDNLS}) and (\ref{RDNLS-EOM}) are derived e.~g.~when
describing two-body interactions in ultracold atomic gases on an optical
lattice within a mean field approximation \cite{oberthaler}, but also when
describing the propagation of light through networks of coupled optical
waveguides in Kerr media \cite{yskgpa03}.

Alternatively we also refer to results for the Hamiltonian of the quartic Klein-Gordon lattice (KG) 
\begin{equation}
\mathcal{H}_{K}= \sum_{l}  \frac{p_{l}^2}{2} +
\frac{\tilde{\epsilon}_{l}}{2} u_{l}^2 + 
\frac{1}{4} u_{l}^{4}+\frac{1}{2W}(u_{l+1}-u_l)^2,
\label{RQKG}
\end{equation}
where $u_l$ and $p_l$ are respectively the generalized coordinates and
momenta, and $\tilde{\epsilon}_{l}$ are
chosen uniformly from the interval $\left[\frac{1}{2},\frac{3}{2}\right]$. 
The equations of motion are $\ddot{u}_{l} = - \partial \mathcal{H}_{K}
/\partial u_{l}$ and yield
\begin{equation}
\ddot{u}_{l} = - \tilde{\epsilon}_{l}u_{l}
-u_{l}^{3} + \frac{1}{W} (u_{l+1}+u_{l-1}-2u_l)\;.
\label{KG-EOM}
\end{equation}
Equations (\ref{KG-EOM}) conserve the energy (\ref{RQKG}). They serve e.g. as
simple models for the dissipationless dynamics of anharmonic optical lattice
vibrations in molecular crystals \cite{aaovchinnikov}.  The energy of an
initial state $E \geq 0$ serves as a control parameter of nonlinearity similar
to $\beta$ for the DNLS case.
For small amplitudes the equations of motion of the KG chain can be
approximately mapped onto a DNLS model
\cite{KG-DNLS-mapping,blskf11}. For the KG model with given parameters $W$ and $E$, the corresponding
DNLS model (\ref{RDNLS}) with norm $S=1$, has a nonlinearity parameter
$\beta\approx 3WE$. 
The norm density of the DNLS model corresponds to the normalized energy
density of the KG model \cite{blskf11}.

The theoretical considerations will be performed within the DNLS framework.
It is straightforward to adapt them to the KG case.

\section{Anderson localization}
\label{sec3}

For $\beta=0$ with $\psi_{l} = A_{l}
\exp(-i\lambda t)$ Eq.~(\ref{RDNLS})
is reduced to the linear eigenvalue problem
\begin{equation}
\lambda A_{l} = \epsilon_{l} A_{l} 
- A_{l-1}-A_{l+1}\;.
\label{EVequation}
\end{equation}
The normal modes (NM) are characterized by the 
normalized eigenvectors $A_{\nu,l}$ ($\sum_l A_{\nu,l}^2=1)$.
The eigenvalues $\lambda_{\nu}$ are the frequencies of the NMs.  The width
of the eigenfrequency spectrum $\lambda_{\nu}$ of (\ref{EVequation}) is
$\Delta=W+4$ with $\lambda_{\nu} \in \left[ -2 -\frac{W}{2}, 2 + \frac{W}{2}
\right] $. While the usual ordering principle of NMs is with their increasing eigenvalues, here we adopt
a spatial ordering with increasing value of the center-of-norm coordinate $X_{\nu}= \sum_l l A^2_{\nu,l}$.

The asymptotic spatial decay of an eigenvector is given by $A_{\nu,l} \sim
{\rm e}^{-|l|/\xi(\lambda_{\nu})}$ where 
$\xi(\lambda_{\nu})$ is the localization length and
$\xi(\lambda_{\nu}) \approx
24(4-\lambda_{\nu}^2)/W^2$ for weak disorder $W \leq 4$ \cite{KRAMER}. 

The NM participation number
$p_{\nu} = 1/\sum_l A_{\nu,l}^4$ measures the number of strongly excited lattice sites in a given wave density distribution.
It is one possible way to quantize the spatial extend $V_{\nu}$
(localization volume) of a NM. However fluctuations of the density distribution inside a given NM lead to an underestimate of $V_{\nu}$ when using
$p_{\nu}$. A better way to estimate the distance between the two exponential tails of an eigenvector is to use the second moment of its density distribution
$m_2^{(\nu)} = \sum_l (X_{\nu}-l)^2A_{\nu,l}^2$. It follows that the estimate $V_{\nu} = \sqrt{12m_2^{(\nu)}}$ is highly precise and sufficient for most purposes \cite{dksf10}.
The localization volume $V$ is on average of the order of $3 \xi$
for weak disorder, and tends to $V=1$ in the limit of strong disorder.

Consider an eigenstate $A_{\nu,l}$ for a given disorder realization. How many of the neighboring eigenstates will have non-exponentially small amplitudes inside its localization volume $V_{\nu}$?
Note that there is a one-to-one correspondence between the number of lattice sites, and the number of eigenstates. Therefore, on average the number of neighboring eigenstates
will be simply $V_{\nu}$. Let us consider sets of neighboring eigenstates. Their eigenvalues will be in general different, but confined to the interval $\Delta$ of the spectrum.
Therefore the average spacing $d$ of eigenvalues of neighboring NMs
within the range of a localization volume is of the order of $d \approx \Delta / V$,
which becomes $d \approx \Delta W^2 /300 $ for weak disorder.
The two scales $ d \leq \Delta $ are expected to determine the
packet evolution details in the presence of nonlinearity.

Due to the localized character of the NMs, any localized wave packet with size $L$ which is launched into
the system for $\beta=0$ , will stay localized for all times. If $L \ll V$, then the wave packet will expand
into the localization volume. This expansion will take a time of the order of $\tau_{lin}=2\pi/d$. 
If instead $L \geq V$, no substantial expansion will be observed in real space.
We remind that Anderson localization is relying on the phase coherence of waves. Wave packets which are trapped due
to Anderson localization correspond to trajectories in phase space evolving on tori, i.e. they evolve quasi-periodically in time.

Finally, the linear wave equations constitute an integrable system with conserved actions where the dynamics happens to be on quasiperiodic tori in phase space.
This can be safely stated for any finite, whatever large, system.

\section{Adding nonlinearity}
\label{sec4}

The equations of motion of (\ref{RDNLS-EOM})  in normal mode space read
\begin{equation}
i \dot{\phi}_{\nu} = \lambda_{\nu} \phi_{\nu} + \beta \sum_{\nu_1,\nu_2,\nu_3}
I_{\nu,\nu_1,\nu_2,\nu_3} \phi^*_{\nu_1} \phi_{\nu_2} \phi_{\nu_3}\;
\label{NMeq}
\end{equation}
with the overlap integral 
\begin{equation}
I_{\nu,\nu_1,\nu_2,\nu_3} = 
\sum_{l} A_{\nu,l} A_{\nu_1,l} 
A_{\nu_2,l} A_{\nu_3,l}\;.
\label{OVERLAP}
\end{equation}
The variables $\phi_{\nu}$ determine the complex time-dependent amplitudes of
the NMs.

The frequency shift of a single site oscillator induced by the nonlinearity is
$\delta_l = \beta |\psi_l|^{2}$. If instead a single mode is excited, its
frequency shift can be estimated by $\delta_{\nu} = \beta |\phi_{\nu}|^2/
p_{\nu}$. 

As it follows from (\ref{NMeq}), nonlinearity induces an interaction between NMs.
Since all NMs are exponentially localized in space, each normal mode is effectively coupled to a finite
number of neighboring NMs, i.e. the interaction range is finite. However the strength of the coupling
is proportional to the norm density $n = |\phi|^2$. Let us assume that a wave packet spreads.
In the course of spreading its norm density will become smaller. Therefore the effective coupling strength
between NMs decreases as well. At the same time the number of excited NMs grows.
One possible outcome would be: (I) that after some time the coupling will be weak enough
to be neglected. If neglected, the nonlinear terms are removed, the problem is reduced to an integrable linear wave equation,
and we obtain again Anderson localization. That implies that the trajectory happens to be on a quasiperiodic torus -
on which it must have been in fact from the beginning. It also implies that the actions of the linear wave equations are not strongly varying in the nonlinear case,
and we are observing a kind of anderson localization in action subspace. 
Another possibility is: (II) that spreading continues for all times. That would imply that the trajectory does not
evolve on a quasiperiodic torus, but instead evolves in some chaotic part of phase space.
This second possibility (II) can be subdivided further, e.g. assuming that the wave packet will exit, or enter, a Kolmogorov-Arnold-Moser (KAM) regime of mixed phase space, or stay all the time outside such a perturbative KAM regime. In particular if the wave packet dynamics will
enter a KAM regime for large times, one might speculate 
that the trajectory will
get trapped between denser and denser torus structures in phase space after some spreading, leading again
to localization as an asymptotic outcome, or at least to some very strong slowing down of the spreading process. We will not go into details of such possible
scenaria, but want the reader to be aware of the fact that the rather innocent set of questions at stake can quickly lead into highly sophisticated mathematical fields.

Consider a wave packet with size $L$ and norm density $n$. Replace it by a {\sl finite} system of size $L$ and
norm density $n$. 
Such a finite system will be in general nonintegrable. Therefore the only possibility
to generically obtain a quasiperiodic evolution is to be in the regime where the KAM theorem holds.
Then there is a finite fraction of the available phase space volume which is filled with KAM tori.
For a given $L$ it is expected that there is a critical density $n_{KAM}(L)$ below which the KAM regime will hold.
We do not know this $L$-dependence. Computational studies may not be very conclusive here, since it is hard to distinguish
a regime of very weak chaos from a strict quasiperiodic one on finite time scales.

The above first possible outcome (I) (localization) will be realized if the packet is launched in a KAM regime.
Whether that is possible at all for an infinite system is an open issue.
The second outcome (II) (spreading) implies that we start in a chaotic regime and remain there. Since 
the packet density is reduced and is proportional to its inverse size $L$ at later times, 
this option implies that the critical density $n_{KAM}(L)$ decays faster than $1/L$, possibly faster than any
power of $1/L$.

 Let us discuss briefly one example of an integrable system, for which Anderson localization will not be destroyed.  
Consider a
Hamiltonian in NM representation using actions $J_{\nu}$ and angles
$\theta_{\nu}$ as coordinates:
\begin{equation}
\mathcal{H}_{int} = \sum_{\nu} \lambda_{\nu} J_{\nu} + \beta
\sum_{\nu_1,\nu_2,\nu_3,\nu_4} I_{\nu_1,\nu_2,\nu_3,\nu_4} \sqrt{ J_{\nu_1}
J_{\nu_2} J_{\nu_3} J_{\nu_4}}\;.
\label{int1}
\end{equation}
We assume that the set of eigenfrequencies $\{ \lambda_{\nu} \}$ and the
overlap integrals $I_{\nu_1,\nu_2,\nu_3,\nu_4}$ are identical with those
describing the DNLS model (\ref{NMeq}), (\ref{OVERLAP}). The equations of
motion $\dot{J}_{\nu} = -\partial \mathcal{H}_{int}/ \partial \theta_{\nu}$
and $\dot{\theta}_{\nu} = \partial \mathcal{H}_{int}/ \partial J_{\nu}$ yield
$\dot{J}_{\nu} = 0$ since the integrable Hamiltonian (\ref{int1}) depends only
on the actions.  Therefore, any localized initial condition
(e.~g.~$J_{\nu}(t=0) \propto \delta_{\nu,\nu_0}$) will stay localized, since
actions of modes which are at large distances will never get  excited.
Thus, any observed spreading of wave packets, which we will study in detail in
the present work, is presumably entirely due to the nonintegrability of the considered
models, at variance to (\ref{int1}).

\subsection{The secular normal form}
\label{sec41}

Let us perform a further transformation $\phi_{\nu} = {\rm e}^{-i \lambda_{\nu} t} \chi_{\nu}$ and insert
it into Eq. (\ref{NMeq}):
\begin{equation}
i \dot{\chi}_{\nu} = \beta \sum_{\nu_1,\nu_2,\nu_3}
I_{\nu,\nu_1,\nu_2,\nu_3} \chi^*_{\nu_1} \chi_{\nu_2} \chi_{\nu_3} {\rm e}^{i(\lambda_{\nu} + 
\lambda_{\nu_1}-\lambda_{\nu_2}-\lambda_{\nu_3})t}
\;.
\label{NMeqchi}
\end{equation}
The right hand side contains oscillating functions with frequencies
\begin{equation}
\lambda_{\nu,\vec{n}} \equiv \lambda_{\nu} + 
\lambda_{\nu_1}-\lambda_{\nu_2}-\lambda_{\nu_3}\;,\;\vec{n} \equiv (\nu_1,\nu_2,\nu_3)\;.
\label{dlambda}
\end{equation}
For certain values of $\nu,\vec{n}$ the value $\lambda_{\nu,\vec{n}}$ becomes exactly zero. These secular terms
define some slow evolution of (\ref{NMeqchi}). Let us perform an averaging over time of all terms
in the rhs of (\ref{NMeqchi}), leaving therefore only the secular terms. The resulting secular
normal form equations (SNFE) take the form
\begin{equation}
i \dot{\chi}_{\nu} = \beta \sum_{\nu_1}
I_{\nu,\nu_,\nu_1,\nu_1} |\chi_{\nu_1}|^2 \chi_{\nu}
\;.
\label{NMeqRNF}
\end{equation}
Note that possible missing factors due to index permutations can be absorbed into the overlap integrals,
and are not of importance for what is following.
The SNFE can be now solved for any initial condition $\chi_{\nu}(t=0)=\eta_{\nu}$ and yields
\begin{equation}
\chi_{\nu}(t) = \eta_{\nu} {\rm e}^{-i \Omega_{\nu} t}\;,\; \Omega_{\nu} = \beta \sum_{\nu_1}
I_{\nu,\nu_,\nu_1,\nu_1} |\eta_{\nu_1}|^2 
\;.
\label{SNFE}
\end{equation}
Since the norm of every NM is preserved in time for the SNFE, it follows that Anderson localization
is preserved within the SNFE. The only change one obtains is the renormalization of the eigenfrequencies
$\lambda_\nu$ into $\tilde{\lambda}_{\nu} = \lambda_{\nu}+\Omega_{\nu}$. Moreover, the phase coherence
of NMs is preserved as well. Any different outcome will be therefore due to the nonsecular terms,
neglected within the SNFE. We note that $I_{\nu,\nu_,\nu,\nu} \equiv p_{\nu}^{-1}$. Then the sum in  (\ref{OVERLAP}) contains only
nonnegative terms. By normalization $A_{\nu,l} \sim 1/\sqrt{V}$ inside its localization volume, and therefore $I_{\nu,\nu_,\nu,\nu} \sim 1/V$.
Similar argumentation leads to $I_{\nu,\nu_,\nu_1,\nu_1} \sim 1/V$ if both modes reside in the same localization volume.

Let us discuss several different initial states.
(a) If only one normal mode is initially excited to norm $n$, then it follows from (\ref{SNFE}) that 
its frequency renormalization $\Omega_{\nu} = \beta n p_{\nu}^{-1} \sim \beta n /V$
where $V$ is a typical localization volume of a normal mode. Comparing this value to the spacing $d \sim \Delta /V$ we conclude
that a perturbation approach (and therefore Anderson localization) might survive up to finite values of $\beta n \sim \Delta$.
(b) If however a large group of normal modes is excited inside a wave packet such that all normal modes have norm $n$, then
the sum in (\ref{SNFE}) will change the frequency renormalization to $\Omega_{\nu} \sim \beta n$ for each of the participating modes.
Comparing that to the spacing $d$ we now find that perturbation approaches might break down at sufficiently weaker nonlinearities
$\beta n \sim \Delta / V$. (c) Finally assume that only one lattice site is initially excited with norm $n$. That means that $V$ normal modes
are excited each with norm $n/V$. And that is is also what we will see in a dynamical evolution of the linear wave equation - after some short 
transient time the wave packet will occupy a localization volume region and stay in there. Then the frequency normalization for each
participating mode becomes $\Omega_{\nu} \sim \beta n /V$ as in (a), and perturbation theory is expected to break down again at 
$\beta n \sim \Delta$.

All of the considered lattices allow for selftrapped states in the regime of strong nonlinearity. These are well known as discrete breathers,
intrinsic localized modes, and discrete solitons \cite{DB} which are time-periodic but spatially localized exact solutions of the equations 
of motion. They exist for any sign of
nonlinearity due to the underlying lattice, which generates finite bounds on the spectrum of the linear wave equation. 
Discrete breathers appear because the nonlinear terms renormalize (shift) frequencies completely out of the linear wave spectrum.
In the limit of strong nonlinearity these states are essentially single site excitations, with very little amplitudes present on neighboring sites.
Therefore, the natural basis for the physics of selftrapping is the original lattice itself, rather than the normal modes of the linear wave
equation. This becomes evident when considering a lattice without any disorder, for which the normal modes of the linear wave equation
are extended states, yet selftrapping and discrete breathers are perfectly present as well within the nonlinear wave equation.
Selftrapping and discrete breathers are examples of a {\it nonperturbative} physics of strong nonlinearity.  
For the above cases of initial conditions, selftrapping can be effectively predicted whenever a single oscillator on one
site renormalizes its frequency $\epsilon_l+\beta |\psi_l|^2$ such that it exits the linear wave spectrum.
For the above initial state case (a) this happens when $\beta n  \sim V (\Delta/2-\lambda_{\nu})$, about $V$ times larger than the
perturbation threshold. For case (b) the norm $n$ per normal mode is also the norm $n$ per lattice site. Therefore selftrapping 
is expected at $\beta n \sim \Delta /2 $, again about $V$ times larger than the corresponding perturbation threshold.
However, case (c) is different. Here we place a norm $n$ {\it initially} on one site. If the selftrapping condition for that site holds, the dynamics
will stay from scratch in the nonperturbative discrete breather regime, without any chance to spread into a localization volume region
set by the linear wave equation. Therefore the selftrapping threshold reads $\beta n \sim \Delta /2 - \epsilon_l$ and becomes of the same
order as the perturbation threshold. Single site excitations will be thus launched either in a perturbative regime, or in a self trapped one.
The other initial states allow for a third regime - outside the perturbative regime, but well below the selftrapping one.
For reasons to come, we coin this additional regime {\it strong chaos regime}, and the perturbative regime {\it weak chaos regime}.
We recapitulate again, that single site excitations are expected to be either in the regime of weak chaos, or selftrapping. Other initial states
allow for another intermediate regime of strong chaos.

\subsection{Expected dynamical regimes}
\label{sec42}

Consider a wave packet at $t=0$ which has norm density $n$ and size $L$.
Let us wrap the above discussion into expected dynamical regimes \cite{sf10}. Note that due to the above ambiguities, 
the following estimates are at the best semi-quantitative.

{\sc Single site excitations} with norm $n$ and $\epsilon_l=0$ at the excitation site:
\begin{eqnarray}
\beta n  < \Delta / 2 \;:\; {\rm weak\; chaos}
\nonumber
\\
{\rm strong \; chaos \; not \; present}
\label{sser}
\\
\beta n > \Delta / 2 \;:\; {\rm selftrapping}
\nonumber
\end{eqnarray}
{\sc Single mode excitations} with norm $n$ and $\lambda_{\nu}=0$ for the excited mode:
\begin{eqnarray}
\beta n < \Delta \;:\; {\rm weak\; chaos}
\nonumber
\\
\Delta < \beta n  < V \Delta /2 \;:\; {\rm strong\; chaos}
\label{smer}
\\
V \Delta / 2 < \beta n \;:\; {\rm selftrapping}
\nonumber
\end{eqnarray}
{\sc Multi site/mode wave packet} with norm density $n$ per site/mode and size $V$:
 \begin{eqnarray}
\beta n < \Delta /V \;:\; {\rm weak\; chaos}
\nonumber
\\
\Delta / V < \beta n  < \Delta /2 \;:\; {\rm strong\; chaos}
\label{wper}
\\
\Delta / 2 < \beta n \;:\; {\rm selftrapping}
\nonumber
\end{eqnarray}

\subsection{Beyond the secular normal form}
\label{sec43}

The time-averaged secular norm form (\ref{NMeqRNF}) keeps the integrability of the nonlinear wave equation, and therefore also 
keeps Anderson localization. Any deviation from Anderson localization is therefore due to the omitted time-dependent oscillating terms
in (\ref{NMeqchi}). Let us isolate one of the many terms in the rhs sum in (\ref{NMeqchi})
\begin{equation}
 \dot{\chi}_{\nu} = \beta 
I_{\nu,\vec{n}} \chi^*_{\nu_1} \chi_{\nu_2} \chi_{\nu_3} {\rm e}^{i\lambda_{\nu,\vec{n}} t}
\;.
\label{oneterm}
\end{equation}
Assume a solution of the secular normal form equations (\ref{NMeqRNF}) in the limit of weak nonlinearity which we coined 
weak chaos. Consider the solution of (\ref{oneterm}) as a first order correction. This correction has an amplitude
\begin{equation}
|\chi_{\nu}^{(1)}| = |\beta \eta_{\nu_1}\eta_{\nu_2}\eta_{\nu_3}|
R_{\nu,\vec{n}}^{-1}\;,\; R_{\nu,\vec{n}} \sim
\left|\frac{\lambda_{\nu,\vec{n}}}{I_{\nu,\vec{n}}}\right| \;,
\label{PERT1}
\end{equation}
The
perturbation approach breaks down, and resonances set in, when $|\eta_{\nu}|
< |\chi_{\nu}^{(1)}|$ for at least one triplet $\vec{n}$, and for at least one excited reference mode $\nu$:
\begin{equation}
|\eta_{\nu}| < |\eta_{\nu_1}\eta_{\nu_2}\eta_{\nu_3}|
\frac{\beta}{R_{\nu,\vec{n}}}\;.
\label{PERT2}
\end{equation}
Let us discuss this result. The eigenfrequencies contribute through the quadruplet  $\lambda_{\nu,\vec{n}}$ (\ref{dlambda}). This quantity can be also
interpreted as the difference of two eigenvalue differences. Resonances will be triggered for small quadruplets. However, for this to hold we do not need to request
that two of the participating eigenvalues are close. In fact, since we consider only participating states from one localization volume, level repulsion between neighboring eigenvalues will
be present anyway, such that the level spacing of nearest neighbor eigenvalues shows signatures of Wigner-Dyson distributions characteristic for random matrices (Fig.4 in \cite{dksf10}).
This means in particular, that the probability density function (PDF) and therefore the probability of finding weakly separated (well beyond $d$) eigenvalues tends to zero
for vanishing separation.
However, the above quadruplet can become small for eigenvalues which are separated way beyond $d$. An extreme example is an equidistant spectrum which allows for
exact zeros of quadruplets.  In the disordered case with $V \gg 1$, for one reference mode $\nu$ we consider $V$ states in its localization volume, which allow for about
$V^3$ quadruplet combinations. It is reasonable to assume that the set of $V$ eigenvalues will show correlations on energy separations of the order of $d$ (level spacing), but a decay
of these correlations at larger energy distances. Therefore, for most of the $V^3$ combinations, the participating eigenvalues can be considered to be uncorrelated.
With that assumption, the PDF $\mathcal{W}_{\lambda}(\lambda_{\nu,\vec{n}})$, which is a sum of four random numbers, 
can be expected to be close to a normal distribution due to the central limit theorem, i.e.
\begin{equation}
\mathcal{W}_{\lambda}(x) \approx \frac{1}{\sqrt{2\pi} \sigma } {\rm e}^{-\frac{x^2}{2\sigma^2}}\;,\;\sigma^2 = \frac{\Delta^2}{12}\;.
\label{gauss}
\end{equation}
In a recent study of a one-dimensional ladder geometry \cite{xysf14} the closeness of the normal distribution to $\mathcal{W}_{\lambda}$ was numerically confirmed.
Since we are interested in small quadruplet values, we stress that the normal distribution has a finite value at zero argument, i.e. 
\begin{equation}
\mathcal{W}_{\lambda}(0) \approx \frac{\sqrt{3}}{\sqrt{2\pi} \Delta }\;.
\label{gausszero}
\end{equation}
Again the predicted value is only a factor of two off the actual numbers computed in \cite{xysf14}.

The second important quantity which enters (\ref{PERT2}) through the definition of $R_{\nu,\vec{n}}$ in (\ref{PERT1}) are the overlap integrals $I_{\nu,\vec{n}}$. Much less is known about
these matrix elements (however see \cite{dksf10}). It is instructive to mention that the same overlap integrals play a crucial role when estimating the localization length of two interacting particles (e.g. within
a Bose-Hubbard chain) with onsite disorder \cite{TIP,dokrksf11,mivtvlsf14} and are the main reason for the absence of any consensus on the scaling properties of this localization length.
This is mainly due to the strong correlations between eigenvectors of states residing in the same localization volume but having sufficiently well separated eigenvalues. Let us ignore those difficulties for the moment, and assume that we can operate with one characteristic (average)
overlap integral $\langle I \rangle$. 
Then the PDF $\mathcal{W}_R$ of $R$ becomes 
\begin{equation}
\mathcal{W}_R (x) = \langle I \rangle \mathcal{W}_{\lambda}(\langle I \rangle x)\;,\; \mathcal{W}_R(0) = \frac{\sqrt{3} \langle I \rangle}{\sqrt{2\pi} \Delta }\;.
\label{pdfR}
\end{equation}
With the additional assumption that all amplitudes $\eta \sim \sqrt{n}$ (note that this excludes a systematic consideration of a single normal
mode excitation) we arrive at the resonance condition
\begin{equation}
\beta n < R_{\nu,\vec{n}}\;.
\label{resonance_R}
\end{equation}
For a given set $\{\nu,\vec{n}\}$ the probability of meeting such a resonance is given by
\begin{equation}
\mathcal{P}_{\nu,\vec{n}} = \int_0^{\beta n} \mathcal{W}_R (x) dx \;,\; \mathcal{P}_{\nu,\vec{n}}|_{\beta n \rightarrow 0} \rightarrow \frac{\sqrt{3} \langle I \rangle}{\sqrt{2\pi} \Delta } \beta n\;.
\label{indres}
\end{equation}
For a given reference mode $\nu$ there are $V^3$ combinations of quadruplets. The probability that at least one of these quadruplets satisfies the resonance condition
is equivalent to the probability that the given mode violates perturbation theory:
\begin{equation}
\mathcal{P}_{\nu} = 1 - \left(  1 -  \int_0^{\beta n} \mathcal{W}_R (x) dx           \right) ^{V^3}\;,\; \mathcal{P}_{\nu} |_{\beta n \rightarrow 0} 
\rightarrow \frac{\sqrt{3} V^3 \langle I \rangle}{\sqrt{2\pi} \Delta } \beta n\;.
\label{resonance}
\end{equation}
The main outcome is that the probability of resonance is proportional to $\beta n$ for weak nonlinearity. Moreover, within the disorder interval $1 \leq W \leq 6$
a numerical evaluation of the average overlap integral $\langle I \rangle \approx 0.6 \; V^{-1.7}$ \cite{dksf10}. This yields 
$\mathcal{P}_{\nu} |_{\beta n \rightarrow 0}  \approx 0.43\;V^{0.3} (\beta n/d)$. The uncertainty of the correct estimate of the overlap integral average, and the restricted
studied disorder range may well address the weak dependence $V^{0.3}$. What remains however is evidence that the resonance probability for weak nonlinearity
is proportional to the ratio $(\beta n ) / d$. Therefore a practical outcome is that the average spacing $d$ sets the energy scale - for $\beta n \ll d$ the resonance probability
$\mathcal{P} \sim (\beta n) /d$, while for $\beta n \gg d$ the resonance probability $P \approx 1$. As already anticipated at the end of the previous subsection, 
two regimes of weak and strong chaos can be defined depending on the ratio $(\beta n)/d$. In the regime of strong chaos, any normal mode within an excited wave packet
will be resonant and not obeying perturbation theory. In the regime of weak chaos, this will be true for a fraction of modes.

A straightforward numerical computation of the above probability can be performed avoiding a number of the above assumptions.
For a
given NM $\nu$ we define $ R_{\nu,\vec{n}_0} = \min_{\vec{n} }
R_{\nu,\vec{n}}$.  Collecting $R_{\nu,\vec{n}_0}$ for many $\nu$ and many
disorder realizations, we can obtain the probability density distribution
$\mathcal{W}(R_{\nu,\vec{n}_0})$. 
The probability $\mathcal{P}$ for a mode, which is excited to a norm $n$ (the
average norm density in a packet of modes), to be resonant with at least one triplet
of other modes at a given value of the interaction parameter $\beta$ is again given
by \cite{dksf10,skkf09}
\begin{equation}
\mathcal{P} = \int_0^{\beta n} \mathcal{W}(x) {\rm d}x\;.
\label{resprob}
\end{equation} 
Therefore again $\mathcal{W}(R_{\nu,\vec{n}_0} \rightarrow 0)
\rightarrow C(W) \neq 0$ \cite{skkf09}.  For the cases studied, the constant $C$ drops with
increasing disorder strength $W$, in agreement with (\ref{resonance}), which suggests
$C = \frac{\sqrt{3} V^3 \langle I \rangle}{\sqrt{2\pi} \Delta }$.

The large power $V^3$ in (\ref{resonance}) allows to make a simple exponential approximation
\begin{equation}
\mathcal{W} (R) \approx C {\rm e}^{-CR}\;,\; C = \frac{\sqrt{3} V^3 \langle I \rangle}{\sqrt{2\pi} \Delta }\;.
\label{approxp}
\end{equation}
which in turn can be expected to hold also for the case of weak disorder.
It leads to the approximative result
\begin{equation}
\mathcal{P} = 1-{\rm e}^{-C\beta n}\;.
\label{approxpp}
\end{equation}
Therefore the probability for a mode in the packet to be resonant is
proportional to $C \beta n$ in the limit of small $n$ \cite{fks08,skkf09}.  

We stress again that the discussed uncertainty  in the definition of an average overlap integral, and the fact that the distribution of quadruplets is expected to be controlled
by the {\it stiffness} of the set of eigenvalues of the normal mode set $\{ \nu,\vec{n} \}$ rather than its spacing $d$, might be related. This does become evident if assuming an equidistant
set. But then again, for a disordered system discussed here, the only scale on which the quadruplets can fluctuate close to zero, is the spacing $d$.

\subsection{Measuring properties of spreading wave packets}
\label{sec44}

We remind that the ordering of NMs is chosen to be by increasing value of the center-of-norm coordinate
$X_{\nu}$.  We will analyze normalized distributions $n_{\nu}
\geq 0$ using the second moment $m_2= \sum_{\nu}
(\nu-\bar{\nu})^2 n_{\nu}$, which quantifies the wave packet's degree of
spreading and the participation number $P=1 / \sum_{\nu} n_{\nu}^2$, which
measures the number of the strongest excited sites in $n_{\nu}$.  Here
$\bar{{\nu}} = \sum_{\nu} \nu n_{\nu}$.  We follow norm density
distributions $n_{\nu}\equiv |\phi_{\nu}|^2/\sum_{\mu} |\phi_{\mu}|^2$.
The second moment $m_2$ is sensitive to the distance of the tails of
a distribution from the center, while the participation number $P$ is a measure of the
inhomogeneity of the distribution, being insensitive to spatial
correlations. Thus, $P$ and $m_2$ can be used to quantify the
sparseness of a wave packet through the compactness index 
\begin{equation}
\zeta=\frac{P^{2/D}}{m_2}
\label{eq:ci}
\end{equation} 
where $D$ is the dimension of the lattice.

In order to have a scale for $\zeta$, we can consider a system of harmonic oscillators (normal modes) which are weakly interacting through 
nonlinear couplings. The contribution of the nonlinear interaction to the overall energy $E$ of the system is assumed to be small and negligible.
However it is essential in order to assume that the considered system is ergodic, i.e. we can replace time averages by suitable ensemble 
distribution averages. We also assume for simplicity that the distribution is of Boltzmann type. Therefore with good accuracy each
oscillator is characterized by its own distribution $\rho (E_{\nu}) = {\rm e}^{-\beta_B E_{\nu}}/\beta_B$ where $\beta_B$ is the inverse temperature, and $E_{\nu}$ is the energy of an oscillator with average $1/\beta_B$. We consider 
a lattice bounded by a $D$-dimensional sphere with radius $R \gg 1$ which contains $N$ lattice sites, and therefore $N$ oscillators.
For $D=1$ we have $N=2R$, for $D=2$ it follows $N=\pi R^2$ and for $D=3$ we have $N=4\pi R^3/3$. We now evaluate the normalized
energy distribution $\beta_B E_{\nu}/N$. Due to ergodicity the inverse of the participation number 
$1/P = (\beta_B/N)^2 \sum_{\nu} E_{\nu}^2 =  \beta_B^2/N \int \rho (E) E^2= 2/N$.   
The second moment can be estimated at any time to be $m_2=R^2/2$ (for $D=2$) and $m_2 = 3R^2/5$ (for $D=3$) since enough oscillators
at large but constant distance from the center allow for an ensemble average. In the one-dimensional case such an estimate can be performed
only after a time average over times larger than the equipartition times (equivalently the correlation decay times) and yields 
$m_2 = L^2/3$ (for $D=1$). Finally we neglect correlations between $P$ and $m_2$ and find with the definition of (\ref{eq:ci})
that the compactness index of a thermal cloud of weakly interacting oscillators $\zeta = 3$ (for $D=1$), $\zeta = 2\pi \approx 6.28$ (for $D=2$) and $\zeta = (4\pi/3)^{2/3}5/3 \approx 4.33$ 
(for $D=3$). Such a result can be straightforwardly used for the KG lattice.
For norm density distributions of the DNLS model a $W$-dependent correction can be expected, however the numerical data show
that this is not of central importance. 
If we assume that density distributions experience large gaps between isolated fragments of the wave packet, then the compactness
index will be lowered down from its equipartition value. In particular, for the above discussed case of selftrapping, we expect that at least a part of the initial
state stays localized, while another part might spread. Then the second moment $m_2$ is expected to grow, the participation number $P$ will stay approximately constant,
and consequently the compactness index $\zeta$ will drop substantially down from its equipartition value.

In order to probe the spreading, we can also compute higher order moments $m_{\eta} = \sum_{\nu} (\nu - \bar{\nu})^{\eta} n_{\nu}$.
In particular the kurtosis $\gamma  = m_4/m_2^2 -3$ is useful as an indicator of the overall shape of the probability distribution profile.
Large values correspond to profiles with sharp peaks and long extending tails. Low values are obtained for profiles with rounded/flattened peaks and steeper tails. For example, the Laplace distribution has $\gamma = 3$, while a compact uniform distribution has $\gamma = -1.2$ \cite{dodge03}.

\section{Computing spreading wave packets: collecting evidence }
\label{sec5}

We will present results on long time numerical simulations. We therefore first
discuss the methods and particularities of our computations (see \cite{skkf09} for more details). For both models,
symplectic integrators were used. These integration schemes replace the original
Hamiltonian by a slightly different (and time-dependent) one, which is integrated exactly.  The
smaller the time steps, the closer both Hamiltonians. Therefore, the computed
energy (or norm) of the original Hamiltonian function will fluctuate in time,
but not grow. The fluctuations are bounded, and are due to the fact, that the
actual Hamiltonian which is integrated, has slightly different energy.

Another possible source of errors is the roundoff procedure of the actual
processor, when performing operations with numbers. Sometimes it is referred
to as `computational noise' although it is exactly the opposite, i.~e.~purely
deterministic and reproducible.The influence of roundoff
errors on the results was discussed in \cite{skkf09}.

The KG chain was integrated with the help of a symplectic integrator of order
$\mathrm{\cal{O}}(\tau^4)$ with respect to the integration time step $\tau$,
namely the SABA$_2$ integrator with corrector (SABA$_2$C), introduced in
\cite{LR01}.  A brief presentation of the integration scheme, as well as its
implementation for the particular case of the KG lattice (\ref{RQKG}) is given
in Appendix \cite{skkf09}. The SABA$_2$C integration scheme proved to be
very efficient for long integrations (e.~g.~up to $10^{10}$ time units) of
lattices having typically $N=1000$ sites, since it kept the required computational time to feasible
levels, preserving at the same time quite well the energy of the system. For
example, an integration time step $\tau=0.2$ usually kept the relative error
of the energy smaller than $10^{-4}$.

The DNLS chain was integrated with the help of the SBAB$_2$ integrator \cite{LR01} 
which introduces an error in energy conservation
of the order $\mathrm{\cal{O}}(\tau^2)$. The number of sites used in 
computations varied from $N=500$ to $N=2000$, in order to exclude finite size
effects in the evolution of the wave packets. For $\tau=0.1$ the
relative error of energy was usually kept smaller than $10^{-3}$. It is worth
mentioning that, although the SBAB$_2$ integrator and the commonly used
leap--frog integrator introduce errors of the same
order, the SBAB$_2$ scheme exhibits a better
performance since it requires less CPU time, keeping at the same time the
relative energy error to smaller values than the leap--frog scheme.

We remind that we order the NMs in space by increasing value of the center-of-norm coordinate
$X_{\nu}=\sum_l l A_{\nu,l}^2$.  We analyze normalized distributions $z_{\nu}
\geq 0$ using the second moment $m_2= \sum_{\nu}
(\nu-\bar{\nu})^2 z_{\nu}$, which quantifies the wave packet's degree of
spreading and the participation number $P=1 / \sum_{\nu} z_{\nu}^2$, which
measures the number of the strongest excited sites in $z_{\nu}$.  Here
$\bar{{\nu}} = \sum_{\nu} \nu z_{\nu}$.  For DNLS we follow norm density
distributions $z_{\nu}\equiv |\phi_{\nu}|^2/\sum_{\mu} |\phi_{\mu}|^2$.  For
KG we follow normalized energy density distributions $z_{\nu}\equiv
E_{\nu}/\sum_{\mu} E_{\mu}$ with $E_{\nu} =
\dot{A}^2_{\nu}/2+\omega^2_{\nu}A_{\nu}^2/2$, where $A_{\nu}$ is the amplitude
of the $\nu$th NM and $\omega^2_\nu=1+(\lambda_{\nu}+2)/W$.

\subsection{Single site excitations}
\label{sec51}

%&&&&&&&&&&&&&&&&&&&&&&&&&&&&&&&&&&&&&&&&&
\begin{figure}
\sidecaption
\includegraphics[angle=0,width=0.64\columnwidth]{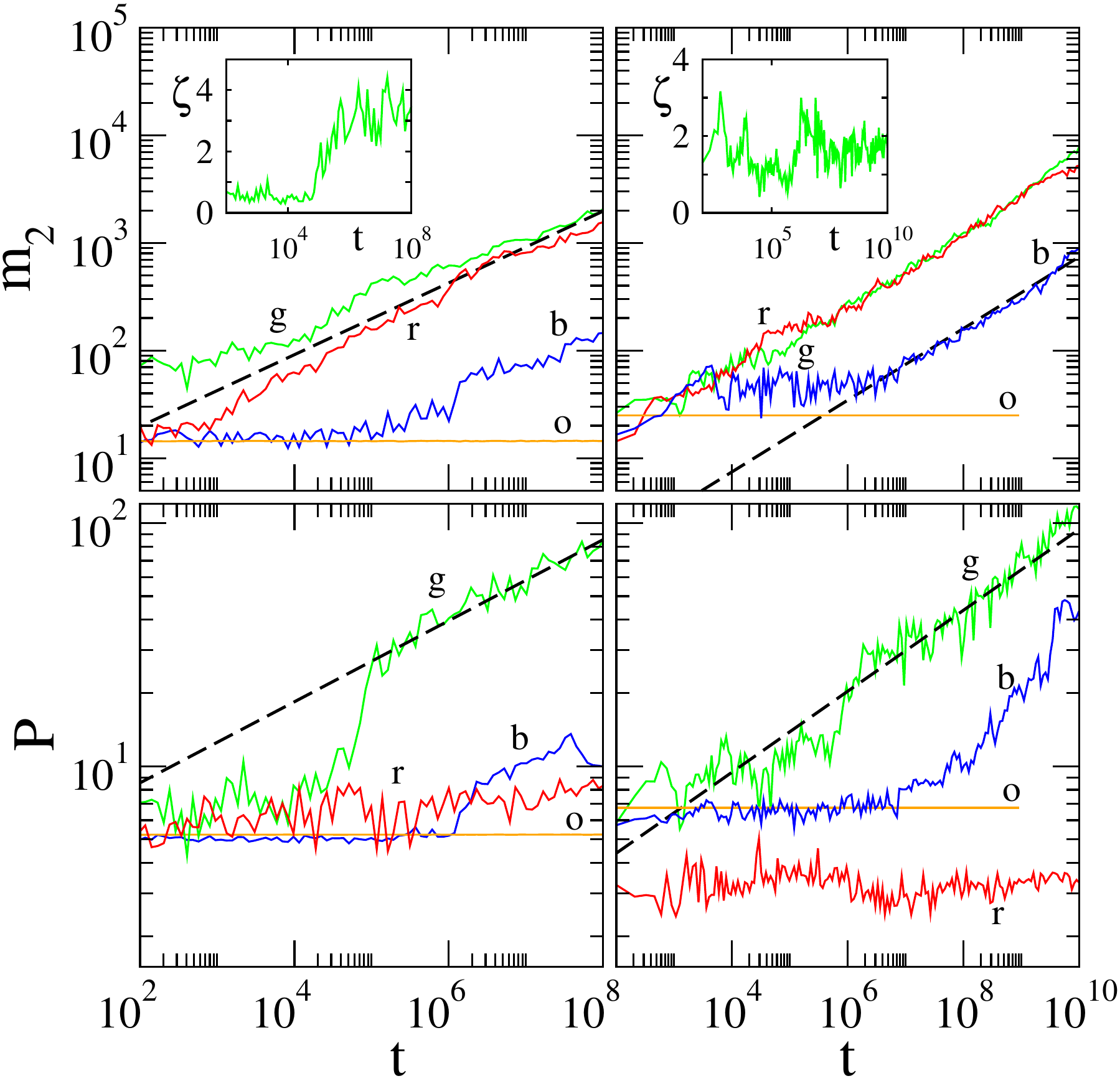}
\caption{Single site excitations.  $m_2$ and $P$ versus time in
log--log plots.  Left plots: DNLS with $W=4$, $\beta=0,0.1,1,4.5$ [(o),
orange; (b), blue; (g) green; (r) red].  Right plots: KG with $W=4$ and
initial energy $E=0.05,0.4,1.5$ [(b) blue; (g) green; (r) red].  
(o) Orange curves for the linear equations of motion, where
the term $u_l^3$ in (\ref{KG-EOM}) was absent.  The disorder realization is
kept unchanged for each of the models.  Dashed straight lines guide the eye
for exponents 1/3 ($m_2$) and 1/6 ($P$) respectively. Insets: the compactness
index $\zeta$ as a function of time in linear--log plots for $\beta=1$ (DNLS)
and $E=0.4$ (KG). Adapted from \cite{skkf09}}
\label{fig1}
\end{figure}
%&&&&&&&&&&&&&&&&&&&&&&&&&&&&&&&&&&&&&&&&&
We first show results for single site excitations from \cite{skkf09} in Fig.\ref{fig1} with $W=4$, $n=1$
and $\epsilon_l=0$ at the excitation site. We plot the time dependence of the second moment $m_2$, the participation number $P$ and
the compactness index $\zeta$.
Let us discuss the DNLS model (left plots in Fig.\ref{fig1}). The outcome for the KG model (right plots in
Fig.\ref{fig1}) is impressively similar.
For $\beta=0$ both $m_2$ and $P$ are constant in time respecting Anderson localization.
For $\beta=0.1$ the quantities fluctuate around their $\beta=0$ values up to $t \approx 10^6$ and start to grow for larger times,
signaling a spreading of the wave packet and a departure from Anderson localization. For $\beta=1$ the spreading is observable
already at shorter times. Note that the compactness index $\zeta$ tends to its equipartition value $\zeta \approx 3$. 
Finally, deep in the selftrapping regime $\beta = 4.5$ the participation number $P$ stays finite, since a significant part of the wave packet
stays localized. Nevertheless, a part of the wave packet spreads with the second moment $m_2$ again growing in time.
This growth appears to follow a subdiffusive law $m_2 \sim t^{1/3}$. 
For single site excitations strong chaos is not expected to be observed (\ref{sser}).
Note that the observed crossover from weak chaos to selftrapping happens for $1 < \beta < 4.5$ which compares well with the 
expected value $\beta \approx 4$ using (\ref{sser}).
Repeating the simulations for 20 different disorder realizations in the regime of weak chaos, with subdiffusive growth of $m_2 \sim t^{\alpha}$ starting around $t=10^2$,
an average $\langle \log_{10} m_2 \rangle$ is obtained. Its time dependence over 6 (DNLS) up to 8 (KG) decades in time was fitted with a power law, yielding
$\alpha = 0.33 \pm 0.02$ for DNLS and $\alpha=0.33 \pm 0.05$ for KG \cite{skkf09}.

\subsection{Single mode excitations}
\label{sec52}

\begin{figure}
\sidecaption
\includegraphics[angle=0,width=0.64\columnwidth]{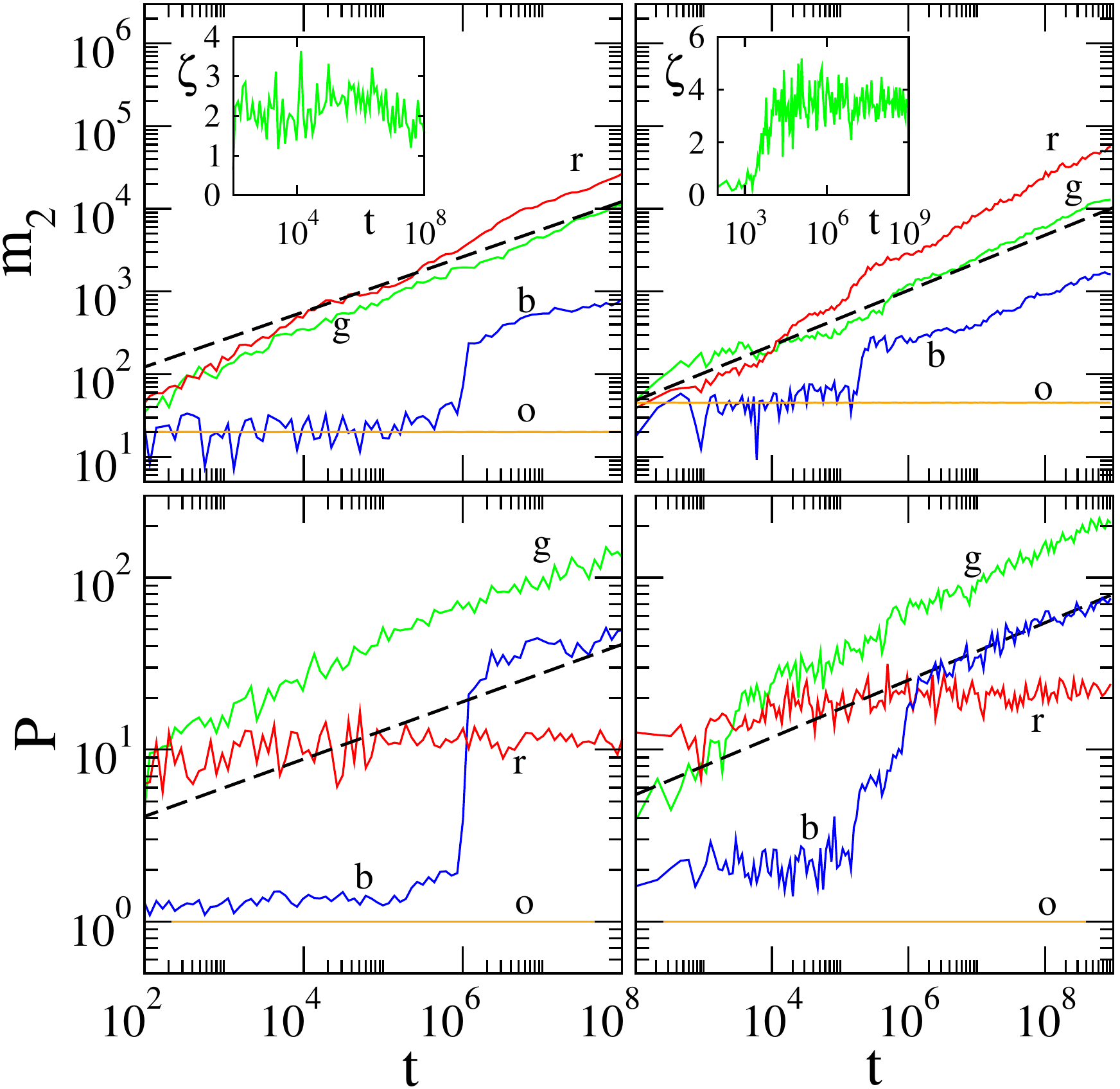}
\caption{Single mode excitations.  $m_2$ and $P$ versus time in
log--log plots.  Left plots: DNLS with $W=4$, $\beta=0,0.6,5,30$ [(o) orange;
(b) blue; (g) green; (r) red].  Right plots: KG with $W=4$ and initial energy
$E=0.17,1.1,13.4$ [(b) blue; (g) green; (r) red].  (o) Orange curves
for the linear equations of motion, where the term
$u_l^3$ in (\ref{KG-EOM}) was absent.  The disorder realization is kept
unchanged for each of the models.  Dashed straight lines guide the eye for
exponents 1/3 ($m_2$) and 1/6 ($P$) respectively. Insets: the compactness
index $\zeta$ as a function of time in linear--log plots for $\beta=5$ (DNLS)
and $E=1.1$ (KG). Adapted from \cite{skkf09}}
\label{fig2}
\end{figure}

For single mode excitations we find a similar outcome, but with rescaled
critical values for the nonlinearity strength which separate the different
regimes.  Examples are shown in
Fig.\ref{fig2} for $W=4$, $n=1$ and $\lambda_{\nu} \approx 0$ for the initially excited mode.
As in the case of single site excitations presented in Fig.~\ref{fig1}, the
compactness index $\zeta$ plotted in the insets in Fig.~\ref{fig2}
remains practically constant for excitations avoiding selftrapping, attaining the
values $\zeta=1.5 $ at $t=10^8$ for the DNLS model and $\zeta=3.3
$ at $t=10^9$ for the KG chain.  
%&&&&&&&&&&&&&&&&&&&&&&&&&&&&&&&&&&&&&&&&&
%&&&&&&&&&&&&&&&&&&&&&&&&&&&&&&&&&&&&&&&&&
According to (\ref{smer}) weak chaos is realized for $\beta < 8$, and selftrapping should set in for $\beta \approx 80$. 
The order of magnitude of these thresholds are well captured by the computations. Moreover, pay attention that the
second moment growth in the strong chaos and selftrapping regimes appears to be subdiffusive $m_2 \sim t^{\alpha}$ but
with an exponent $\alpha > 1/3$. It is hard to make quantitative conclusions about the observed subdiffusive growth laws.
For that to be achieved, we need to perform averaging over disorder realizations.

\begin{figure}
\sidecaption
\includegraphics[angle=0,width=0.6\columnwidth]{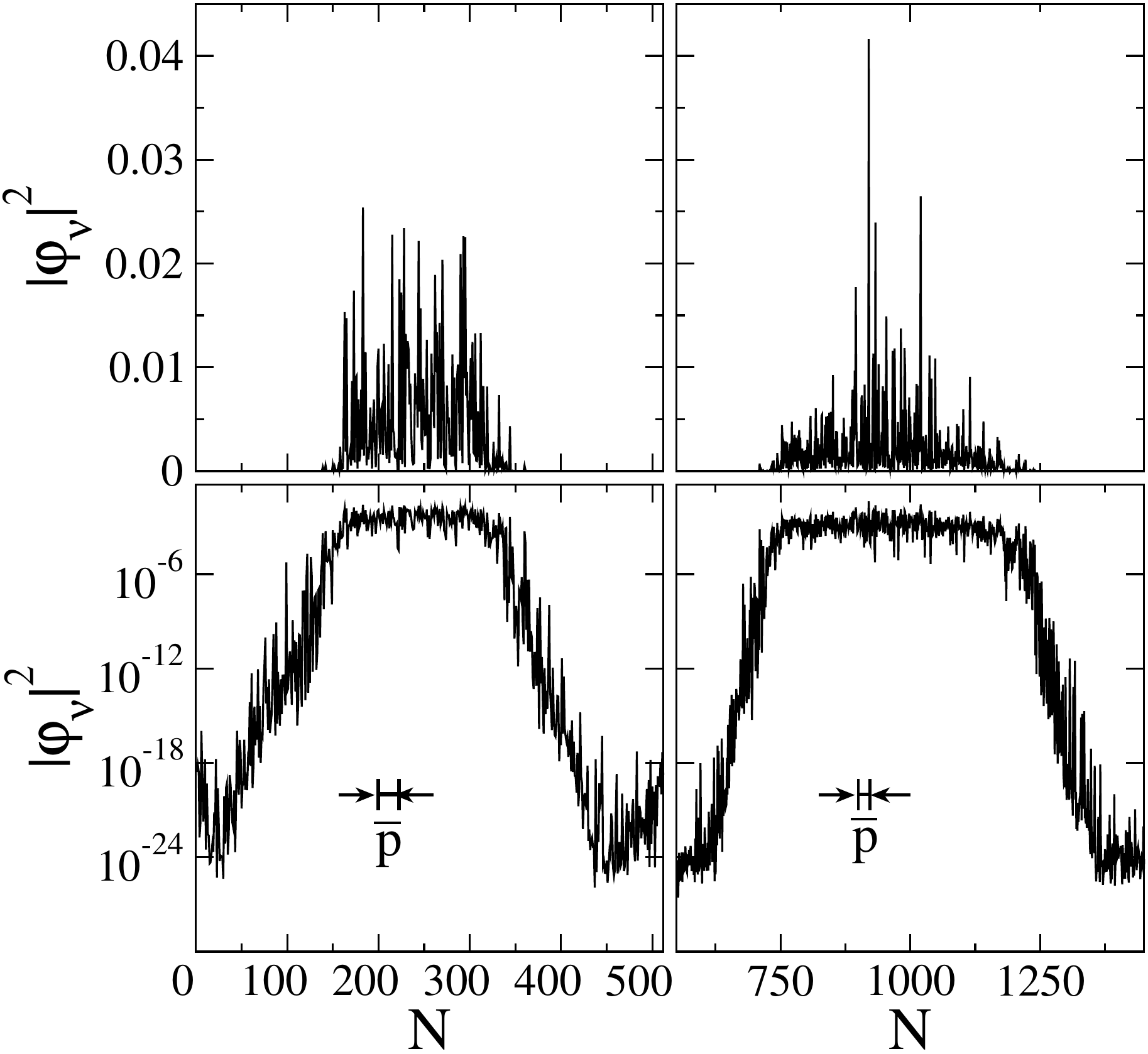}
\caption{Norm density distributions in the NM space at time
$t=10^8$ for the initial excitations of the DNLS model shown
in the left plots of Figs.~\ref{fig1} and \ref{fig2}. Left plots: single
site excitation for $W=4$ and $\beta=1$. Right plots: single mode
excitation for $W=4$ and $\beta=5$.  $|\phi_\nu|^2$ is plotted in
linear (logarithmic) scale in the upper (lower) plots. The maximal mean value
of the localization volume of the NMs $\overline{p}\approx22$ (shown
schematically in the lower plots) is much smaller than the length over
which the wave packets have spread.  Adapted from \cite{skkf09}}
\label{fig3}
\end{figure}

The final norm density distribution for the
DNLS model is plotted in Fig.~\ref{fig3} for both single site and single mode excitations. 
The wave packets grow substantially beyond the maximum size dictated by Anderson localization. 
The wave packets show thermal fluctuations, which are barely seen on logarithmic scales (bottom plots). On these logarithmic
scales the remnants of Anderson localization are nicely observed - these are the exponential tails at the edges of the wave packet.
As time increases, the wave packet spreads further, and the exponential tails are pushed into outer space. 
The average
value $\overline{\zeta}$ of the compactness index over 20 realizations
at $t=10^8$ for the DNLS model with $W=4$ and $\beta=5$ was found to be
$\overline{\zeta}=2.95 \pm 0.39$ \cite{skkf09}. 
The slow subdiffusive spreading is apparently sufficient for a rough thermalization of the wave packet and the formation of exponential
Anderson-localized tails.
%&&&&&&&&&&&&&&&&&&&&&&&&&&&&&&&&&&&&&&&&&
%&&&&&&&&&&&&&&&&&&&&&&&&&&&&&&&&&&&&&&&&&

The observed start of the growth of $m_2$ for weak nonlinearity at times $t\sim 10^6$ in Figs.\ref{fig1},\ref{fig2} can, but must not signal a qualitative change in the
dynamics. Indeed, relaunching wave packets which have spread already substantially (at somewhat stronger nonlinearity) will yield similar transient curves from a constant
to a growing function \cite{skkf09}. Therefore an alternative explanation for the observed transients is a large characteristic diffusion time scale for a given initial state, which
will be observable in the time-dependence of the second moment only beyond a corresponding, potentially large, time.

\subsection{Normal mode dephasing}
\label{sec53}

\begin{figure}
%\sidecaption
\includegraphics[angle=0,width=0.8\columnwidth]{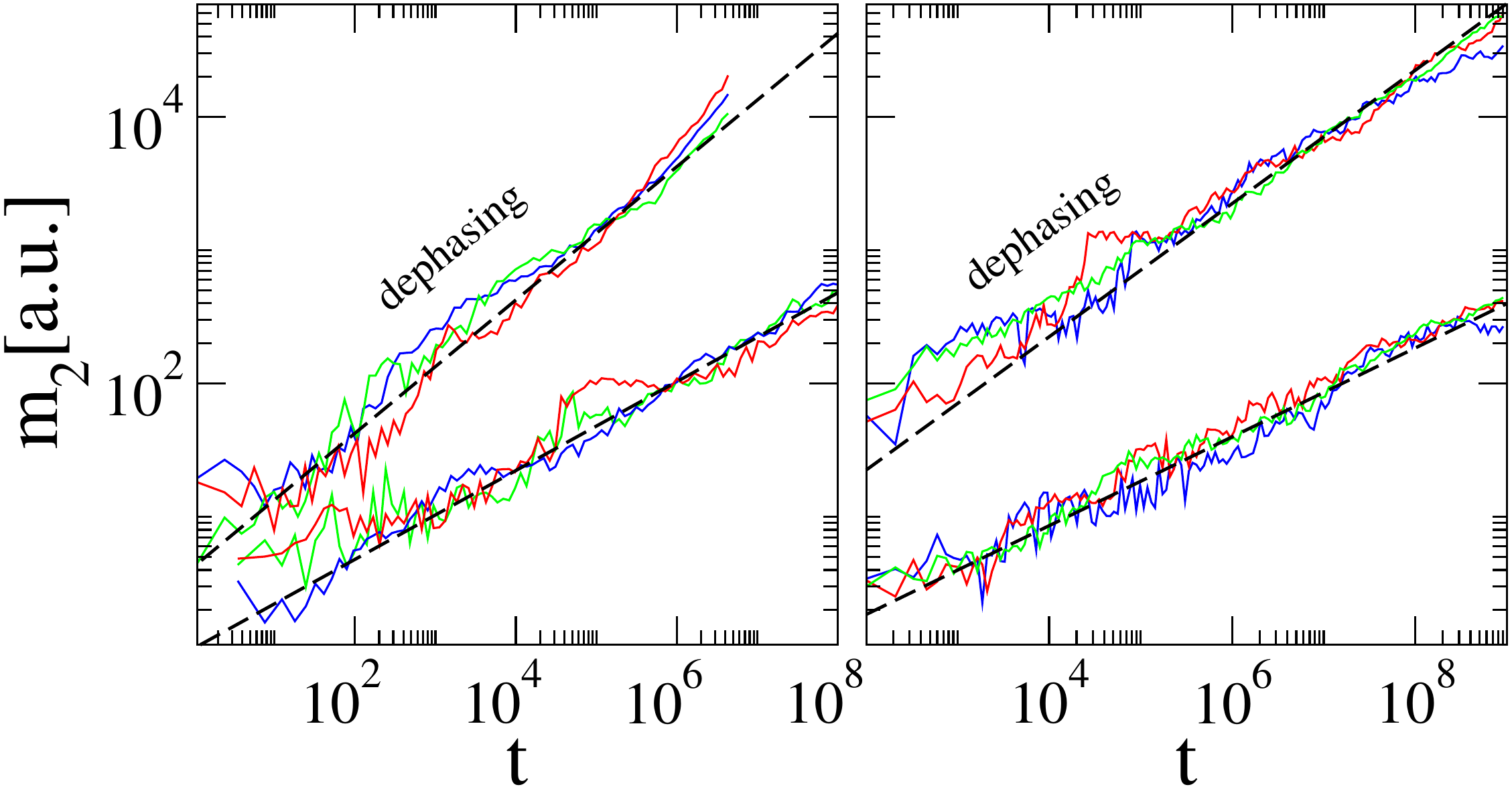}
\caption{Single site excitations.  $m_2$ (in arbitrary units)
versus time in log--log plots for different values of $W$.  Lower
set of curves: plain integration (without dephasing); upper set of curves:
integration with dephasing of NMs.  Dashed
straight lines with exponents 1/3 (no dephasing) and 1/2 (dephasing) guide the
eye.  Left plot: DNLS, $W=4$, $\beta=3$ (blue); $W=7$, $\beta=4$ (green);
$W=10$, $\beta=6$ (red).  Right plot: KG, $W=10$, $E=0.25$ (blue) , $W=7$,
$E=0.3$ (red) , $W=4$, $E=0.4$ (green).  The curves are shifted vertically in
order to give maximum overlap within each group. Adapted from \cite{skkf09} }
\label{fig4}
\end{figure}

For single site
excitations the exponent $\alpha \approx 1/3$ does not appear to depend on $\beta$ in the
case of the DNLS model or on the value of $E$ in the case of KG, as shown in 
Fig.\ref{fig4}.  What is the origin of the observed slow subdiffusive process? If the dynamics is accompanied by
an enforced randomization of phases of the variables $\psi_l$ in real space (respectively the phases of the oscillators of the KG model)
then even for the linear wave equation Anderson localization is destroyed, and instead a process of normal diffusion with $m_2 \sim t$ 
is observed \cite{krgrsf13}, which is much faster than the observed subdiffusion. The above tests of the linear wave equation in Fig.\ref{fig1},\ref{fig2} also
show that the numerical scheme is correctly reproducing Anderson localization. Could it then be that the relative phases of the participating normal modes are
randomized leading to the observed slow spreading? 
We test this by enforcing decoherence of NM phases.  Each 100 time units on
average 50\% of the NMs were randomly chosen, and their phases were shifted by
$\pi$ (DNLS). For the KG case we change the signs of the corresponding NM
momenta.  We obtain $m_2\sim t^{1/2}$ \cite{skkf09} (see Fig.\ref{fig4}).  
This is also a subdiffusive process, yet faster than the observed one with $\alpha=1/3$. Therefore, we can expect that
the numerical integration is rather accurate.
When the NMs dephase completely, the exponent $\tilde{\alpha}=1/2$, {\sl
contradicting} numerical observations {\sl without dephasing}.  Thus, not all
NMs in the packet are randomizing their phases quickly, and dephasing is at best a partial outcome.
%&&&&&&&&&&&&&&&&&&&&&&&&&&&&&&&&&&&&&&&&&
%&&&&&&&&&&&&&&&&&&&&&&&&&&&&&&&&&&&&&&&&&

\section{Nonlinear diffusion}
\label{sec6}

The integrable equations of the secular normal form preserve Anderson localization. It is therefore tempting to assume that the observed departure from
Anderson localization is due to nonintegrability and chaos. Indeed, assume that a wave packet with $V \gg 1$ NMs is excited. Trap it and replace the exponential edges
(see Fig.\ref{fig3}) by fixed walls (boundaries). Then continue to evolve the equations. We are dealing for sure with a nonintegrable system with many degrees of 
freedom (DOF). Will the dynamics be chaotic or regular? That depends on the number of DOF, and on the energy/norm density of the system. The question
touches the range of validity of the Kolmogorov-Arnold-Moser regime for persisting invariant tori with finite measure of a weakly perturbed integrable system.
To the best of our knowledge, no results are known which can help and guide our search. Yet in a huge body of molecular dynamical simulations of various systems,
a large enough number of degrees of freedom usually ensures equipartition down to extremely small temperatures (energy densities), with the only consequence that
decoherence time scales increase with lowering the temperature. Let us therefore take the point that the dynamics inside the trapped wave packet is chaotic.
Then, as we will show below, a removing of the trap (the fixed walls) will inevitably lead to a spreading and increase of the wave packet size. Therefore the participating
number of DOF increases - linearly with the wave packet size. At the same time the densities drop - inversely proportional to the wave packet size.
The nonlinear terms in the equations of motion (\ref{RDNLS-EOM}),(\ref{KG-EOM}) become small compared to the linear ones. It is therefore tempting to skip the nonlinear
terms at some point. But if we skip them, we return to the linear wave equation, restore integrability, and recover Anderson localization. So then, for that enlarged
wave packet, we can again add trapping hard walls, but keep the nonlinear terms, and ask the question whether the dynamics inside the wave packet remains regular, or will be chaotic at large enough times. Again the experience of molecular dynamics tells that the dynamics will stay chaotic with high probability, but the decoherence times increase.
Therefore the possible flaw in the argument when dropping the nonlinear terms is the time scale. For sure, at weak enough nonlinearity, and up to some finite time,
the nonlinear terms can be neglected. But  how will that time scale with weak nonlinearity? If it stays finite, then the dropping of nonlinear terms will be incorrect for
large enough times. Which might be just the times at which we observe the slow subdiffusive wave packet spreading.

\subsection{ Measuring chaos}
\label{sec61}

Michaeli and Fishman studied the evolution of single site excitations for the DNLS model \cite{emsf12}. They considered the rhs of Eq.(\ref{NMeqchi})
as a function of time $i\dot{\chi}_{\nu} =F_{\nu}(t)$ for a mode $\nu=0$ which was strongly excited at time $t=0$. The statistical analysis of the time dependence of 
$F_0(t)$ shows a quick decay of its temporal correlations for spreading wave packets. Therefore the force $F_0(t)$ can be considered as a random noise function
on time scales relevant for the spreading process. This is a clear signature of chaos inside the wave packet.

Vermersch and Garreau (VG) \cite{bvjcg13} followed a similar approach for the DNLS model. They measured the time dependence of the participation number $P(t)$ of a spreading wave packet
(see e.g. the curves in Fig.\ref{fig1},\ref{fig2}). VG then extracted a spectral entropy, i.e. a measure of the number of participating frequencies
which characterize this time dependence. Spectral entropies are convenient measure to discriminate between regular and chaotic dynamics. VG concluded
that the dynamics of spreading wave packets {\it is} chaotic. They also measured short time Lyapunov exponents to support their conclusion.

 The long-time dependence of the largest  Lyapunov exponent $\Lambda$ as chaos strength indicators inside spreading wave packets for KG models was recently tested in \cite{csigsf13}. The crucial point
is that during spreading the energy density {\it is} decreasing, and therefore a weakening of the momentary chaos indicator is expected. Therefore $\Lambda (t)$ will
be not constant in time, but decrease its value with increasing time. Moreover, the calculation of Lyapunov exponents for integrable systems will also yield nonzero numbers
when integrating the system over any finite time. This is due to the method used - one evolves the original trajectory in phase space, and in parallel runs the
linearized perturbation dynamics of small deviations from the original trajectory in tangent space. Since this deviation is nonzero, any computer code
will produce nonzero estimates for the Lyapunov exponent at short times. The crucial point is that for integrable systems the long-time dependence of $\Lambda$
follows $\Lambda \sim 1/t$. This is also the result found in \cite{csigsf13} for the {\it linear} wave equation which obeys Anderson localization.
However the nonlinear case of wave packet spreading yields a dependence 
\begin{equation}
\Lambda (t) \sim \frac{1}{t^{1/4}} \gg \frac{1}{t}\;.
\label{mle}
\end{equation}
%&&&&&&&&&&&&&&&&&&&&&&&&&&&&&&&&&&&&&&&&&
\begin{figure*}
\includegraphics[width=\linewidth]{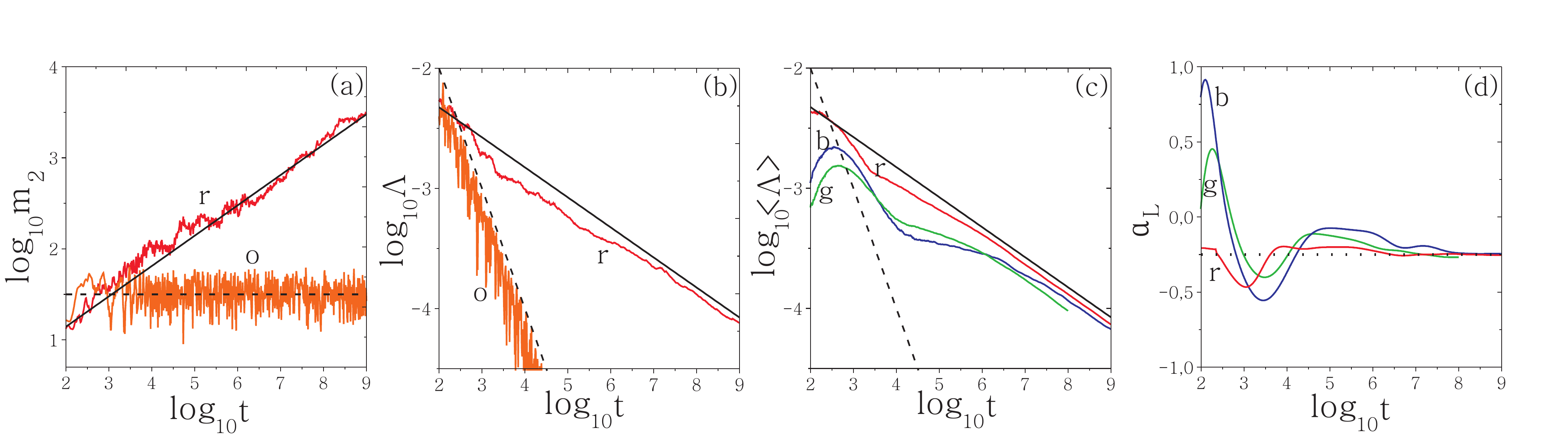}
\caption{(a) Time evolution of the second moment $m_2$
  for one disorder realization of an initially single site excitation
  with $E=0.4$, $W=4$ (Case I), in $\log - \log$ scale (red (r)
  curve). The orange (o) curve corresponds to the solution of the
  linear equations of motion, where the term $u_l^4$ in
  Eq.(\ref{RQKG}) is absent. Straight lines guide the eye for slopes
  $1/3$ (solid line) and $0$ (dashed line).  (b) Time evolution of the
  finite time maximum Lyapunov exponent $\Lambda$ (multiplied by 10
  for the orange (o) curve) for the trajectories of panel (a) in $\log
  - \log$ scale. The straight lines guide the eye for slope $-1$
  (dashed line), and $-1/4$ (solid line). (c) Time evolution of the
  averaged $\Lambda$ over 50 disorder realizations for the `weak
  chaos' cases I, II and III [(r) red; (b) blue; (g) green] (see text
  for more details). Straight lines guide the eye for slopes $-1$ and
  $-1/4$ as in panel (b). (d) Numerically computed slopes $\alpha_L$
  of the three curves of panel (c). The horizontal dotted line denotes
  the value $-1/4$. Adapted from \cite{csigsf13}}.
\label{fig5}
\end{figure*}
%&&&&&&&&&&&&&&&&&&&&&&&&&&&&&&&&&&&&&&&&&
In Fig.\ref{fig5}(a) we first show the result for a trajectory of a single
site excitation with total energy $E=0.4$ and $W=4$ (Case I).  We show the time dependence
of the second moment (red curve) and observe the expected subdiffusive
growth $m_2 \sim t^{1/3}$. The simulation of a single site excitation
in the absence of nonlinear terms (orange curve) corresponds to
regular motion and Anderson localization is observed. In
Fig.\ref{fig5}(b) we plot the time dependence of $\Lambda(t)$ for the
two cases of Fig.\ref{fig5}(a). At variance to the $t^{-1}$ decay for
the regular nonchaotic trajectory (orange curve), the observed decay
for the weak chaos orbit is much weaker and well fitted with
$\Lambda \sim t^{-1/4}$ (red curve).

These findings are further substantiated by averaging
$\log_{10} \Lambda$ over 50 realizations of disorder and extending to
two more weak chaos parameter cases with initial energy density
$\epsilon=0.01$ distributed evenly among a block of 21 central sites
for $W=4$ (case II) and 37 central sites for $W=3$ (case III). All
cases show convergence towards $\Lambda \sim t^{-1/4}$
(Fig.\ref{fig5}(c)). The curves are further analyzed by estimating 
their slope
$\alpha_L=\frac{d(\log_{10}\Lambda(t))}{d\log_{10}t}$. The
results in Fig.\ref{fig5}(d) underpin the above findings.

The authors of \cite{csigsf13} further compare  the obtained chaoticity time scales $1\Lambda$ with the time scales governing the
slow subdiffusive spreading and conclude, that the assumption about persistent and fast enough chaoticity needed for thermalization
and inside the wave packet is correct. The dynamics inside the spreading wave packet is chaotic, and remains chaotic up to the largest simulation times,
without any signature of a violation of this assumption for larger times (no visible slowing down).

A further very important result concerns the seeds of
deterministic chaos and their {\sl meandering} through the packet in
the course of evolution. Indeed, assume
that their spatial position is fixed. Then such seeds will act as
spatially pinned random force sources on their surrounding. The noise
intensity of these centers will decay in time. At any given time the
exterior of the wave packet is then assumed to be approximated by the
linear wave equation part which enjoys Anderson localization. However,
even for constant intensity it was shown \cite{aubry2009} that the
noise will not propagate into the system due to the dense discrete
spectrum of the linear wave equation. Therefore the wave packet can
only spread if the nonlinear resonance locations meander in space and
time.

The motion of these chaotic seeds was visualized by following the
spatial evolution of the deviation vector distribution (DVD) used for the computation of
the largest Lyapunov exponent \cite{csigsf13}. This vector tends to align with the most unstable direction
in the system's phase space. Thus, monitoring how its components on
the lattice sites evolve allow to identify the most chaotic spots. Large
DVD values tell at which sites the sensitivity on initial
conditions (which is a basic ingredient of chaos) is larger.

\begin{figure}
\sidecaption
\includegraphics[scale=.4]{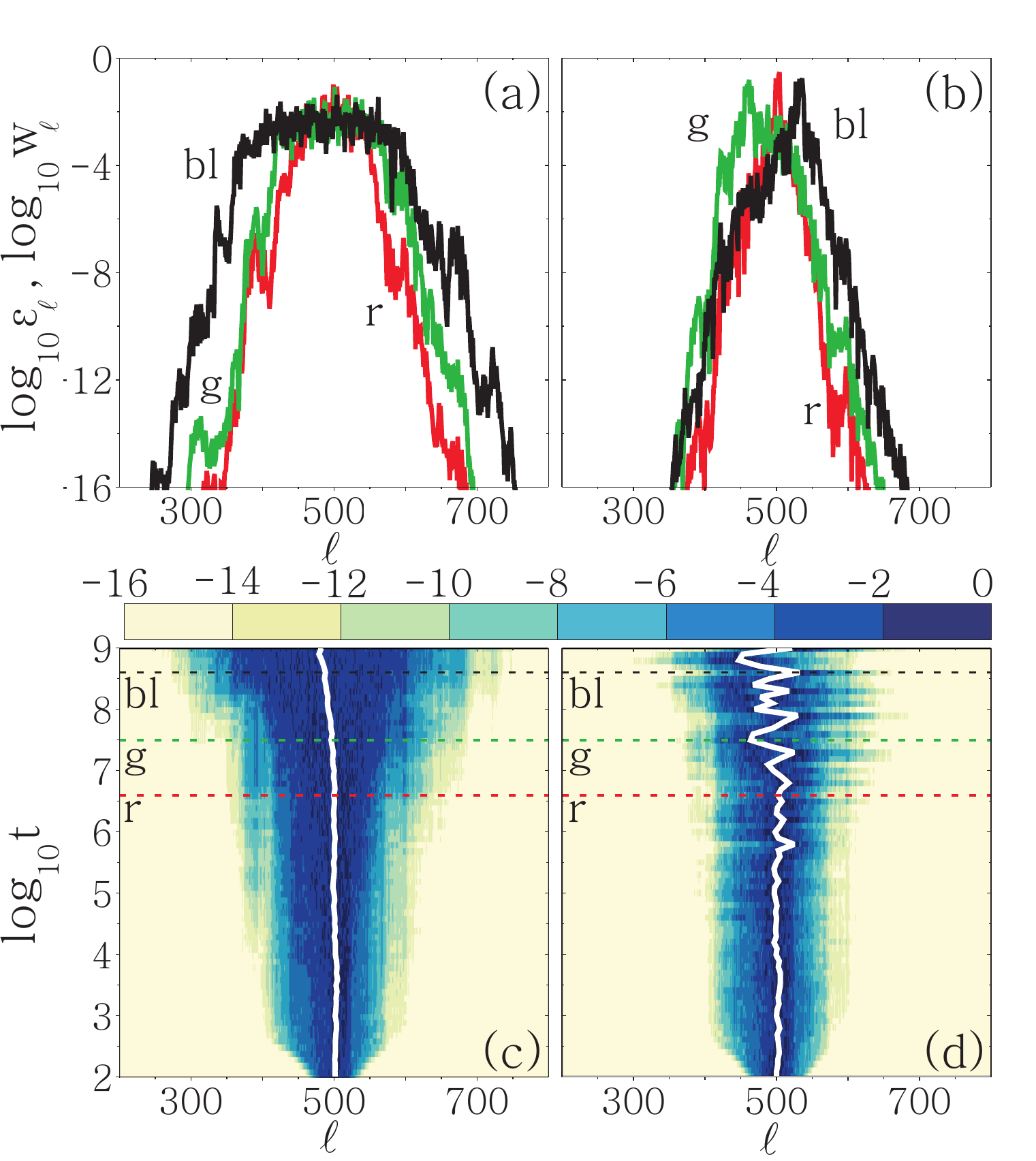}
\caption{The dynamics of an individual trajectory of
  case I. Normalized (a) energy ($\epsilon_l$) and (b) deviation
  vector ($w_l$) distributions at $t=4\times 10^{6}$, $t=3\times
  10^{7}$, $t=4 \times 10^8$ [(r) red; (g) green; (bl) black].  Time
  evolution of (c) the energy distribution and (d) the DVD for the
  realization of panel (a) in $\log_{10}$ scale.  The position of the
  distribution's mean position is traced by a thick white curve. The
  times at which the distributions of panels (a) and (b) are taken are
  denoted by straight horizontal lines in (c) and (d).  Adapted from \cite{csigsf13}}
\label{fig6}
\end{figure}

In Fig.\ref{fig6}(a) we plot the energy density distribution for an
individual trajectory of case I (cf. Fig.\ref{fig5}) at three different times
$t\approx10^6,\,10^7,\,10^8$ and in Fig.\ref{fig6}(b) the
corresponding DVD.  We observe that the energy densities spread more
evenly over the lattice the more the wave packet grows. At the same
time the DVD stays localized, but the peak positions clearly meander
in time, covering distances of the order of the wave packet width. The
full time evolution of the energy density and the DVD is shown in
Figs.\ref{fig6}(c,d) together with the track of the distribution's
mean position (central white curve). While the energy density
distribution shows a modest time dependence of the position of its
mean, the DVD mean position is observed to perform fluctuations whose
amplitude increases with time.
%&&&&&&&&&&&&&&&&&&&&&&&&&&&&&&&&&&&&&&&&&
%&&&&&&&&&&&&&&&&&&&&&&&&&&&&&&&&&&&&&&&&&

\subsection{Effective noise theory}
\label{sec62}

Having established that the dynamics inside a spreading wave packet is chaotic, let us proceed to construct an effective noise theory for spreading.
For that we replace the time dependence on the rhs of Eq. (\ref{NMeqchi}) by a random function in time:
\begin{equation}
i\dot{\chi}_{\nu} = F(t)\;,\;\langle F \rangle = 0\;,\; \langle F^2(t) \rangle = f^2;.
\label{ent1}
\end{equation}
Assume that the norm density (norm per site/mode) inside the wave packet is $n$.
What are the consequences? Consider a normal mode $\mu$ which is outside the wave packet, but in a boundary layer of one of its edges. For
obvious reasons the boundary layer thickness is of the order of $V$.  The equation of motion for this mode is given by (\ref{ent1}).
At some initial time $t_0$ assume that the norm of the considered mode is close to zero $|\chi_{\mu}(t_0)|^2 = n_{\mu}(t_0) \ll n$.
Then the solution of the stochastic differential equation (\ref{ent1}) is yielding a diffusion process in norm/energy space of the considered
NM:
\begin{equation}
n_{\mu} (t) \sim f^2 t\;.
\label{ent2}
\end{equation}
The considered mode will reach the packet norm $n$ after a time $T$ whose inverse will be proportional to the
momentary diffusion rate of the wave packet $D \sim 1/T$:
\begin{equation}
D \sim \frac{f ^2}{n} \;.
\label{ent3}
\end{equation}
Let us estimate the variance $f$ for the nonlinear wave equation. It follows from estimating the absolute value of the rhs of (\ref{NMeqchi}) 
which corresponds to the absolute value of the stochastic force $F(t)$ in (\ref{ent1}). We find
that
$f \sim \beta n^{3/2} \langle I \rangle$. With that, we arrive at $D \sim (\beta n \langle I \rangle )^2$. 
The main point here is that the diffusion coefficient is proportional to $n^2$, therefore the more the packet spreads, the lower its density, and the smaller $D$.
We obtain a time-dependent diffusion coefficient, and a tendency to spread slower than within a normal diffusion process. 
What are the consequences? A quick first argumentation line is to observe that
the second moment $m_2$ of a wave packet is inverse proportional to its squared norm density $m^2 \sim 1/n^2$. At the same time it should obey
$m_2 \sim D t$. Since $D \sim 1/m_2$ it follows $m_2 \sim t^{1/2}$. 

The second way is to write down a nonlinear diffusion equation \cite{NLD} for the norm density distribution
(replacing the lattice by a continuum for simplicity, see also \cite{ark10}):
\begin{equation}
\partial_t n = \partial_{\nu}(D \partial_{\nu} n) \;,\; D \sim n^{\kappa}\;.
\label{ent4}
\end{equation}
The solution $n(\nu,t)$ obeys the scaling $n(\nu,t/a) = bn(c\nu,t)$  with $b=c=a^{1/(\kappa+2)}$ if $n(\nu \pm \infty,t) \rightarrow 0$. Therefore the second moment
\begin{equation}
m_2 \sim t^{\alpha}\;,\;\alpha = \frac{2}{\kappa+2}\;.
\label{ent4b}
\end{equation}
Notably an explicit self-similar solution was calculated by Tuck in 1976 \cite{tuck76} which has
the following spatial profile:
\begin{equation}
n(\nu)=\left( B - \frac{\kappa \nu^2}{2(\kappa+2)}\right)^{1/\kappa}\;.
\end{equation}
Here $B$ is an integration constant (see also \cite{nonlineardiffusion}).

With $\kappa=2$ we obtain the subdiffusive law $m_2 \sim t^{1/2}$ again. We do arrive at a subdiffusive spreading. 
Note that the above nonlinear diffusion equation can be derived through a master equation and a Fokker-Planck equation for both norm and energy densities \cite{dmb14}, or Boltzmann equations \cite{gsamf13}.
However the exponent is $1/2$ {\sl and not} $1/3$.
Furthermore, recall that an enforcing of the randomization of NM phases during the spreading {\sl does yield the exponent} $1/2$. Therefore, we are on the right track -
enforcing the assumption of random NM phases, both numerics and effective noise theory approaches coincide. What is then the reason for the even slower subdiffusion
with $\alpha=1/3$ ? We recall that perturbation theory in Sec.\ref{sec42} leads to a probability $\mathcal{P}$ of a given NM being resonant which is 
small for small densities (\ref{resonance}): 
$\mathcal{P}_{\nu} |_{\beta n \rightarrow 0} 
\rightarrow \frac{\sqrt{3} V^3 \langle I \rangle}{\sqrt{2\pi} \Delta } \beta n$. In case when this probability is equal to one, the above diffusion constant assumption 
should make sense, since in that case every degree of freedom participating in the wave packet evolves chaotically, i.e. randomly in time. In the case when the resonance probability is zero, perturbation theory should be applicable, the resonance normal form from Sec.\ref{sec41}
yields Anderson localization, and spreading should stop. In that case $f=0$ and then $D=0$. In best traditions of phenomenology we assume that another factor is missing in the
expression of $f$. This factor shall be a function of $\mathcal{P}$ such that the factor becomes one when $\mathcal{P}=1$ and zero when
$\mathcal{P}=0$. The simplest prefactor is $\mathcal{P}$ itself. Let us test whether this works (recalling $d=\Delta / V$):
\begin{equation}
f \sim \mathcal{P} \beta n^{3/2} \langle I \rangle\;,\; D \sim (\mathcal{P} \beta n \langle I \rangle )^2\;,\; \mathcal{P} = 1-{\rm e}^{-C\beta n}\;,\;
C=\frac{\sqrt{3} V^2 \langle I \rangle}{\sqrt{2\pi} d }\;.
\label{ent5}
\end{equation} 
Then the solution of the nonlinear diffusion equation (\ref{ent4}) reads
\begin{eqnarray}
m_2 &\sim& (\beta \langle I \rangle V)^{4/3} d^{-2/3} t^{1/3} \;,\; C\beta n \ll  1\; : \; {\rm weak \; chaos}\;,
\label{ent6a}
\\
m_2 &\sim& \beta \langle I \rangle t^{1/2}\;\;\;\;\;\;\;\;\;\;\;\;\;\;\;\;\;\;\;\;\;\;\;,\; C\beta n \gg  1\; : \; {\rm strong \; chaos}\;.
\label{ent6b}
\end{eqnarray}
We arrived at a construction which results in the correct weak chaos exponent $\alpha = 1/3$ \cite{fks08}. We also predict that there must be an intermediate regime
of strong chaos for which $\alpha = 1/2$ - {\sl without any enforcing of the randomization of NM phases} \cite{sf10}.  It has to be intermediate, since with an assumed further
spreading of the wave packet, the density $n$ will decrease, and at some point satisfy the weak chaos condition (\ref{ent6a}) instead of the strong chaos condition (\ref{ent6b}).
Therefore, a potentially long lasting regime of strong chaos has to cross over into the asymptotic regime of weak chaos \cite{sf10}. That crossover is not a sharp one in
the time evolution of the wave packet. It might take several orders of magnitude in time to observe the crossover. Therefore, instead of fitting the numerically
obtained time dependence $m_2(t)$ with power laws, it is much more conclusive to compute derivatives $d\langle \log_{10} m_2 \rangle / d \log_{10} t$ in order
to identify a potentially long lasting regime of strong chaos, crossovers, or the asymptotic regime of weak chaos.

The conditions for weak and strong chaos in (\ref{ent6a},\ref{ent6b}) match those in Eqs. (\ref{sser}-\ref{wper}) if the constant $C$ is replaced by $1/d$. Although this is not
strictly correct according to Eq. (\ref{ent5}), numerical data \cite{fks08} suggest that both estimates yield the same order of magnitude in a wide range of weak and intermediate disorder
strength.

\subsection{Generalizations}
\label{sec63}

Let us consider $\vec{D}$-dimensional lattices with
nonlinearity order $\sigma > 0$:
\begin{equation}
i\dot{\psi_{\vec{l}}}= \epsilon_{\vec{l}} \psi_{\vec{l}}
-\beta |\psi_{\vec{l}}|^{\sigma}\psi_{\vec{l}}
-\sum\limits_{\vec{m}\in
D(\vec{l})}\psi_{\vec{m}}\;.
\label{RDNLS-EOMG}
\end{equation}
Here $\vec{l}$ denotes a $\vec{D}$-dimensional lattice vector with
integer components, and $\vec{m}\in
D(\vec{l})$ defines its set of nearest neighbor lattice sites.
We assume that (a) all NMs are spatially localized (which can be obtained for strong
enough disorder $W$), (b) the property $\mathcal{W}(x \rightarrow 0) \rightarrow const \neq 0$
holds, and (c) the probability of resonances on the edge surface of a wave packet is tending
to zero during the spreading process. A wavepacket with average norm $n$ per excited mode has a second moment
$m_2 \sim 1/n^{2/\vec{D}}$. The nonlinear frequency shift is proportional to $\beta n^{\sigma/2}$.
The typical localization volume of a NM is still denoted by $V$, and the average spacing by $d$.

Consider a wave packet with norm density $n$ and volume $L < V$. A straightforward generalization of
the expected regimes of spreading leads to the following:
\begin{eqnarray}
\beta n^{\sigma/2} \left( \frac{L}{V}\right)^{\sigma/2} V   < \Delta \;:\; {\rm weak\; chaos}\;,
\nonumber
\\
\beta n^{\sigma/2} \left( \frac{L}{V}\right)^{\sigma/2} V > \Delta \;:\; {\rm strong\; chaos}\;,
\nonumber
\\
\beta n^{\sigma/2} > \Delta \;:\; {\rm selftrapping}\;.
\nonumber
\end{eqnarray}
The regime of strong chaos, which is located between selftrapping and weak chaos,
can be observed only if 
\begin{equation}
L > L_c = V^{1-2/\sigma}\;,\; n > n_c = \frac{V}{L} \left( \frac{d}{\beta}\right)^{2/\sigma}\;.
\end{equation}
For $\sigma =2$ we need $L>1$, for $\sigma \rightarrow \infty$ we need $L > V$,
and for $\sigma < 2$ we need $L \geq 1$. Thus the regime of strong chaos can be observed
e.g. in a one-dimensional system with a single site excitation and $\sigma < 2$.

If the wave packet size $L > V$ then the conditions for observing different regimes simplify to
\begin{eqnarray}
\beta n^{\sigma/2} < d \;:\; {\rm weak\; chaos}\;,
\nonumber
\\
\beta n^{\sigma/2} > d \;:\; {\rm strong\; chaos}\;,
\nonumber
\\
\beta n^{\sigma/2} > \Delta \;:\; {\rm selftrapping}\;.
\nonumber
\end{eqnarray}
The regime of strong chaos can be observed if 
\begin{equation}
n > n_c = \left( \frac{d}{\beta}\right)^{2/\sigma}\;.
\end{equation}

Similar to the above we obtain a diffusion
coefficient
\begin{equation}
D \sim \beta^2 n^{\sigma} (\mathcal{P}(\beta n^{\sigma/2}))^2
\;.
\label{ggeneralizeddiffusion}
\end{equation}
In both regimes of strong and weak chaos the spreading is subdiffusive
\cite{fks08,sf10}:
\begin{eqnarray}
m_2 \sim (\beta^2 t)^{\frac{2}{2+\sigma \vec{D}}}\;,\;{\rm strong}\;{\rm chaos}\;,
\label{sigma_strong}
\\
m_2 \sim (\beta^4 t)^{\frac{1}{1+\sigma \vec{D}}}\;,\;{\rm weak}\;{\rm chaos}\;.
\label{sigma_weak}
\end{eqnarray}
Note that the strong chaos result was also obtained within a Boltzmann theory approach \cite{gsamf13}.

Let us calculate the number of resonances in the wave packet volume ($N_{RV}$) and on its surface ($N_{RS}$)
in the regime of weak chaos:
\begin{equation}
N_{RV} \sim \beta n^{\sigma/2-1}\;,\; N_{RS} \sim \beta n^{\frac{\vec{D}(\sigma-2)+2}{2\vec{D}}}\;.
\end{equation}
We find that there is a critical value of nonlinearity power $\sigma_c = 2$ such that
the number of volume resonances grows for $\sigma < \sigma_c$ with time, drops for $\sigma > \sigma_c$ and
stays constant for $\sigma=\sigma_c$. Therefore subdiffusive spreading is expected to be more effective for $\sigma < \sigma_c$.

We also find that the number of surface resonances will grow with time for
\begin{equation}
\vec{D} > \vec{D_c}=\frac{1}{1-\sigma/2}\;,\; \sigma < 2\;.
\end{equation}
Therefore, for these cases, the wave packet surface might not stay compact. Instead surface resonances may lead to
a resonant leakage of excitations into the exterior. This process can increase the surface area, and therefore
lead to even more surface resonances, which again increase the surface area, and so on. 
The wave packet could even fragmentize, perhaps get a fractal-like structure, and
lower its compactness index. The spreading of the wave packet would speed up, but not anymore be due to
pure incoherent transfer, instead it might even become a complicated mixture of incoherent and coherent transfer processes.

\section{Testing the predictions}
\label{sec7}

In this chapter we will review numerical results which test the above predictions. We will in particular discuss the crossover from strong to weak chaos,
the scaling of the density profiles, the impact of different powers of nonlinearity and different lattice dimensions, and the temperature dependence of
heat conductivity. We will also extend the discussion to
quasiperiodic Aubry-Andre localization, dynamical localization with kicked rotors, Wannier-Stark localization, and time-dependent ramping protocols of the nonlinearity
strength which speed up the slow subdiffusive spreading process up to normal diffusion.

\subsection{The crossover from strong to weak chaos}
\label{sec71}

\begin{figure}[htb]
\includegraphics[width=0.49\columnwidth]{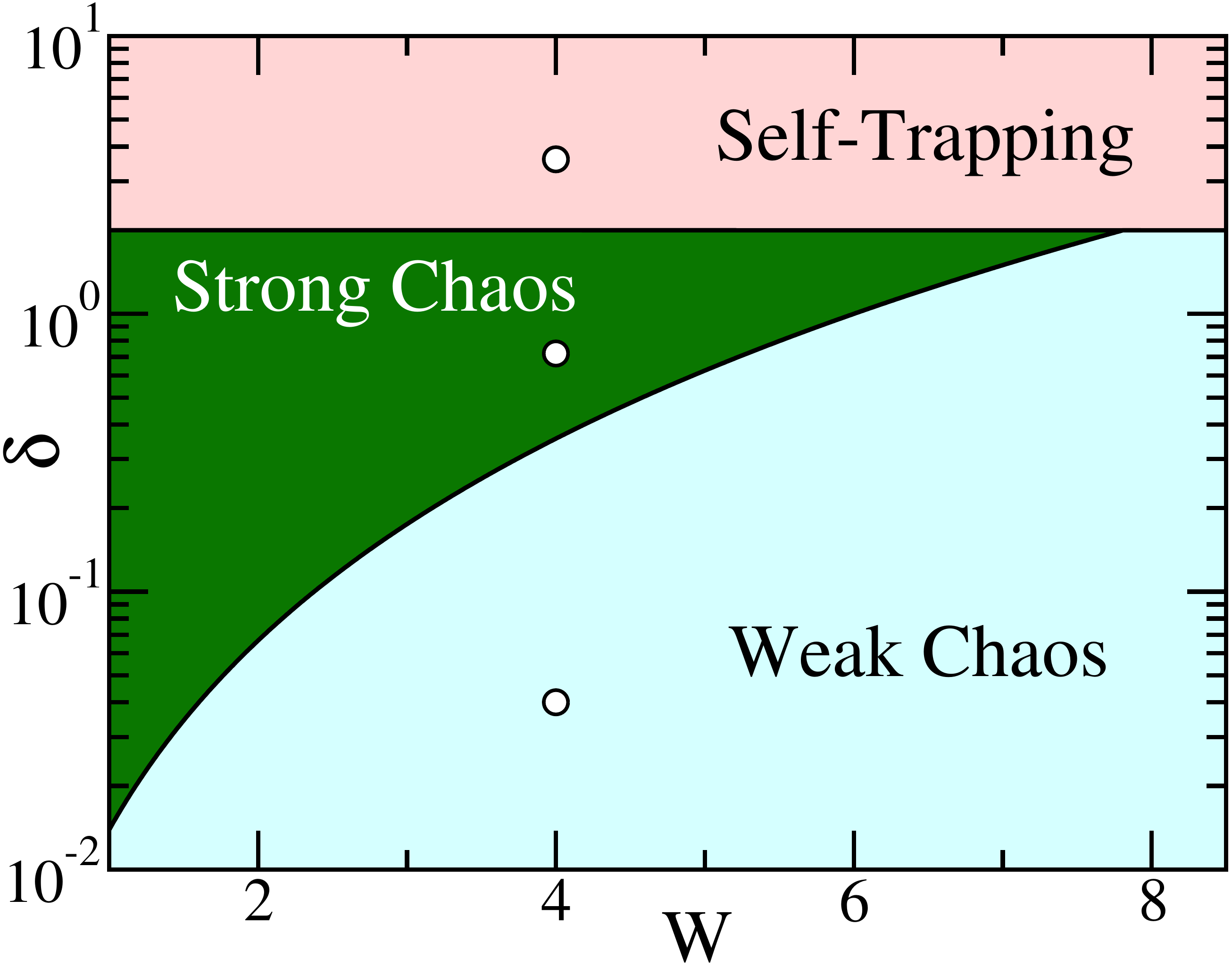} \hspace*{2mm}
\includegraphics[width=0.49\columnwidth]{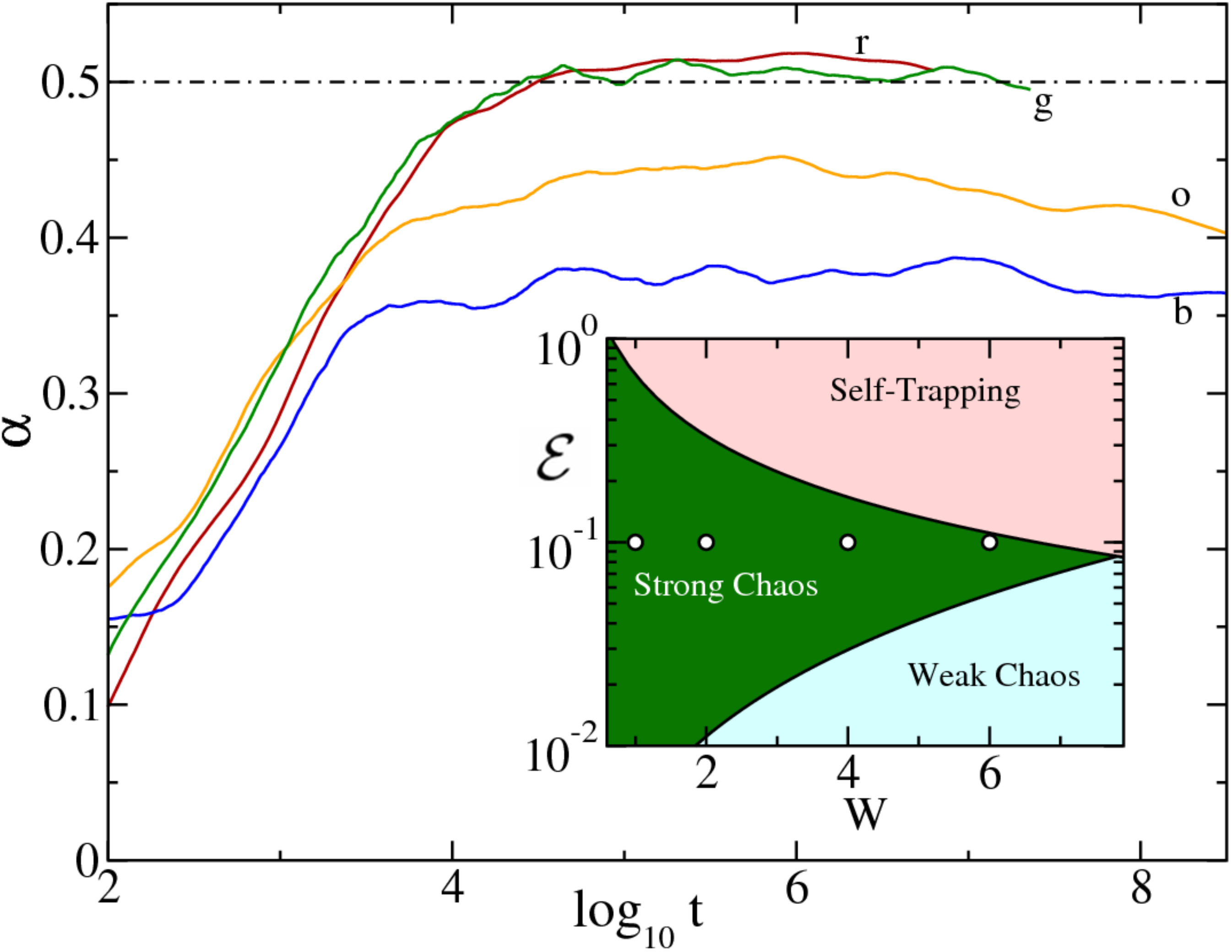}
\caption{ {\sc Left plot:} Parametric space of disorder, $W$, vs. the frequency shift induced by nonlinearity, $\delta$, for the DNLS model. 
Three spreading regimes are shown for dynamics dictated by: (i) weak chaos (pale blue), (ii) strong chaos (green), and 
(iii) the onset of self-trapping (pale red). 
The three circles show the initial numerical values used in Fig.\ref{fig8}.
{\sc Right plot:}
Spreading behavior in the strong chaos regime for the KG model, with an initial energy density of 
$\mathcal{E}=0.1$. The four curves are for the disorder strengths of: $W=1$ - (r)ed, $W=2$ - (g)reen, $W=4$ - (o)range,
$W=6$ - (b)lue. Inset: the KG analog of the DNLS parametric space. 
It is obtained by the small amplitude mapping $\mathcal{E} \rightarrow 3W\delta$. 
The four points correspond to the disorder
strengths used in the main portion of the figure. Adapted from \cite{tvljdbdokcssf10}
}
\label{fig7}
\end{figure}
The first prediction concerns the possibility to observe subdiffusive spreading of wave packets in the intermediate regime of strong chaos (\ref{wper}),
and the crossover to the asymptotic regime of weak chaos. The discussed results were obtained by Laptyeva et al \cite{tvljdbdokcssf10}.
We consider compact wave packets at $t=0$ spanning a width $V$ centered in the lattice, such that within $V$ there is a constant 
initial norm density of $n$ and a random phase at each site (outside the volume $V$ the norm density is zero). 
In the KG case, this equates to exciting each site in the width $V$ with the same energy density, $\mathcal{E}=E/V$, 
\textit{i.e.} initial momenta of $p_l = \pm \sqrt{2\mathcal{E}}$ with randomly assigned signs. 
Fig.\ref{fig7} (left plot - DNLS, inset right plot - KG) summarizes the predicted regimes, in which lines represent 
the regime boundaries.  It should be noted that the regime boundaries are NOT sharp, rather there 
is some transitional width between the regimes. The weaker the strength of disorder, the larger the window of strong chaos. Inversely, for 
$W \geq 8$ the strong chaos window closes almost completely. Ideally, one should utilize the smallest possible value of $W$.
Computational limits restrict this, so a reference of $W=4$ was chosen. It is important to note that $\delta$ will be reduced in time, 
since a spreading wave packet increases in size and drops its 
norm (energy) density. This gives the following interpretation of Fig.\ref{fig7}: given an initial norm density, the packet is in 
one of the three regimes (for example, the three circles in Fig.\ref{fig7}). A packet launched in the weak chaos regime stays in
this regime, while one launched in the strong chaos regime spreads to the point that it eventually crosses over into the asymptotic regime 
of weak chaos at later times.

For DNLS, an initial norm density of $n=1$ was used, so that initially $\delta \sim \beta$. Nonlinearities ($\mathcal{E}$ for KG) 
were chosen within the three spreading regimes (see Fig.\ref{fig7}), respectively  $\beta \in \left\lbrace 0.04, 0.72, 3.6\right\rbrace$ and 
$\mathcal{E} \in \left\lbrace 0.01, 0.2, 0.75\right\rbrace$.
\begin{figure}
\sidecaption
\includegraphics[width=0.64 \columnwidth]{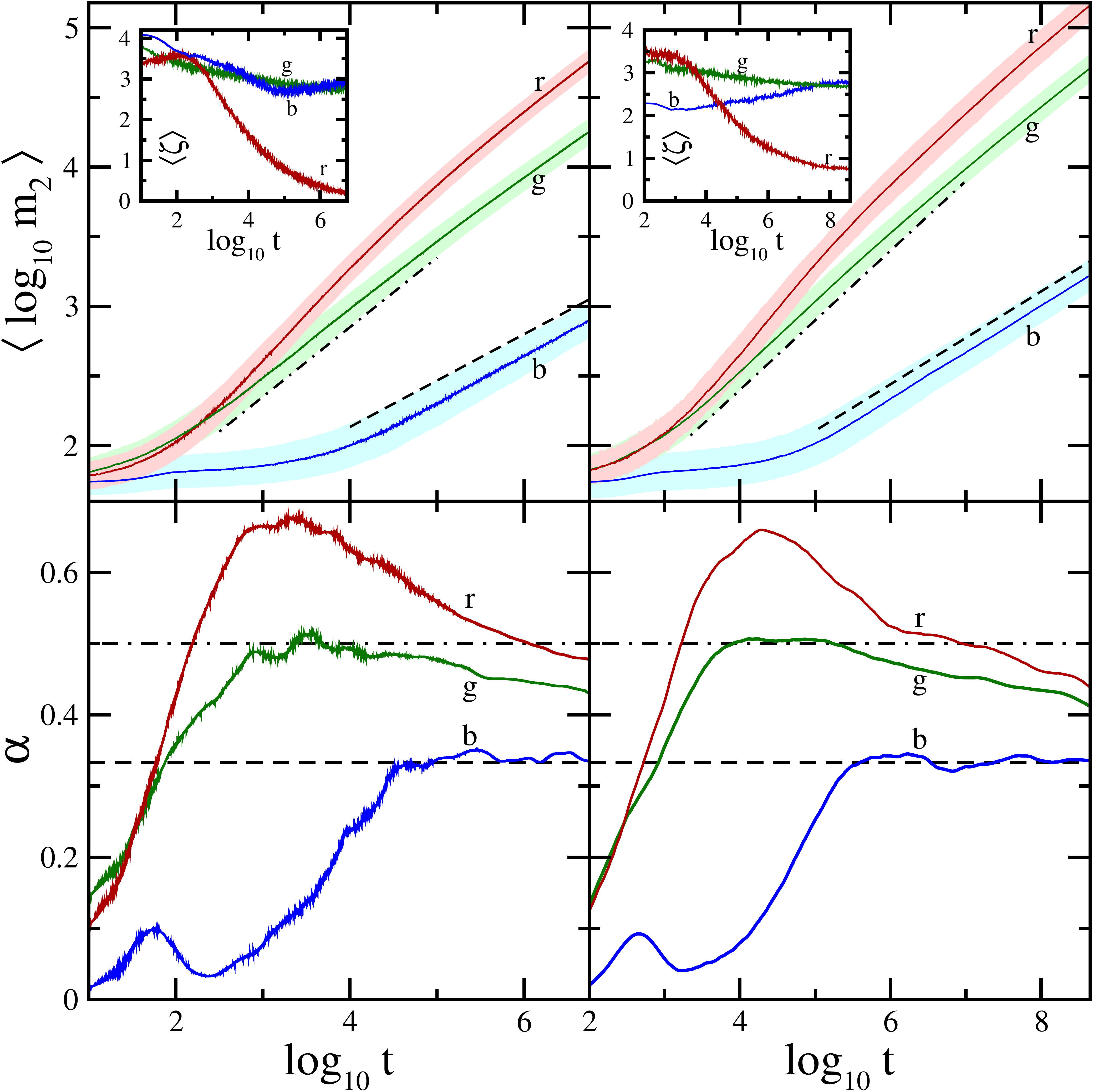}
\caption{
Upper row: Average log of second moments (inset: average compactness index) vs. log time for the DNLS (KG) on the 
left (right), for $W=4, L=21$. Colors/letters correspond the three different regimes: 
(i) weak chaos - (b)lue, $\beta =0.04 \, (\mathcal E=0.01)$, 
(ii) strong chaos - (g)reen, $\beta = 0.72 \, (\mathcal E=0.2)$,
(iii) self-trapping - (r)ed, $\beta = 3.6 \, (\mathcal E=0.75)$.
The respective lighter surrounding areas show one standard deviation error. Dashed lines are to guide the eye to $\sim t^{1/3}$, 
while dotted-dashed guides for $\sim t^{1/2}$.
Lower row: Finite difference derivatives for the smoothed $m_2$ data respectively from above curves.
 Adapted from \cite{tvljdbdokcssf10}
}
\label{fig8}
\end{figure}

Ensemble averages over disorder were calculated for $1000$ realizations and are shown in Fig.\ref{fig8} (upper row).
In the regime of weak chaos we find a subdiffusive growth of $m_2$ at large times according to $m_2 \sim t^\alpha, \, \alpha \le 1$,
with a compactness index $\zeta \approx 3$. Note that the subdiffusive growth is difficult to see initially in Fig.~\ref{fig8}
for two reasons. Firstly, the logarithmic scaling hides any small initial growth, and secondly, there is a
characteristic time scale for the packet to spread from its initial preparation. 
In the regime of strong chaos we observe a faster subdiffusive growth of $m_2$, with an additional slowing down at larger times, as expected 
from the predicted crossover. The compactness index is also $\zeta \approx 3$, as in the weak chaos regime. Finally, in the regime of partial 
self-trapping $m_2$ grows, but the compactness index $\zeta$ decreases in time substantially. This indicates that a part of the wave packet 
is arrested, and another part is spreading.

In order to quantify these findings,  smoothed data $\langle \log m_2 \rangle $ were produced \cite{tvljdbdokcssf10},  with a locally weighted regression algorithm \cite{CD88}, and 
a subsequently applied central finite difference
to calculate the local derivative 
\begin{equation}
\alpha(\log t) = \frac{{\rm d} \langle \log m_2 \rangle }{ {\rm d} \log t}\;.
\label{alpha(t)}
\end{equation}
The outcome is plotted in the lower row in Fig.\ref{fig8}.
In the weak chaos regime the exponent $\alpha(t)$ increases up to $1/3$ and stays at this value for later times.
In the strong chaos regime $\alpha(t)$ first rises up to $1/2$, keeps this value for one decade, and then drops down, as predicted. Finally, in the self-trapping
regime we observe an even larger rise of $\alpha(t)$. Additionally, we also mention numerics for $W \in \left\lbrace 1,2,6\right\rbrace$ with respective initial 
packet widths of $L=V \in \left\lbrace 361, 91, 11\right\rbrace $ \cite{tvljdbdokcssf10}. Results are qualitatively similar to those shown in Fig.~\ref{fig8}, and thus omitted
for graphical clarity.

The duration of the strong chaos regime with $\alpha = 1/2$ (and thus the location of the crossover) is largely dependent on how deep in the strong chaos regime the state is initially. Since the boundaries between different regimes are NOT sharp, but rather have some characteristic width, ideally one should utilize the smallest possible value of $W$. This is shown in Fig.\ref{fig7} (right plot) for the KG model. For $W \in \left\lbrace 1,2\right\rbrace $, a long plateau at $\alpha = 1/2$ is clearly observed. For $W \in \left\lbrace 4,6 \right\rbrace$, the initial energy density approaches one of the boundary lines and likely crosses into a boundary window, in which $\alpha < 1/2$. 

\subsection{Density profile scaling}
\label{sec72}

\begin{figure}[ht]
\begin{center}
\includegraphics[width=0.49\columnwidth,keepaspectratio,clip]{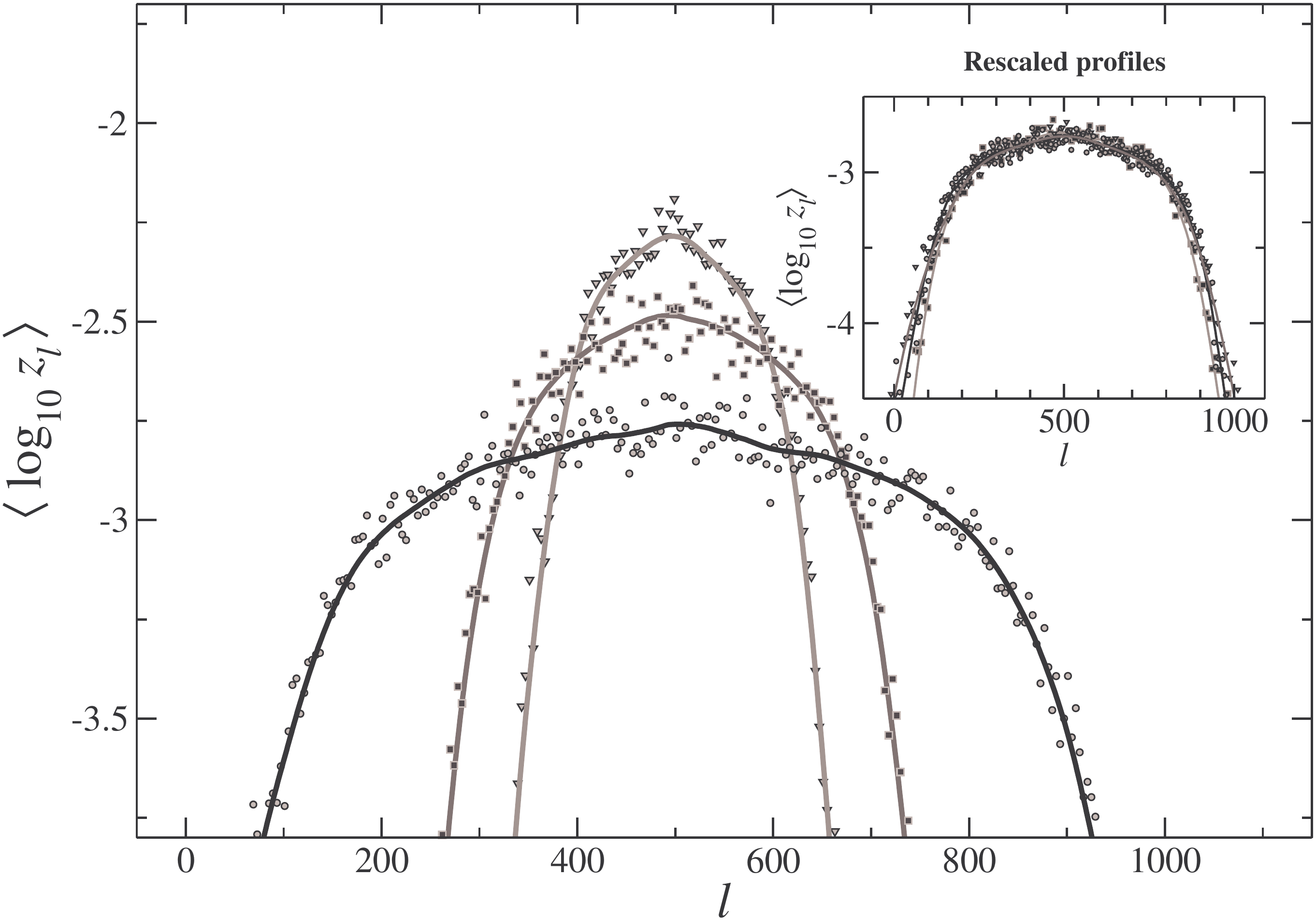} \hspace*{2mm}
\includegraphics[width=0.44\columnwidth,keepaspectratio,clip]{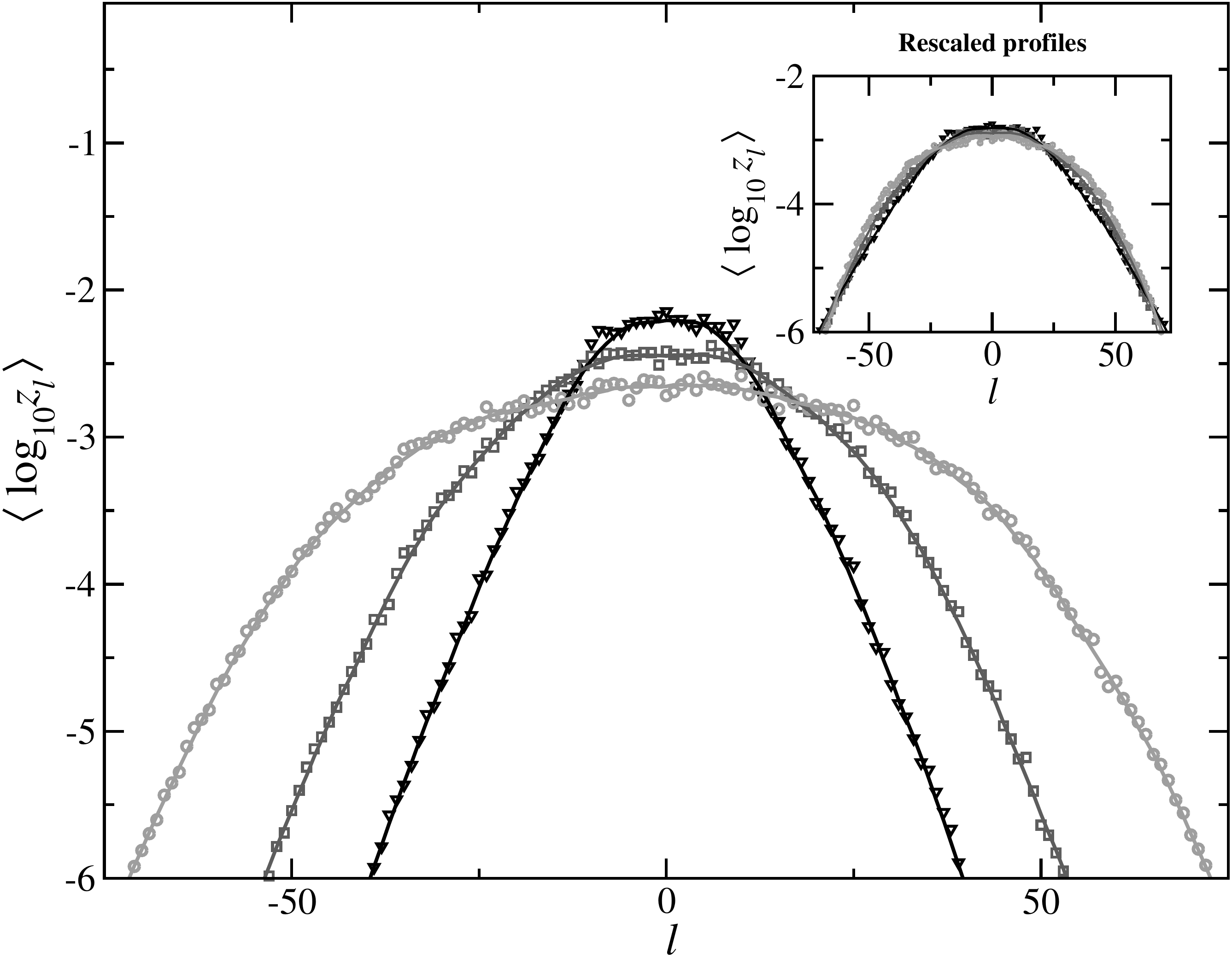}
\caption{{\sc Left Plot:} KG. The log of the normalized energy density distribution $\left\langle \log_{10}z_l \right\rangle$ at three different times (from top to bottom $t \approx 10^4$, $t \approx 10^7$, $t \approx 10^8$). The initial parameters are $E=0.2$, $W=4$ and $V=21$. Symbols correspond to 
the average over $10^3$ disorder realizations, and solid lines correspond to an additional smoothing. 
Inset: Rescaled distributions (see text). 
{\sc Right Plot:} DNLS. The log of the normalized norm density distribution $\left\langle \log_{10}z_l \right\rangle$ at three different times (from top to bottom $t \approx to^5$,
$t \approx 10^6$, $t\approx 10^7$). The initial parameters are $\beta = 0.04$, $W = 4$, and $V = 21$. Symbols correspond to the average over $10^3$ 
disorder realizations, and solid lines correspond to an additional smoothing. Inset: Rescaled distributions (see text).
Adapted from \cite{tvljdbsf13}
}
\label{fig9}
\end{center}
\end{figure}

If the effective noise theory (Sec.\ref{sec62}) applies, then the density distribution (energy for KG, norm for DNLS) should obey the nonlinear diffusion equation (\ref{ent4}).
In the weak chaos regime we have $\kappa = 4$. A numerical study was performed by Laptyeva et al \cite{tvljdbsf13} to test whether the scaling properties  of 
the solutions (see Sec.\ref{sec62}) hold. The main results are shown in Fig.\ref{fig9} (for details we refer to \cite{tvljdbsf13}).
The evolution of the averaged energy density profiles (KG) $\left\langle E \right\rangle$ in the course of spreading is illustrated in the left plot in Fig.\ref{fig9}. The peaked initial distribution
profiles transform into more flat ones as time evolves. The most striking result is obtained by rescaling the profiles in Fig.\ref{fig9} according to the scaling laws of the nonlinear
diffusion equation (\ref{ent4}). 
The rescaled densities are plotted  in the inset of the left plot of Fig.\ref{fig4}. We observe very good scaling behavior. 
For the DNLS with $\beta = 0.04$ similar data are shown in the right plot in Fig.\ref{fig9} for the times $t \approx to^5$,
$t \approx 10^6$, $t\approx 10^7$. The data are rescaled similar to the KG case. The result is shown in the inset of the right plot of Fig.\ref{fig9} and shows again very good agreement.
Together with the proper scaling of the edge of the wave packets, which was tested in \cite{Mul10}, this is the
strongest argument to support the applicability of NDE and MNDE to the spreading of wave packets in nonlinear disordered systems. It also strongly supports that the spreading
process follows the predicted asymptotics and does not slow down or even halt.

\subsection{Tuning the power of nonlinearity and the lattice dimension}
\label{sec73}

Let us consider a generalization of DNLS model (gDNLS) by tuning the power of nonlinearity, which corresponds to the case $\vec{D}=1$ in (\ref{RDNLS-EOMG})
\begin{equation}
i\dot{\psi_{l}}= \epsilon_{l} \psi_{l}
+\beta |\psi_{l}|^{\sigma}\psi_{l}
-\psi_{l+1} - \psi_{l-1},
\label{gDNLS}
\end{equation}
where $\sigma$ is a positive real number. We want to test the predictions presented in Sec.\ref{sec63}. Note that the previous DNLS and KG models
had $\sigma=2$ which correspond to cubic nonlinearities in the equations of motion, quartic anharmonicities in the Hamiltonian, 
and are related to two-body interactions in quantum many-body systems. Some other integer values of $\sigma$ might well have physical relevance,
e.g. $n=\sigma/2+1$ corresponds to $n$-body interactions, and $\sigma=1$ relates to quadratic Kerr media in nonlinear optics.

\begin{figure}
\includegraphics[width=0.49\columnwidth]{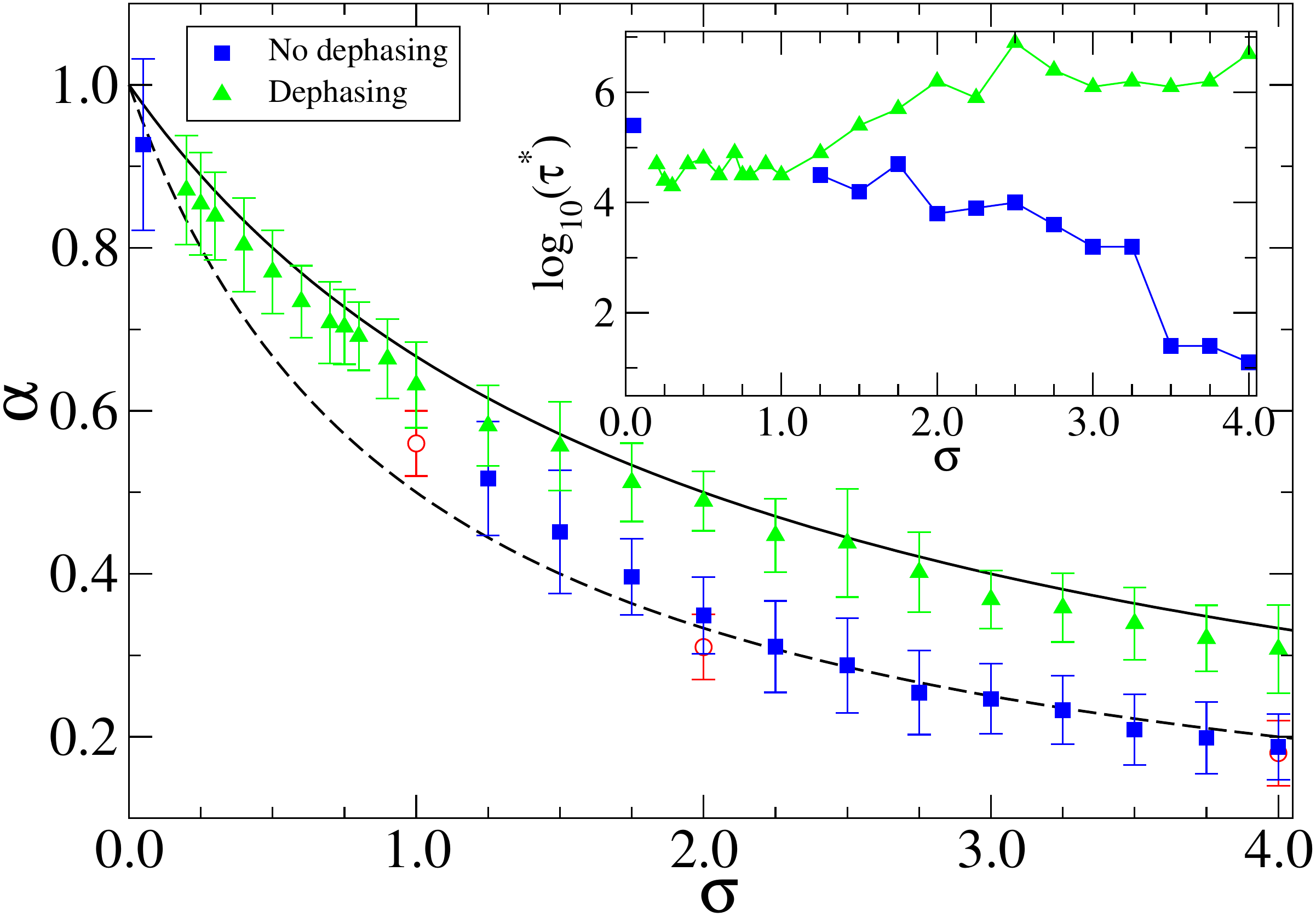}
\hspace*{2mm}
\includegraphics[width=0.48\columnwidth]{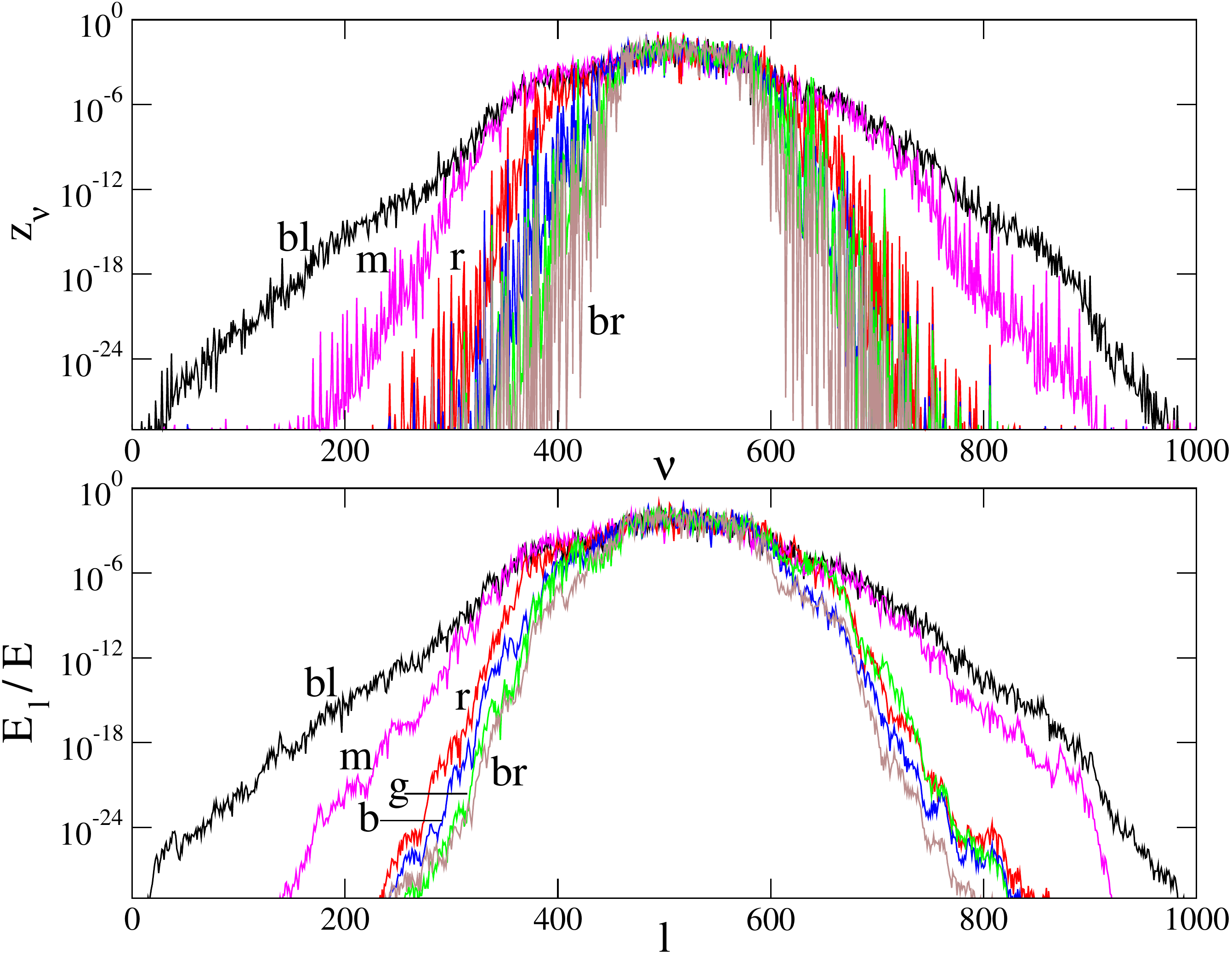}
\caption{{\sc Left Plot:} Exponent $\alpha$ ($m_2 \sim t^{\alpha}$)
  versus the nonlinearity order $\sigma$ for plain integration without
  dephasing (filled squares) and for integration with dephasing of NMs
  (filled triangles). Results without dephasing obtained in \cite{M09}
  are plotted with empty circle symbols. The theoretically predicted
  functions $\alpha=1/(1+\sigma)$ (weak chaos) and
  $\alpha=2/(2+\sigma)$ (strong chaos) are plotted by dashed and solid
  lines respectively. Inset: The logarithm of the minimum time
  $\tau^*$ for which the evolution of $m_2$ can be numerically fitted
  by a function of the form $t^{\alpha}$ versus $\sigma$ for
  integration with (filled triangles) and without (filled squares)
  dephasing.
{\sc Right Plot:}
Normalized energy distributions in NM (upper plot) and real (lower plot) space for $\sigma=0.05, 0.2, 0.8, 1.25,
  2.0, 3.0$ [(bl) black; (m) magenta; (r) red; (b) blue; (g) green;
  (br) brown] at times $t=3.6\times 10^5, 1.3 \times 10^5, 2.5 \times
  10^5, 1.4 \times 10^6, 3 \times 10^7, 10^9$ respectively. The second
  moment of each distribution is $m_2\approx10^3$. In the upper plot
  the distributions for $\sigma=1.25, 2.0$ are not clearly visible as
  they are overlapped by the distribution for $\sigma=3.0$.
Adapted from \cite{cssf10}
}
\label{fig10}
\end{figure}

Mulansky \cite{M09}
presented numerical simulations of the gDNLS model for a few integer
values of $\sigma$ and single site excitations, and fitted the dependence $m_2(t)\sim t^{\alpha}$ with exponents $\alpha$ which depend on
$\sigma$  (see open circle data in left plot in Fig.\ref{fig10}).  
In \cite{VKF09} numerical simulations of the
gDNLS model were performed for non integer values of $\sigma$ on
rather short time scales, leaving the characteristics of the
asymptotic ($t \rightarrow \infty$) evolution of wave packets aside.

The corresponding generalized KG model (gKG) follows the equations of motion
\begin{equation}
\ddot{u}_{l} = - \tilde{\epsilon}_{l}u_{l}
-|u_{l}|^{\sigma} u_{l} + \frac{1}{W} (u_{l+1}+u_{l-1}-2u_l)\;.
\label{H_EM}
\end{equation}
and was studied by Skokos et al \cite{cssf10}, again for single site excitations, and a whole range of different values of $0.02 \leq \sigma \leq 4$.
The dependence $m_2(t)\sim t^{\alpha}$ was again fitted with exponents $\alpha$ which depend on
$\sigma$. In order to emulate strong chaos from scratch, an additional normal mode dephasing (see Sec.\ref{sec53} and Fig.\ref{fig4}) was performed, and again 
the data were fitted with $\sigma$-dependent values of $\alpha$. The outcome is shown in the left plot in Fig.\ref{fig10}.
The data with dephasing (filled triangles) are nicely following the prediction from strong chaos (\ref{sigma_strong}) $\alpha=2/(2+\sigma)$ in the 
range $0.2 \leq \sigma \leq 4$. The data without dephasing (filled squares) show very good agreement with the prediction
from weak chaos (\ref{sigma_weak}) $\alpha=1/(1+\sigma)$ in the 
range $2 \leq \sigma \leq 4$. However for $1 \leq \sigma \leq 1.8$ the numerical results overestimate the weak chaos prediction, and tend towards
the strong chaos ones. The reason for that is simply, that for $\sigma < 2$ a single site excitation {\sl can} be launched in the strong chaos regime \cite{sf10}.
Therefore fitting procedures will average over the strong chaos region, crossover region, and weak chaos region, and result in a number which is located somewhere 
between the two theoretical lines. 
Instead of fitting the numerically
obtained time dependence $m_2(t)$ with power laws, one should compute derivatives $d\langle \log_{10} m_2 \rangle / d \log_{10} t$ in order
to identify a potentially long lasting regime of strong chaos, crossovers, or the asymptotic regime of weak chaos. This is a task yet to be accomplished for the above
cases.

The order of nonlinearity $\sigma$ influences not only the spreading
rate of wave packets, but also the morphology of their profiles. In the right plot in 
Fig.~\ref{fig10} we plot the normalized energy distributions of
initial single site excitations, for different $\sigma$ values in NM
(upper plot) and real (lower plot) space. Starting from the outer,
most extended wave packet we plot distributions for $\sigma=0.05$
(black curves), $\sigma=0.2$ (magenta curves), $\sigma=0.8$ (red
curves), $\sigma=1.25$ (blue curves), $\sigma=2$ (green curves) and
$\sigma=3$ (brown curves). All wave packets were considered for the
same disorder realization but at different times of their evolution
when they have the same value of second moment $m_2\approx10^3$. These
times are $t=3.6\times 10^5$ for $\sigma=0.05$, $t=1.3\times 10^5$ for
$\sigma=0.2$, $t=2.5\times 10^5$ for $\sigma=0.8$, $t=1.4\times 10^6$
for $\sigma=1.25$, $t=3\times 10^7$ for $\sigma=2$ and $t=10^9$ for
$\sigma=3$ and increase for $\sigma \geq 0.2$ since the spreading
becomes slower for larger $\sigma$. When $\sigma\rightarrow0$ wave packets remain
localized for very large time intervals before they start to
spread \cite{cssf10}. This is why for $\sigma=0.05$ the second moment becomes
$m_2\approx10^3$ at a larger time than in cases with $\sigma=0.2$ and
$\sigma=0.8$.
From the results of Fig.~\ref{fig10} we see that for large enough
values of $\sigma$ ($0.8 \leq \sigma \leq 3$), the distributions on a
logarithmic scale have a chapeau-like shape consisting of a highly
excited central part and exponential tails having practically the same
slope. Contrarily, the distributions for $\sigma =0.2$ and
$\sigma=0.05$ become more extended having different slopes in the
tails.

A characteristic of the NM space distributions in the right plot in Fig.\ref{fig10} 
for  $\sigma \geq 0.8 $
is that they exhibit very large value fluctuations (up to 5-10 orders of magnitude)
in their tails,
contrarily to the corresponding distributions in real space.
Tail NMs are driven by the core of the wave packet, but may also interact with neighboring
tail NMs. The presence of large tail amplitude fluctuations signals that neighboring
tail NMs do not interact significantly (otherwise we would expect a tendency towards equipartition).
Tail NMs are then excited only by the core. The further
away they are, the weaker the excitation. But within a small tail
volume, NMs with larger localization length will be more strongly
excited than those with smaller localization length, hence the large
observed fluctuations, which on a logarithmic scale are of the order
of the relative variation of the localization length. 
Therefore Anderson
localization is preserved in the tails of the distributions over very
long times (essentially until the given tail volume becomes a part of
the core).  But the NM space distributions for $\sigma=0.05$ and
$\sigma =0.2$ exhibit less fluctuations in their tail values with
respect to the other distributions in the upper right plot of
Fig.~\ref{fig10}, implying that tail NMs are now interacting with each
other on comparatively short time scales and reach a visible level of
local equipartition. Therefore we observe for these cases a
destruction of Anderson localization even in the tails of the
spreading wave packets. 

How is Anderson localization restored in the limit $\sigma \rightarrow 0$, since we
obtain a linear wave equation for $\sigma = 0$ ? Both weak and strong chaos exponents yield $\alpha(\sigma \rightarrow 0 ) \rightarrow 1$ in this case,
i.e. normal diffusion. The answer is in the prefactor of the subdiffusive law $m_2 = C t^{\alpha}$. The only possibility is to assume $C(\sigma \rightarrow 0) \rightarrow 0$.
The diverging waiting times for single site excitations in this limit, which have to pass before spreading is observed, are a good confirmation of the above assumption \cite{cssf10}.

\begin{figure}
\sidecaption
\includegraphics[width=0.64\columnwidth]{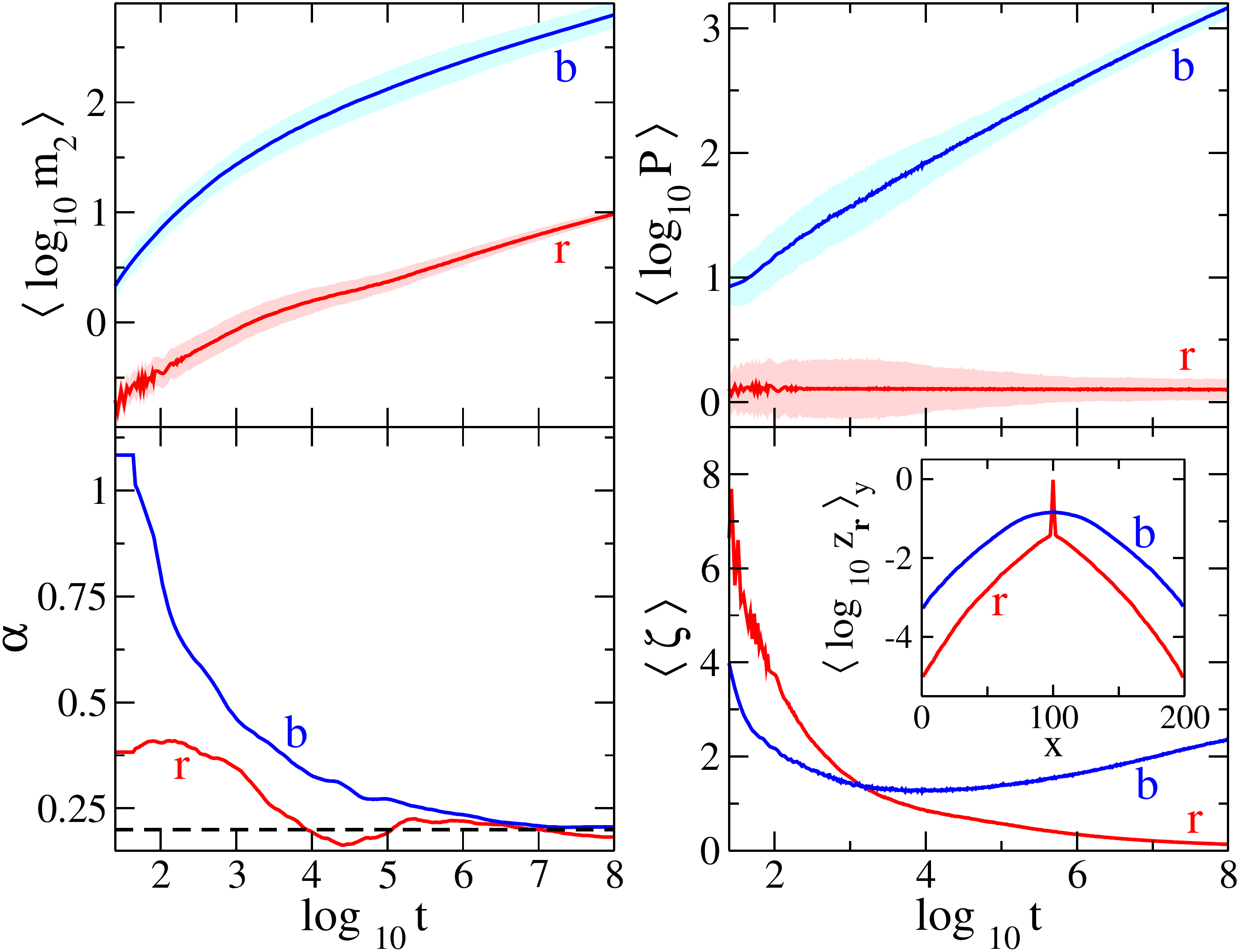}
\caption{The parameters $(\sigma,\mathcal{E})=(2, 0.3), (2, 2.0)$ correspond to the weak chaos ((b)lue) and self-trapping ((r)ed). 
\textit{Left column}: average log of second moment (upper) and its power-law exponent (lower) vs. log time. The dashed line is the theoretical expectation for the weak chaos $\alpha=0.20$.
\textit{Right column}: average log of participation number (upper) and average compactness index (lower) vs. log time. In both columns of the upper row the lighter clouds correspond to a standard deviation.
\textit{Inset:} normalized radial density distributions at $t=10^8$.
Adapted from \cite{tvljdbsf12}
}
\label{fig11}
\end{figure}
Let us move on to two-dimensional cases. The two-dimensional DNLS case yields the equations of motion
\begin{equation}
i \dot{\psi}_{\vec{b}} = \epsilon_{\vec{b}} \psi_{\vec{b}} + \beta \left|\psi_{\vec{b}} \right|^\sigma \psi_{\vec{b}} - \sum_{\vec{n}} \psi_{\vec{n}}. \label{eq:HDEOM}
\end{equation}
Here $\vec{b}=\left(l, m\right)$ denotes a two-dimensional lattice vector with integer components, 
and $\vec{n}$ runs over nearest neighbors.
Garcia-Mata et al \cite{Shep08} studied (\ref{eq:HDEOM}) with $\sigma=2$. Single site excitations were launched and
the numerically
obtained time dependence of $m_2(t)$ was fitted with power laws. With the largest integration time $t=10^6$ and 10 disorder realizations the
fitting result was $\alpha \approx 0.23$. Note that the effective noise theory predicts $\alpha=1/3$ for the strong chaos case (\ref{sigma_strong}), and $\alpha=0.2$ for the asymptotic
weak chaos case (\ref{sigma_weak}). Therefore the result from \cite{Shep08} is again located between both predictions, which might be due to crossover effects, and insufficient
averaging and integration time (see above discussion). 

A further work by Laptyeva et al \cite{tvljdbsf12} studies the two-dimensional KG case for various values of $\sigma$:
\begin{equation}
\ddot{u}_{\vec{b}}=-{\tilde \epsilon}_{\vec{b}} u_{\vec{b}} - \left|u_{\vec{b}}\right|^\sigma u_{\vec{b}} + \frac{1}{W}\sum_{\vec{n}}\left( u_{\vec{n}} - u_{\vec{b}} \right).
 \label{eq:HKGEOM}
\end{equation}
Rather than fitting the numerically
obtained time dependence $m_2(t)$ with power laws as in \cite{Shep08}, Laptyeva et al \cite{tvljdbsf12} computed derivatives $d\langle \log_{10} m_2 \rangle / d \log_{10} t$ in order
to identify a potentially long lasting regime of strong chaos, crossovers, and the asymptotic regime of weak chaos. The number of disorder realizations was as large as 400,
and integration times extended up to $t=10^8$. Initial states were wave packets occupying a typical localization volume $V\sim 30$ of the linear wave equation.
In Fig.\ref{fig11} the results for $\sigma=2$ are shown. The weak chaos exponent measures as $\alpha\approx 0.21$ which is very close to the theoretical prediction $\alpha=0.2$.
\begin{figure}
\includegraphics[width=0.45\columnwidth]{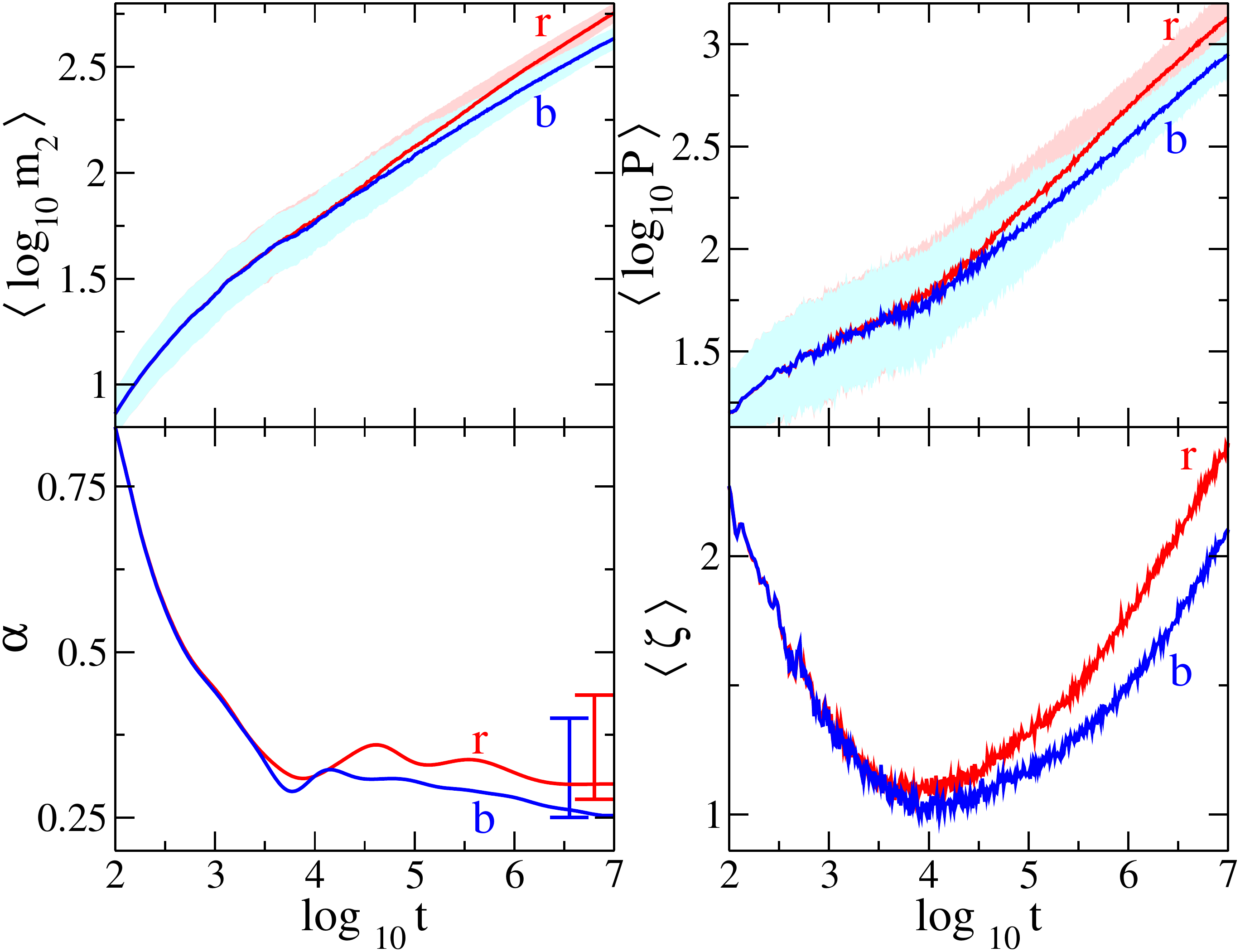}
\hspace*{2mm}
\includegraphics[width=0.45\columnwidth]{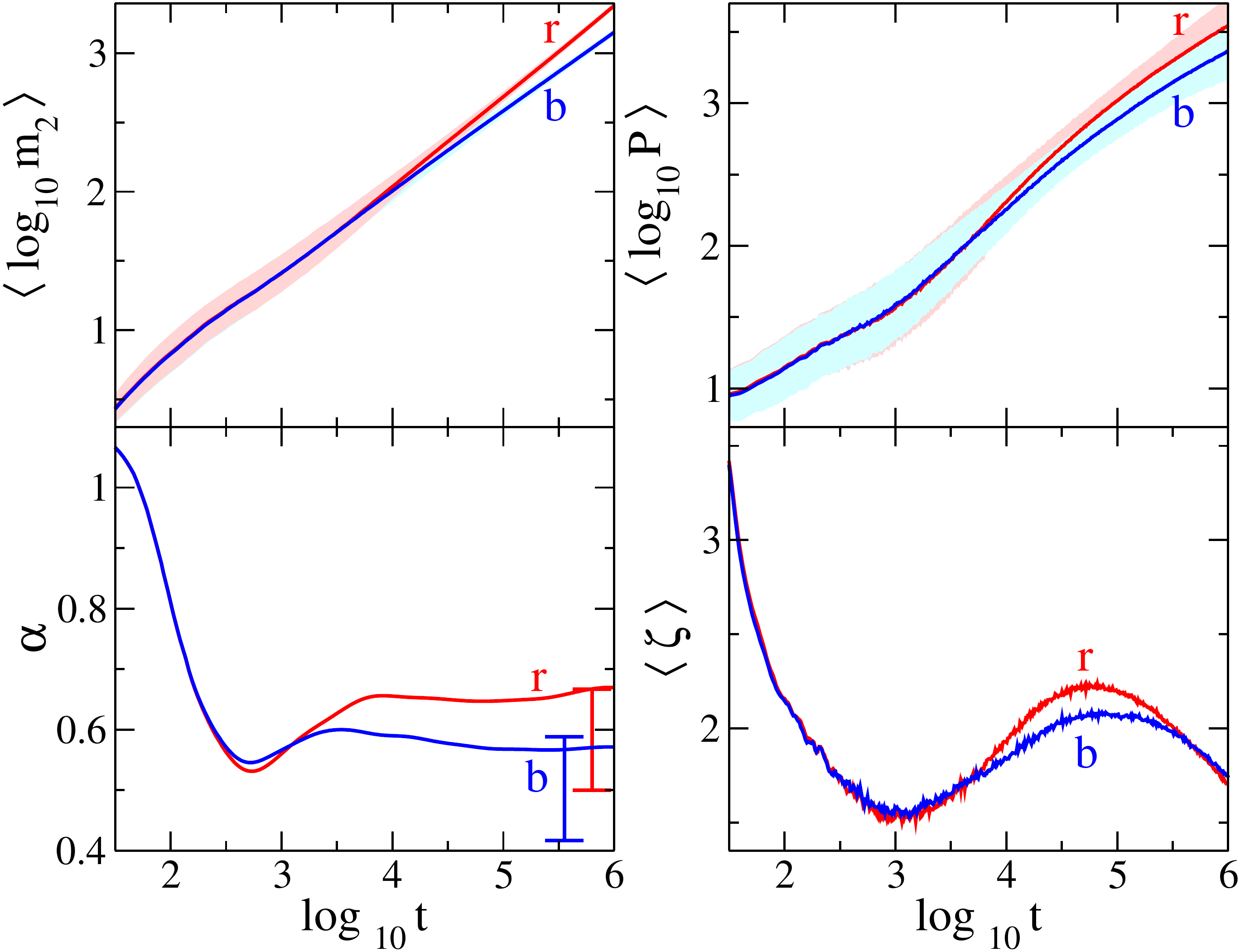}
\caption{{\sc Left Plot:} The parameters $(\sigma,\mathcal{E})=(1.3,0.025),(1.5,0.04)$ are colored respectively as (r)ed and (b)lue.    
\textit{Left column}: average log of second moment (upper) and its power-law exponent (lower) vs. log time. 
\textit{Right column}: average log of participation number (upper) and average compactness index (lower) vs. log time. 
{\sc Right Plot:} The parameters $(\sigma,\mathcal{E})=(0.5,0.005),(0.7,0.03)$ are colored respectively as (r)ed and (b)lue.    
\textit{Left column}: average log of second moment (upper) and its power-law exponent (lower) vs. log time.  \textit{Right column}: average log of participation number (upper) and average compactness index (lower) vs. log time. In both columns of the upper row, the lighter clouds correspond to a standard deviation.
The I-bar bounds denote the theoretical expectations from Eqs.(\ref{sigma_strong},\ref{sigma_weak}) for weak chaos (lower bound) and strong chaos (upper bound).  
Adapted from \cite{tvljdbsf12}
}
\label{fig12}
\end{figure}
Extensions to $\sigma=1.5,1.3$ in the weak chaos regime and to $\sigma=0.7,0.5$ in the strong chaos regime show very good agreement between the 
numerically observed exponents, and the theoretical predictions in Fig.\ref{fig12}. 

We can conclude, that the predictions from effective noise theory and the nonlinear diffusion approach have been impressively confirmed in various numerical
studies. 

\subsection{Heat conductivity}
\label{sec74}

Assuming the validity of  effective noise theory, we arrive at the next prediction that the heat conductivity of a thermalized system at small
temperature (density) must be proportional to the diffusion coefficient (\ref{ent5}) where the density $n$ is replaced by the temperature $T$. While one has to be careful 
in the DNLS case, where two conserved quantities (energy, norm) enforce Gibbs, or non-Gibbs distributions \cite{dmb14}, the KG case might be 
again a better testing ground, where one conserved quantity (energy) can be expected to enforce a Boltzmann distribution.
\begin{figure}
\sidecaption
\includegraphics[width=0.6\columnwidth]{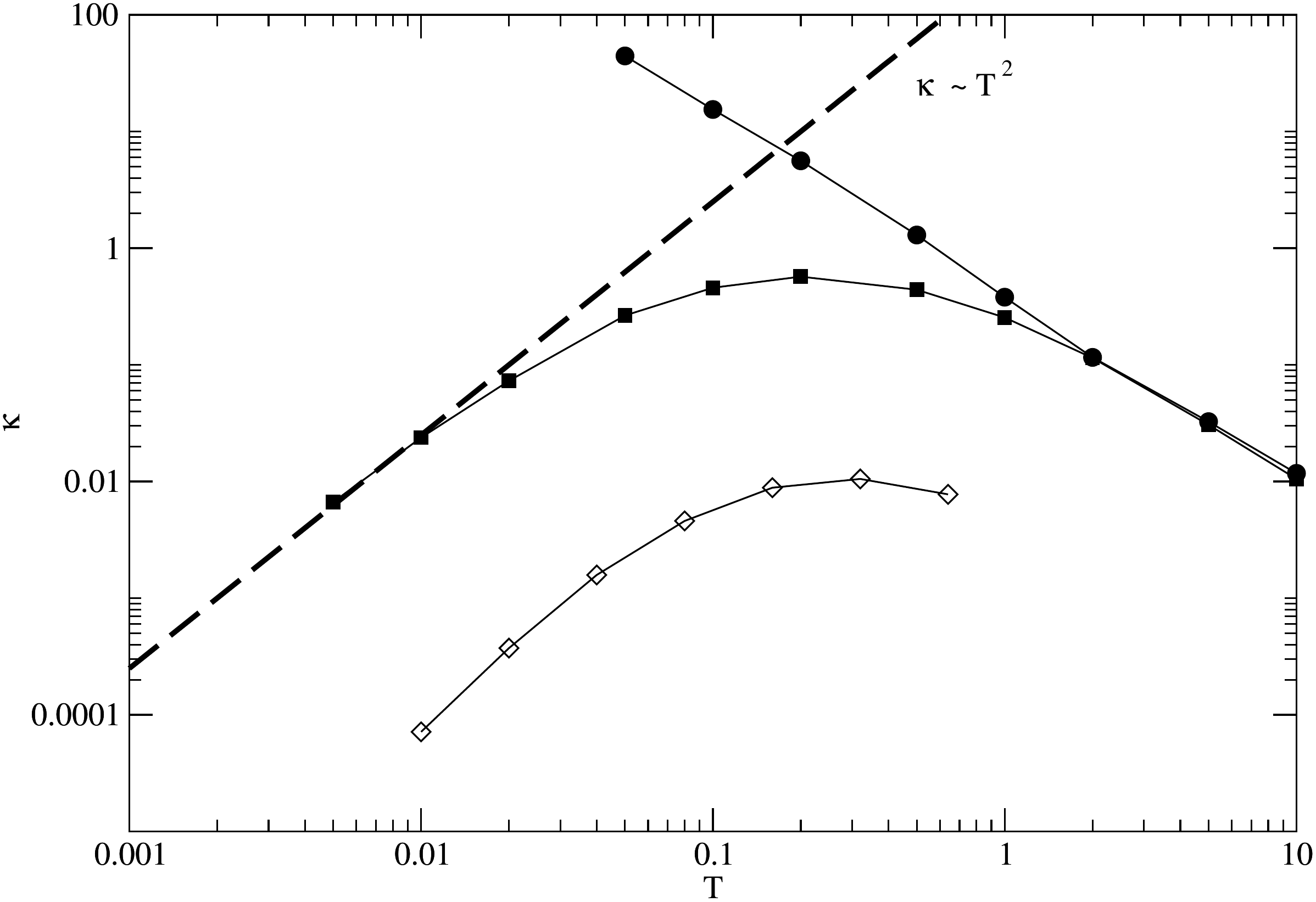}
\caption{KG chain: Heat conductivity $\kappa(T)$ for $W=2$ (filled squares). For comparison we also show the
data for the ordered case  $\tilde{\epsilon}_l\equiv 1$ (filled circles). 
Thin solid lines guide the eye.
The dashed line corresponds to the power law
$T^2$. 
The stronger disorder case $W=6$ corresponds to the open diamond data points.
Adapted from \cite{sfminl11}
}
\label{fig13}
\end{figure}
The calculation of the heat conductivity for (\ref{KG-EOM}) was performed in Ref.\cite{sfminl11}. Its dependence on the temperature is shown
in Fig.\ref{fig13}. The strong chaos scaling $\kappa(T)\sim T^2$ is observed nicely. The expected weak chaos regime was not reachable by the
heavily extensive numerical efforts. Note that the decay of the heat conductivity for large temperatures is due to selftrapping, and observed even for
the ordered chain at $W=0$ (solid circles in Fig.\ref{fig13}).

\subsection{Ramping nonlinearity}
\label{sec75}

Subdiffusion is notoriously slow. This poses problems for numerical studies, especially in two and even more in three space dimensions.
The situation is even more severe with experimental studies of ultracold interacting K atomic clouds, where the conversion of  the maximum time of keeping
the coherence of the macroscopic quantum cloud is about 10 seconds \cite{LDTRZMLDIM}, which turns into $t \approx 10^4...10^5$ dimensionless time units used throughout this
chapter. Consequently the probing of subdiffusion in \cite{LDTRZMLDIM} allowed to conclude qualitatively that the onset of a subdiffusive spreading of the interacting
cloud does take place, but was not sufficient to reliably measure the exponent. In order to fit a power law, we need at least two decades of variation in both variables.
With a weak chaos exponent $1/3$ and two decades in the second moment we arrive at six decades in time - added to $t\approx 10^2$ which is the time
the linear wave equation spreads into the localization volume. Therefore times $\sim 10^8$ are desirable, which turn into experimental times of the order of $10^5$ seconds -
clearly not reachable with nowadays techniques. On the other side, the reader is welcome to reread the above presented numerical data and analysis and welcome to observe
that restricting to maximum integration time $10^5$ will not allow for an accurate estimate of the exponents. At the same time, numerical studies also suffer from
the computational time restriction. While this appears to be no serious issue for most one-dimensional system studies, already two dimensional systems can easily
raise the problem of insufficient computational times.

\begin{figure}
\includegraphics[width=0.48 \columnwidth]{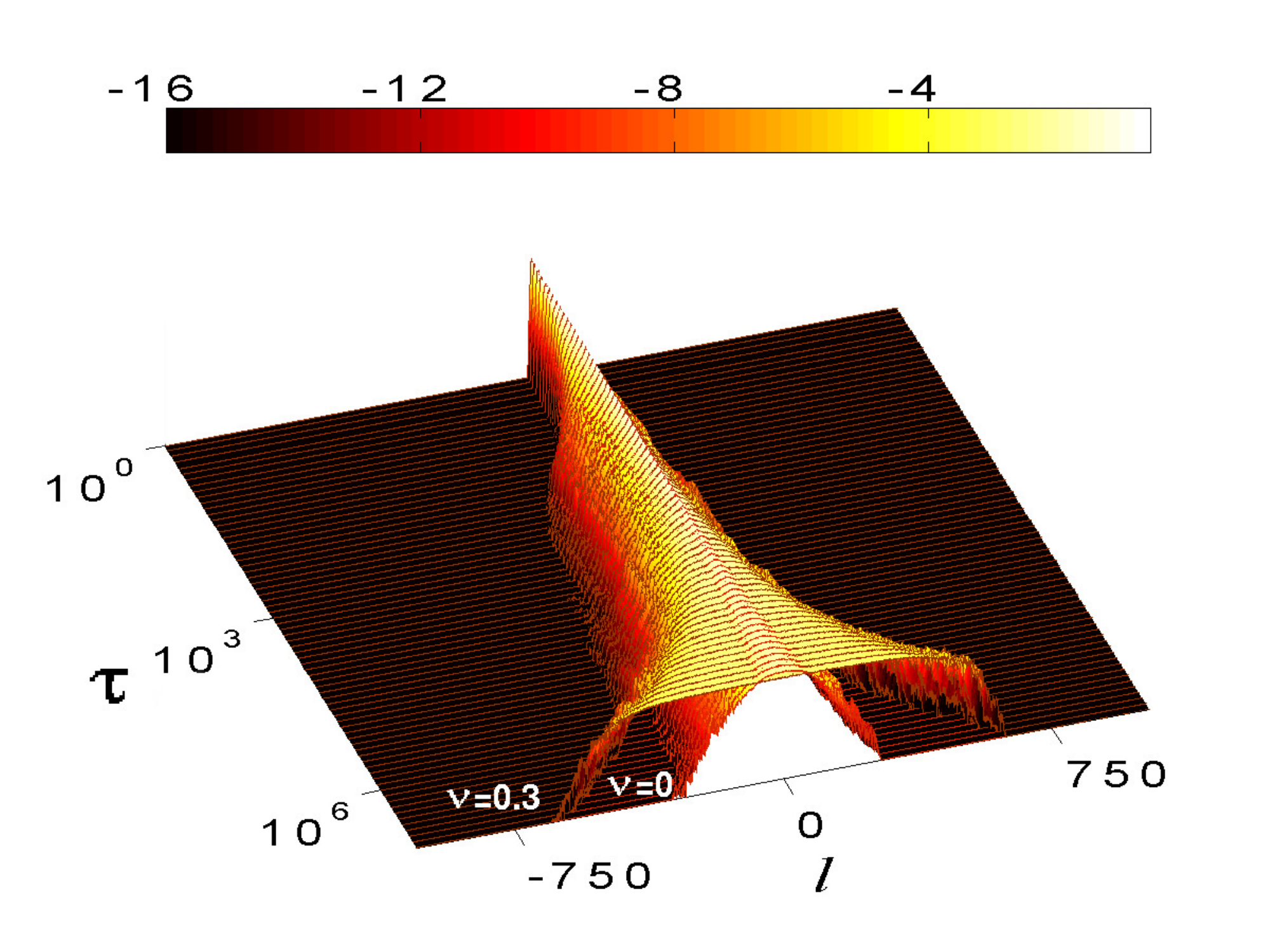}
\hspace*{2mm}
\includegraphics[width=0.48 \columnwidth]{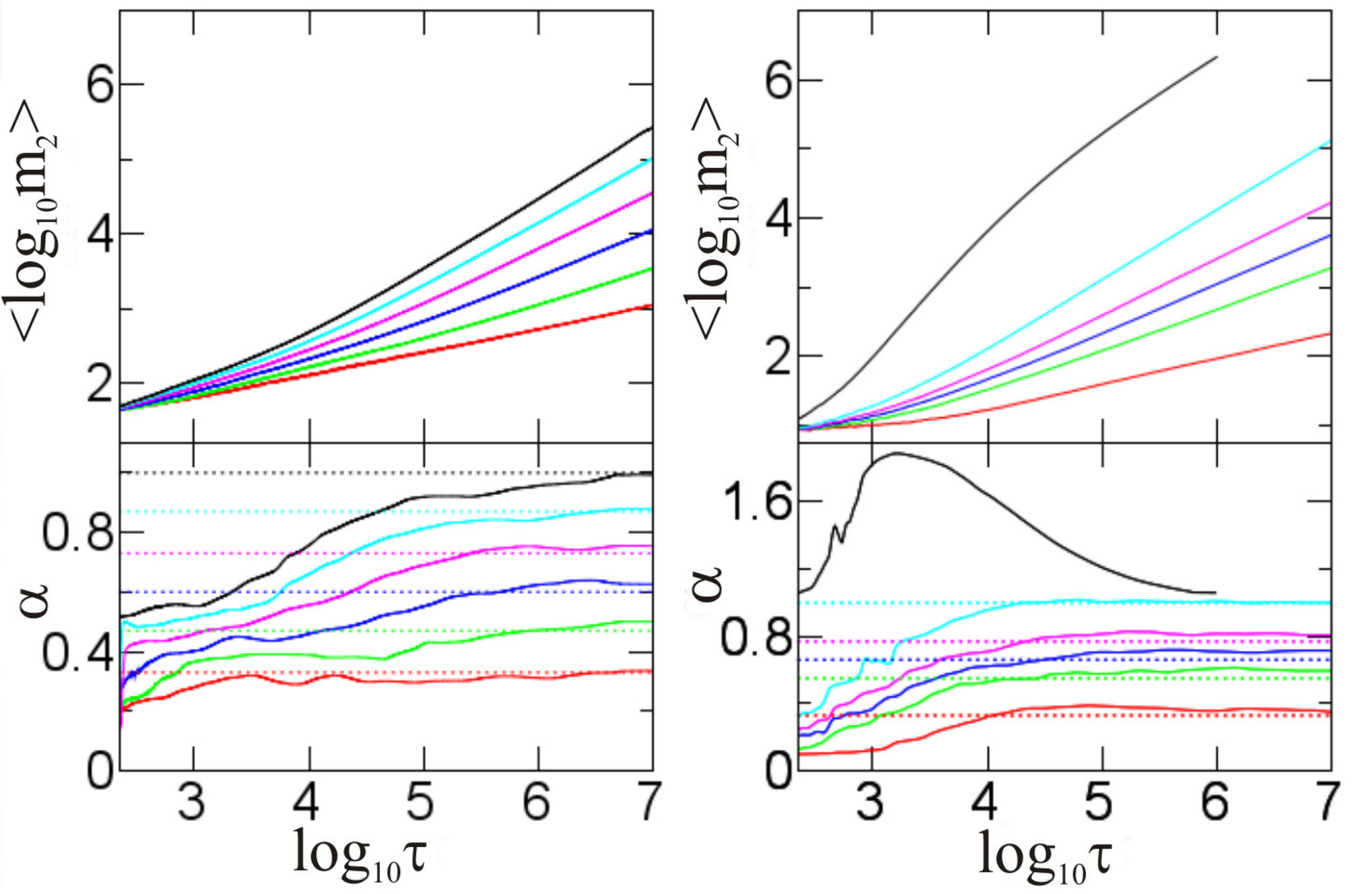}
\caption{{\sc Left Plot:} Evolution of the averaged norm density $<n_{l}(\tau)>$ in the case
without ($\nu=0$) and with ramping ($\nu=0.3$) in log scale for the DNLS
model.
{\sc Right Plot:} \textit{Left column:} the second moments (upper) and their power-law
exponents $\alpha$ (lower) for the DNLS model for $\nu=0$ (red), $\nu=0.1$
(green),
$\nu=0.2$ (blue), $\nu=0.3$ (magenta), $\nu=0.4$ (cyan), and $\nu=0.5$ (black).
\textit{Right column:} the second moments (upper) and their power-law
exponents $\alpha$ (lower) for the NQKR model for $\nu=0$ (red), $\nu=0.17$
(green),
$\nu=0.25$ (blue), $\nu=0.33$ (magenta), $\nu=0.5$ (cyan), and $\nu=1.5$
(black). Dashed colored lines correspond to expected values for exponents in
both cases.
Adapted from \cite{ggkrsf13}
} \label{fig14}
\end{figure}
Gligoric et al \cite{ggkrsf13} suggested a possible way out. 
Instead of trying to substantially increase available time scales, they propose
to speed up the subdiffusive process itself. 
This is done by a temporal ramping of the two-body interaction strength,
which can be varied e.g. for K atoms by three orders of magnitude close to the
Feshbach resonance \cite{RZECMSIM}. Why should that help? The momentary
diffusion rate $D$ of
a spreading packet in one spatial dimension is proportional to the fourth power
of the product of interaction strength $\beta$ and particle density $n$: $D \sim
(\beta
n)^4$ for the asymptotic case of weak chaos (\ref{ent5}). In the course of
cloud spreading the density $n$ decreases, and therefore also $D$. This is the
reason for the predicted subdiffusion process, which is substantially slower
than normal diffusion.
The proposal is to compensate the decrease of the density $n$ with an increase
in the interaction strength $\beta$. Depending on the concrete ramping protocol
$\beta(\tau)$
one can expect different faster subdiffusion processes, and possibly even normal
diffusion. The condition for that outcome to be realized is, that the internal
chaos time scales
(basically the inverse Lyapunov coefficients) will be still short enough so that
the atomic cloud can first decohere, and then spread.  
With that achieved, the cloud spreading will be faster, and one can expect that
the available experimental time will suffice for the precise observation
and analysis of the process.

Let us get into numbers for one spatial dimension. The second moment is $m_2
\sim 1/n^2$ and the momentary diffusion constant $D \sim (\beta n)^4$. For a
constant $\beta$ the 
solution of $m_2 =D t$ yields $m_2\sim 1/n^2 \sim t^{1/3}$, and therefore
$n\sim t^{-1/6}$. Thus we choose now a time dependence $\beta \sim
t^{\nu}$.
Then the resulting spreading is characterized by 
\begin{equation}
m_2 \sim t^{(1+4\nu)/3}\;, \;d=1\;. \label{result1d}
\end{equation}
For $\nu=1/2$ we already obtain normal diffusion $m_2 \sim t$.

Similar for two spatial dimensions, where $m_2 \sim 1/n$, for a constant
$\beta$ the cloud spreading is even slower with $m_2 \sim t^{1/5}$. With a
time dependent ramping
$\beta \sim t^{\nu}$ the resulting speedup is 
\begin{equation}
m_2 \sim t^{(1+4\nu)/5}\;,\; d=2\;. \label{result2d}
\end{equation}
For $\nu=1$ we again obtain normal diffusion.
Note that if numerics confirm the above predictions then also the above conditions for the chaoticity time scales are met with good probability.

Once ramping is too fast, one can
expect to see several different scenaria. Either fragmenting atomic
clouds appear since some parts of the cloud get self-trapped and
some other parts do not.
If self-trapping is avoided, one may also see ramping-induced diffusion: while
the
internal cloud dynamics does not suffice to decohere phases, initial
fluctuations in the
density distribution can lead to considerably different temporal energy
renormalizations in
different cloud spots, and therefore to an effective dephasing similar to a
random noise process
in real time and space. 

The spreading of wave packets in the DNLS model, without and with ramping of
the nonlinearity are shown in the left plot in Fig. (\ref{fig14}) (note that time $t$ is coined $\tau$ in the plots).. Clearly packets 
spread faster when the nonlinearity is ramped in time. To quantify the
spreading exponent, the authors of  \cite{ggkrsf13}
averaged the logs (base 10) of $m_2$ over 1000 different
realizations and smoothened additionally with locally weighted regression
\cite{CD88}. 
The (time-dependent) spreading exponents are obtained through 
central
finite difference method \cite{CD88}, 
$\alpha=\frac{d<\log_{10}(m_2)>}{d(\log_{10}(t))}$.
The results 
for the DNLS model are shown in the right plot in Fig. (\ref{fig14}). The exponents
of
subdiffusive spreading reach the theoretically predicted values. 
Note that the first assumption of the 
asymptotic exponent occurs after similar waiting times for all $\nu$.
Monitoring of the participation
number $P$ for the DNLS model
indicates that self-trapping starts to occur already for $\nu=0.4$. Results for the
nonlinear quantum kicked rotor 
(NQKR) model 
(see Sec.\ref{sec83}), 
are also shown in the right plot in Fig.\ref{fig14}. 
Since self-trapping is avoided in the NQKR model, a
normal diffusion process
for $\nu=0.5$ can be reached, as predicted. 

\section{Correlated potentials}
\label{sec8}

The effective noise and nonlinear diffusion theories need only a few assumptions on input, in particular that i) the linear wave equation
has a regime of localization with finite upper bound on the localization length, and ii) the nonlinear dynamical system should be nonintegrable
to allow for deterministic chaos (and therefore normal mode dephasing). The predicted subdiffusive exponents are controlled only by the lattice dimension,
and the power of nonlinearity.

So far we discussed the resulting nonlinear diffusion for uncorrelated random potentials $\epsilon_l$.
For linear wave equations, a number of other {\sl correlated} potentials are known to result in wave localization
for a corresponding linear wave equation.
 
\subsection{Subdiffusive destruction of Aubry-Andre localization}
\label{sec81}

\begin{figure}
\sidecaption
\includegraphics[width=0.64\columnwidth]{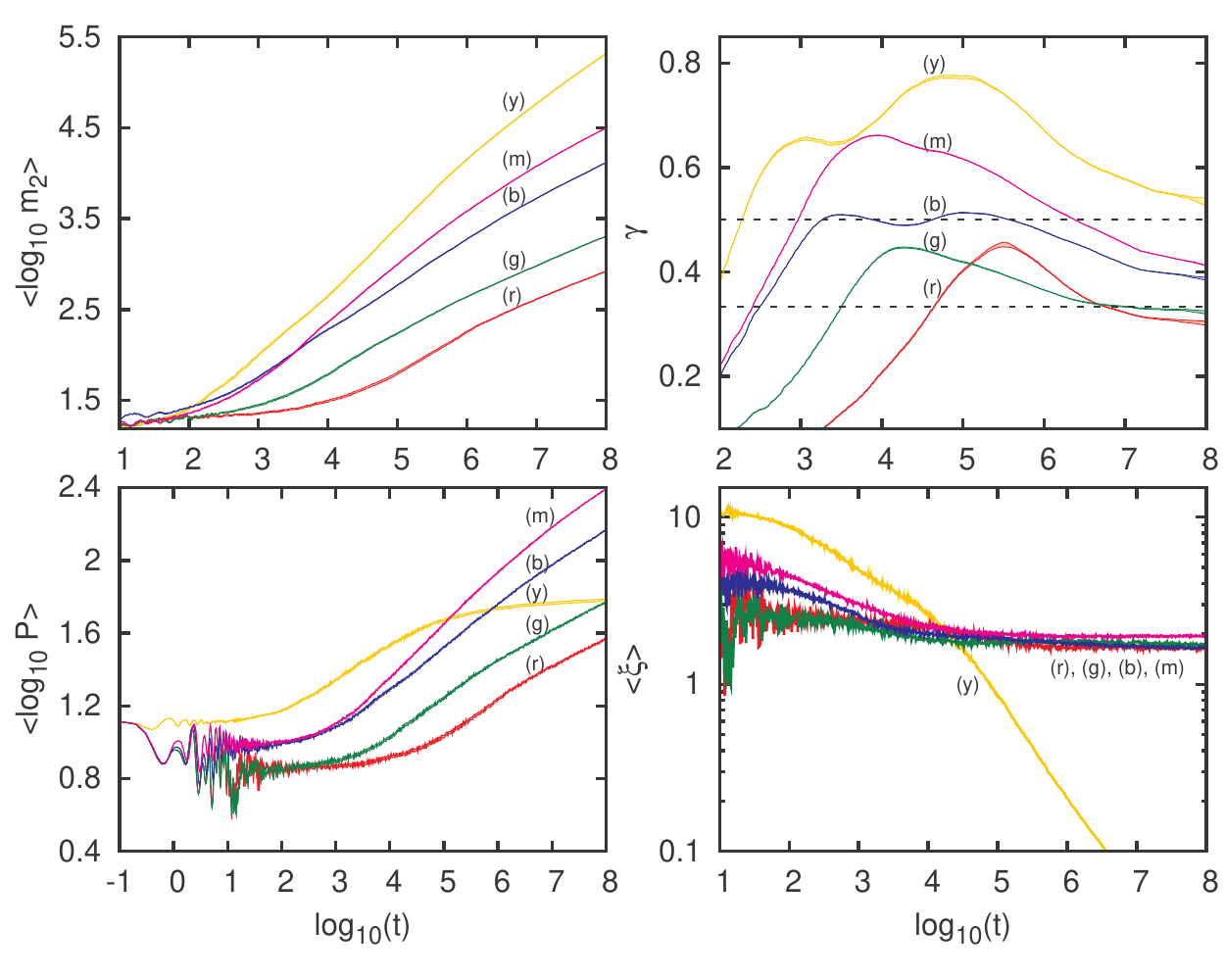}
\caption{Top left panel: time evolution of $\langle\log_{10}\,m_2\rangle$; top right panel:  $\langle\log_{10}\, P\rangle$; 
bottom left panel: spreading 
exponent $\gamma$; bottom right panel: average compactness index $\langle\xi\rangle$. The  nonlinear 
parameter $\beta=0.1, 1, 5, 10, 100$  (red (r), green (g), blue (b), magenta (m) and yellow (y) curves respectively). 
The initial wave packet has $V=13$ sites excited, and  $\lambda=2.5$.
The two dashed lines in the top right panel correspond 
to the values $\gamma=1/3$ and $\gamma=1/2$. 
Adapted from \cite{mltvljdbfdmmsf12}}
\label{fig15}
\end{figure}
Let us replace the uncorrelated disorder potential in Sec.\ref{sec2} by  
\begin{equation}
\epsilon_l=\lambda \cos (2\pi \alpha_{AA} l+ \theta)\;.
\label{qpp}
\end{equation}
For the linear wave equation $\beta=0$ and any irrational choice of $\alpha_{AA}$
this results in the well-known Aubry-Andre localization \cite{aubry1980}. 
Note that the irrationality of $\alpha_{AA}$ implies that the spatial period of (\ref{qpp}) is incommensurate with the lattice spacing $\Delta l = 1$,
and therefore the lattice potential becomes a quasiperiodic one.
For shallow potentials $\lambda < 2$ all eigenstates are extended. At the critical value $\lambda = 2$ a metal-insulator transition takes place,
and for $\lambda > 2$ all eigenstates are localized with localization length $\xi =  1/\ln ( \lambda /2)$, independent of $\alpha_{AA}$ and the eigenenergy of the state \cite{aubry1980}.
One peculiarity of the linear wave equation is that its eigenvalue spectrum is fractal, has a self-similar Cantor set structure and fractal dimension 1 for all $\lambda \neq 0\;,\; 2$.
In particular it displays a self-similar hierarchy of gaps and subgaps, which implies that self-trapped states can be generated at any weak nonlinearity.

Spreading wave packets were studied by Larcher et al \cite{mltvljdbfdmmsf12} in the presence of nonlinearity (see Fig.\ref{fig15}). Again a clear regime of weak chaos $m_2 \sim t^{\gamma}$ was observed,
with the exponent $\gamma \approx 1/3$.  Signatures of strong chaos are also observed, which however might be affected by the presence of selftrapping even
at weak nonlinearities.

\subsection{Subdiffusive destruction of Wannier-Stark localization}
\label{sec82}

\begin{figure}
\sidecaption
\includegraphics[width=0.64\columnwidth]{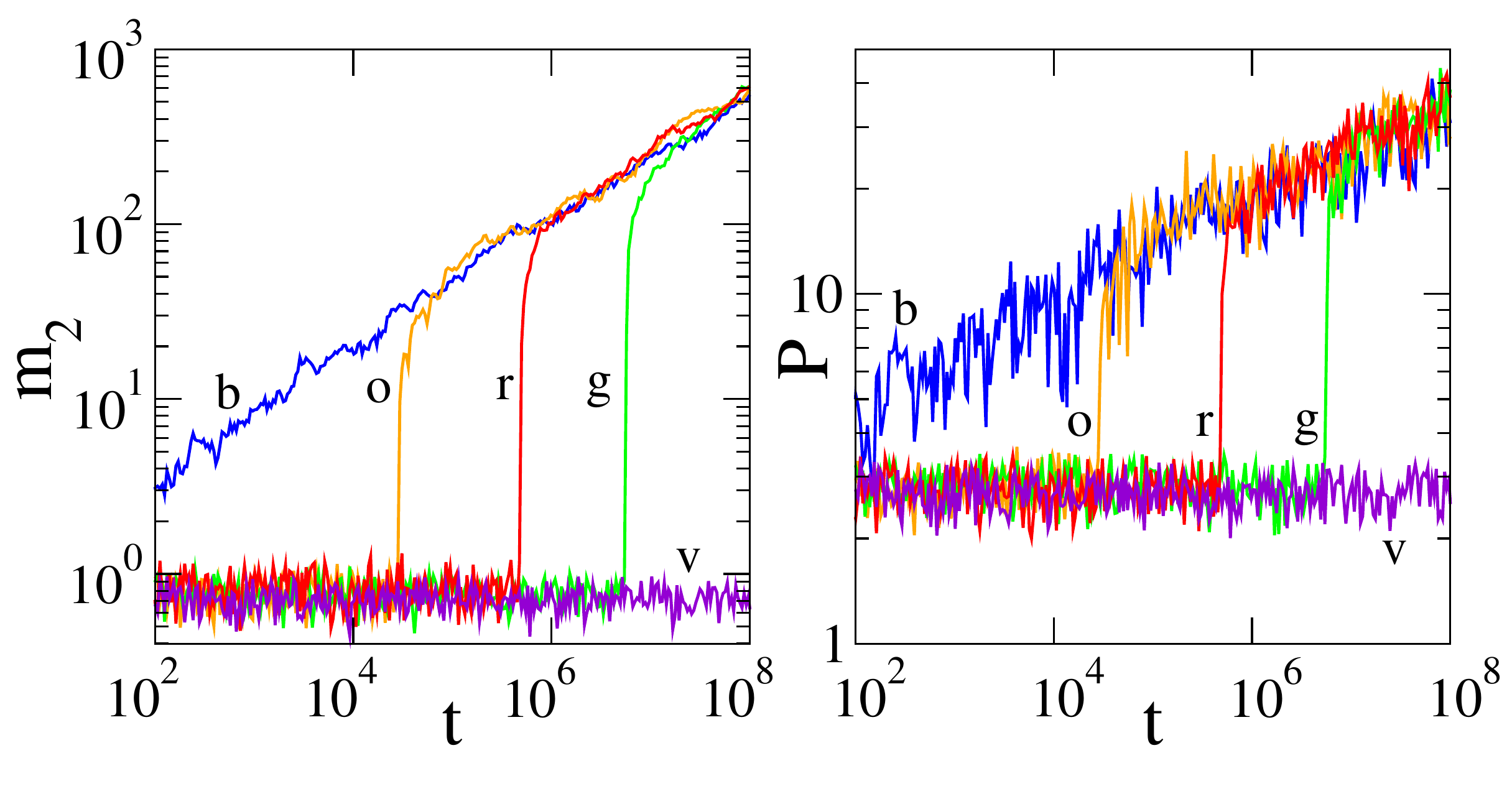}
\caption{Single site excitation for $E=2$. Second moment $m_2$ and participation number $P$ versus time
in log-log plots for different values of $\beta$ inside the interval where an explosive delocalization of the trapped
regime occurs: $\beta=8.15,8.25,8.5$  [(o) orange; (g) green; (r) red]. $\beta=8$ [(b), blue]: intermediate regime.
$\beta=8.9$ [(v), violet]: trapped regime.
Adapted from  \cite{dokrksf09}
}
\label{fig16}
\end{figure}
An even simpler choice of a dc bias potential
\begin{equation}
\epsilon_l = E l
\label{wsp}
\end{equation}
with a constant dc field value $E$  is generating localized states as well. The spectrum of the linear wave equation is an equidistant Wannier-Stark ladder with $\lambda_{\nu} = E \nu$.
All states are localized with localization volume $V \sim | 1/(E \ln E )|$ for weak field strength $E < 1$, and $V (E \rightarrow \infty) \rightarrow 1$. 
These Wannier-Stark states are superexponentially localized
$|A_{\nu, l \rightarrow \infty}^{(0)}| \rightarrow \left(1/E\right)^l/l!$ and therefore very compact in the tails, even for weak dc fields.

Spreading wave packets were studied by Krimer et al \cite{dokrksf09} in the presence of nonlinearity (see Fig.\ref{fig16}). While subdiffusion is observed for a wide range
of parameters, there are distinct differences to the cases discussed so far. Namely, initial staes may be trapped for very long times, but then explosively start to
spread. Further, the subdiffusive growth $m_2\sim t ^{\alpha}$ shows a field dependence of the exponent $\alpha(E)$. Krimer et al \cite{dokrksf09} report $\alpha(E=2) \approx 0.38$,
while Kolovsky et al \cite{arkeaghjk10} report $\alpha (E=0.25) \approx 0.5$. The reason for this dependence might be routed in the fact, that a spreading wave packet
has to excite exterior modes close to its boundary, whose eigenenergies are {\sl outside} of the energy spectrum excited inside the wave packet (due to the Wannier-Stark ladder
spectrum). The larger $E$, the larger is this frequency mismatch. Another interesting feature of this model is, that exact quadruplet resonances exist, which seem to
leave no room for perturbation approaches.  

\subsection{Subdiffusive destruction of dynamical localization}
\label{sec83}

\begin{figure}
\sidecaption
\includegraphics[width=0.4 \columnwidth]{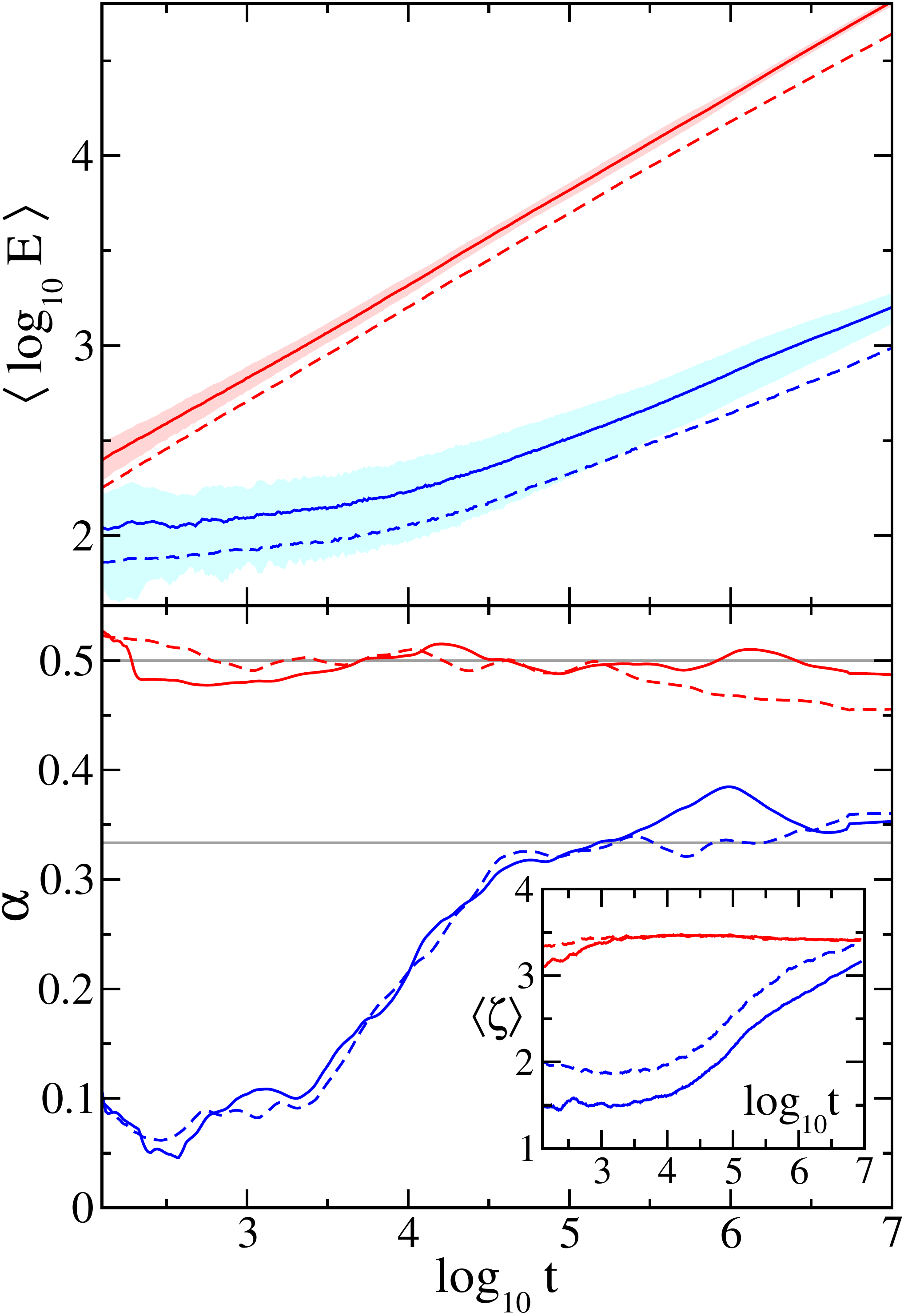}
\caption{Under a kick strength of $k=5$, measures for $\beta=0.3$ (blue) and $\beta=10$ (red), for both quasiperiodic sequences set by $\tau=1$ (solid line), and for random 
sequences (dashed line, see \cite{ggjdbsf11} for details). Upper row: Mean logarithms for energy $<\log_{10} E>$. The clouds around the quasiperiodic sequences correspond to one standard deviation error. 
Lower row: finite-difference derivative of the above. Grey horizontal lines correspond to exponents for weak and strong chaos regimes. Inset: average compactness 
index $\langle \zeta \rangle $ as a function of time.
Adapted from \cite{ggjdbsf11}
}  
\label{fig17}
\end{figure}
Experiments of quantum kicked rotor systems \cite{kr2,kr34} within Bose-Einstein condensates \cite{kr911},
where many-body interactions play a significant role, focus theoretical
attention on dynamical localization in the presence of nonlinear interactions.
In the mean-field approximation, the dynamics of the kicked rotor can be modeled by the
following form of the Gross-Pitaevskii equation
\begin{equation}
i \hbar 
\frac{\partial \psi}{\partial t} = - \frac{\hbar^2}{2 M} 
\frac{\partial^2 \psi}
{\partial \theta^2} + 
\tilde{\beta}
|\psi|^2 \psi + 
{\bar k}
\cos(\theta) \cdot \psi \sum_m \delta(t-mT). \label{eq:NLS}
\end{equation}
Here $\tilde{\beta}$ is the nonlinear strength, which is proportional to the 
tunable scattering length of atoms in a BEC. $M$ is the mass of the atoms,
$\bar k$ is the perturbative kick strength, and $T$ is the period of applied
kicks. 
Note that the analogy between an abstract rotor and the atomic wavefunctions is obtained
when the atoms are loaded into a momentum eigenstate of the lattice with Bloch wavenumber zero,
Spatially homogeneous kicks will keep the Bloch wavenumber invariant, but allow 
to change the momentum.
The solution $\psi(\theta,t)$ can be expanded in an angular momentum basis
\begin{equation}
\psi(\theta,t) = \frac{1}{\sqrt{2\pi}} \sum^{\infty}_{n=-\infty} A_n(t) e^{i n \theta} \label{eq:AngMomMap}
\end{equation}
where the coefficients $A_n(t)$ are Fourier coefficients of the time-dependent wave function $\psi(\theta,t)$. 
The dynamics between two
successive kicks is described by following equation
\begin{equation}
i\frac{\partial A_n}{\partial t} = - \frac{1}{2}\tau n^{2} A_n +
\beta \sum_{n_{1}}\sum_{n_{2}} A^{*}_{n_{1}} A_{n_{2}}
A_{n-(n_{2}-n_{1})}, \label{eq:momNLS} 
\end{equation}
where $\beta = \tilde{\beta}T/2\pi\hbar$. Keeping only the 
diagonal terms in Eq.(\ref{eq:momNLS}) and
integrating over the free motion between two delta kicks, $A_{n}(t)$
evolves according to
\begin{equation}
A_{n}(t+1) = A_{n}(t) e^{-i\frac{\tau}{2}n^2 + i\beta|A_n|^2}, \label{eq:momBK} 
\end{equation}
After additional integration over the
infinitesimal interval over one kick, the map - which now describes the evolution over one whole period -
becomes
\begin{equation}   
A_n(t+1)=\sum_m (-i)^{n-m} J_{n-m}(k) A_m(t) e^{-i\frac{\tau}{2}m^2 + i\beta|A_m|^2}. \label{eq:NLS_Shep}
\end{equation}
This map was first introduced by Shepelyansky in \cite{ds1993}. Comparison of the results of this map with direct numerical 
simulation of the corresponding model, Eq.~(\ref{eq:NLS}), has shown differences on a short time scale, but the same asymptotic behavior in the rotor 
energy \cite{rebuzzini_delocalized_2005}. At the same time, this model allows for more efficient and faster numerical computation. 

For $\beta=0$ all eigenstates are exponentially localized \cite{kr34}. 
The eigenvalues are located on the unit circle, and therefore embedded in a compact space. This implies, that nonlinear frequency shifts, i.e.
shifts of eigenvalues along the unit circle, may shift points out of a cloud, but with increasing nonlinearity the shifted point will return after making one
revolution. Therefore the nonlinear quantum kicked rotor (NQKR) serves as a model which lacks selftrapping. It should thus be an ideal testing
ground not only of weak chaos, but also of strong chaos.

Shepelyansky performed the first pioneering study on subdiffusive spreading 
and destruction of dynamical localization for $\beta \neq 0$ in Ref. \cite{ds1993}. 
Due to the possible presence of strong chaos, the method to extract exponents from fitting power laws to $m_2(t)$ lead to inconclusive
results.
Gligoric et al  \cite{ggjdbsf11} repeated the calculations with more averaging over initial conditions, and computing derivatives 
$\alpha=\frac{d <\log_{10} E>}{d \left( \log_{10} t \right) }$ instead  (note here that the second moment $m_2$ is equivalent to the rotor energy $E$).
The results impressively obtain a regime of weak chaos with $\alpha \approx 1/3$, and also strong chaos with $\alpha \approx 1/2$. 
The original simulations of Shepelyansky \cite{ds1993} were performed in the crossover region between strong and weak chaos, leading to 
incorrect fitting results - which are however between the two weak and strong chaos limits, as expected.

\section{Discussion}
\label{sec9}

If a linear wave equation generates localization with upper bounds on the localization length (degree of localization), then the corresponding nonlinear
wave equation shows destruction of this localization in a broad range of control parameters, and a subdiffusive spreading of initially localized wave packets. This observation holds
for a broad range of wave equations, e.g. with uncorrelated random potentials (Anderson localization), quasiperiodic potentials (Aubry-Andre localization),
dc fields (Wannier-Stark localization), kicked systems (dynamical localization in momentum space).  What is the cause for the observed subdiffusion?
Firstly it is
the nonintegrability of the systems, which leads to generic intrinsic deterministic chaos in the dynamics of the nonlinear system. 
Second, wave localization is inherently based on keeping the phases of participating waves coherent. Chaos is destroying phase coherence, and therefore
destroying localization. Wave packets can spread, but the densities will drop as spreading goes on. Therefore the effective nonlinearity and strength of chaos
decreases, and spreading is slowing down, becoming subdiffusive. The subdiffusive exponents are controlled by very few parameters and therefore rather universal.
Typically we only need to know the dimensionality of the system, and the power of nonlinearity (Anderson, Aubry-Andre, and dynamical localization). 
For Wannier-Stark localization the dc field strength is also becoming a control parameter, probably because the wave packet not only expands in space, but
also in the frequency (energy) domain.
 \begin{figure}
\includegraphics[width=1\columnwidth]{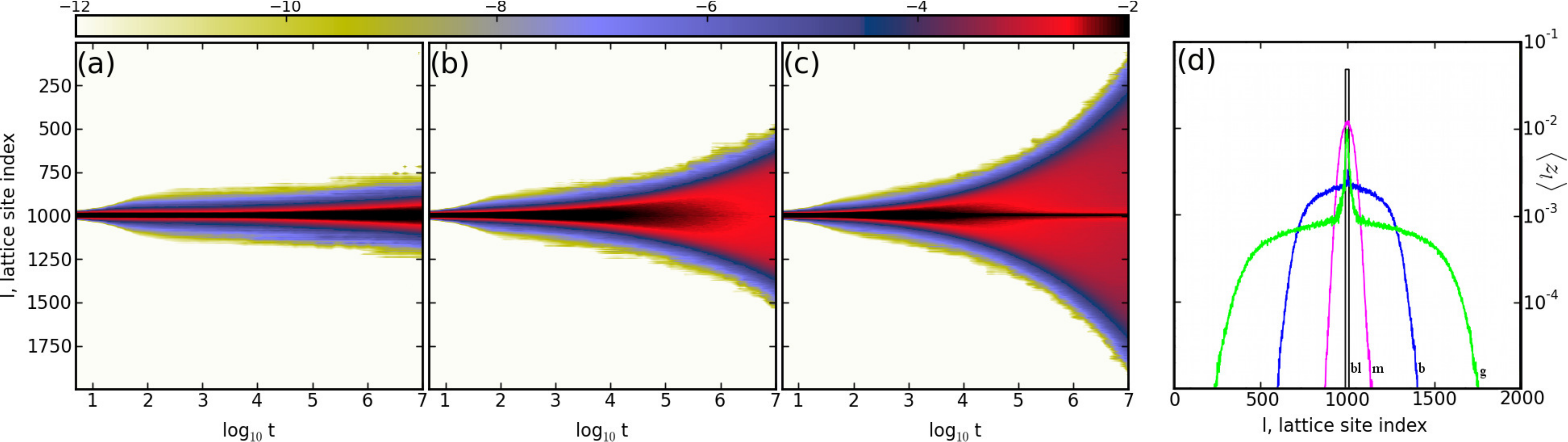} 
\caption{(Color online) DNLS, $W=4$: Time evolution of average norm density
  distributions $\langle z_l \rangle $ in real space for (a)
  $\beta=0.04$, (b) $\beta=0.72$ and (c) $\beta=3.6$. The color scales
  shown on top of panels (a)-(c) are used for coloring
  each lattice site according to its $\log_{10} \langle z_l \rangle$
  value. (d) The values $\langle z_l \rangle $ (in logarithmic scale)
  at the end time $t=10^7$ of numerical simulations for $\beta= 0.04 ,
  \, 0.72, \, 3.6$ [(m) magenta; (b) blue; (g) green]. 
For comparison
the initial norm distribution is also plotted [(bl) black].
Adapted from \cite{jdbtvlcsdoksf11}  
}
\label{fig18}
\end{figure}
A typical evolution outcome for the DNLS chain discussed at length here (see Sec.\ref{sec2}) is shown in Fig.\ref{fig18} with all three regimes of
weak chaos, strong chaos, and selftrapping.
The effective noise theory (which contains a phenomenological twist) and the nonlinear diffusion theory yield a rather coherent and consistent explanation.
Many predictions of this approach were tested, and verified to the extend of current computational possibilities.
 A number of construction places are left unfinished and call for more work. This includes e.g. i) the explanation of the dc-field dependent subdiffusive exponents
for Wannier-Stark localization, ii) the testing of the prefactor (\ref{ent6a},\ref{ent6b}), iii) its complete derivation for higher dimensions and different powers of nonlinearity,
and also iv) for other localization potentials (quasiperiodic, dc field, kicked, etc). A rather unexplored direction concerns the breaking of time-reversal symmetry,
which should lead to an increase of the stiffness of the spectrum of interacting modes, and therefore affect the statistics of interactions. A first work has been
recently finished \cite{xysf14}, but certainly more is needed.

One of the hotly debated questions in the community is whether the subdiffusive spreading will continue forever or eventually slow down, or even stop
(see e.g. \cite{sfekas12} and references therein).
This is an interesting and perhaps mathematically deep question, despite the absence of rigorous results which would fuel the above doubts.
From the perspective of current computational studies, efforts to observe any slowing down directly were not successful  \cite{jdbtvlcsdoksf11}.

Another question concerns the restoring of Anderson localization in the limit of weak disorder. 
The answer appears to depend strongly on the considered initial states. For instance, in an infinite lattice, we have to discuss the temperature dependence
of the conductivities. One possibility is that the conductivities vanish in the zero temperature limit (see Sec.\ref{sec74}), which restores the linear wave equation, and Anderson localization.
Then, Anderson localization will be destroyed at the smallest amount of nonlinearity. But may be there is a small but finite nonzero critical temperature/density/nonlinearity
threshold at which the conductivity vanishes, similar to the quantum many body localization case \cite{dmbilabla06} ?
\begin{figure}
\sidecaption
\includegraphics[width=0.6\columnwidth]{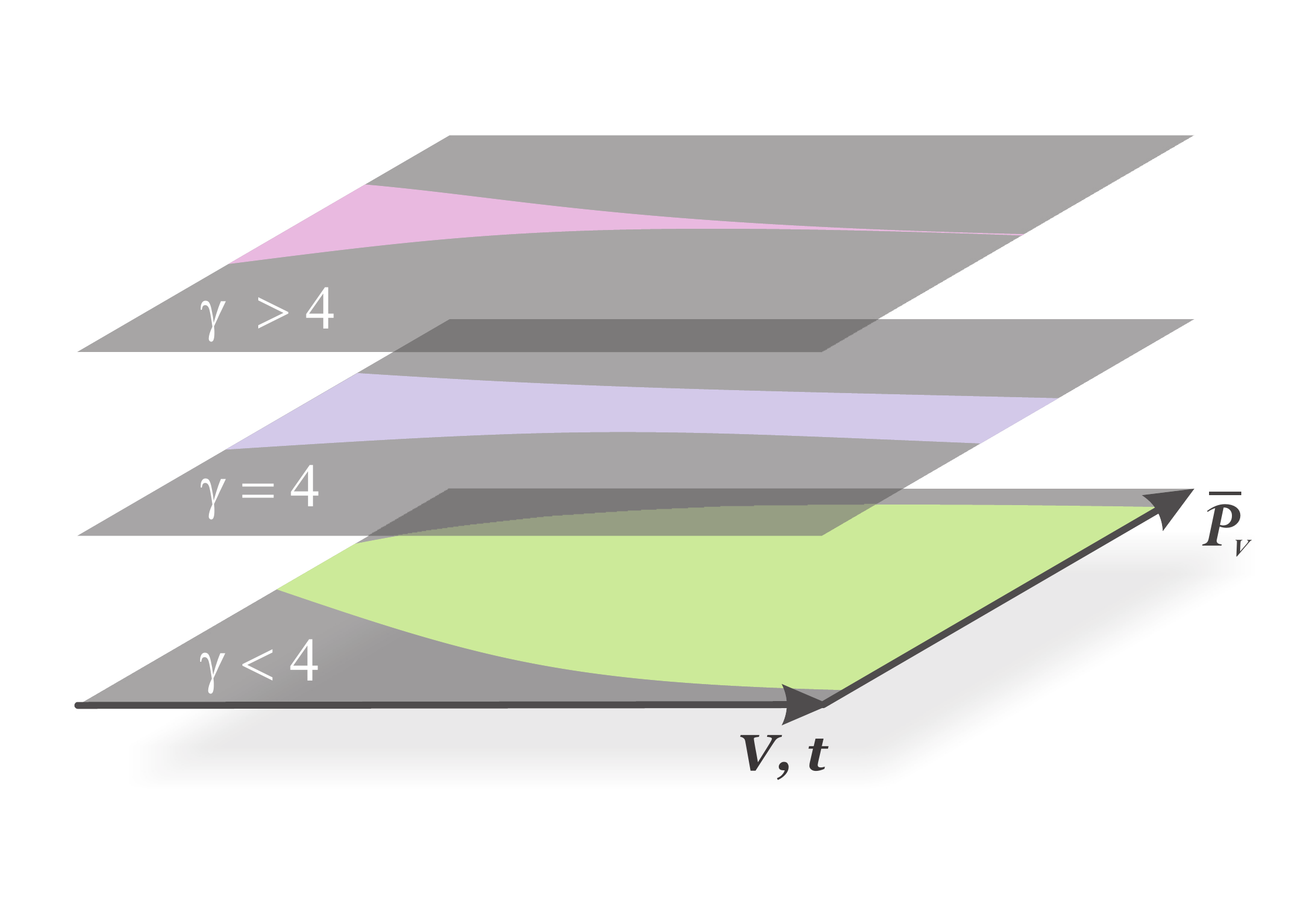}
\caption{Schematic dependence of the probability $\mathcal{P}_V$ for wave packets to stay localized (dark area) together with the complementary
light area of spreading wave packets versus the wave packet volume $V$ (either initial or attained at some time $t$) for three different orders of nonlinearity
$\gamma<4$, $\gamma=4$ and $\gamma > 4$.
Adapted from \cite{mvitvlsf11}
}
\label{fig19}
\end{figure}
Another type of initial states are the ones mostly considered in this chapter - compact localized wave packets in a zero density surrounding.
Then, if nonlinearity is lowered, several papers study the fate of these states \cite{mjgksa10},\cite{sa11},\cite{mvitvlsf11}.
The main outcome appears to be, that for a {\sl given and fixed} initial state, at small enough nonlinearity, the dynamics will be in a KAM regime,
i.e. there will be a finite probability $P_{R}$ that the state is launched on a torus in phase space, dynamics is regular, phase coherence is conserved, and
no spreading will occur. But then, there is the complementary probability $P_{Ch}= 1-P_R$ to miss the torus, and instead to be launched on a chaotic trajectory,
where dynamics is irregular, phase coherence is lost, and spreading may occur. Here probability is meant with respect to the disorder realization (or the
space location of the initial state). The probability $P_R$ increases to one for vanishing nonlinearity, and therefore Anderson localization is restored in this
probabilistic sense. A consequence of the considerations in \cite{mvitvlsf11} is shown in Fig.\ref{fig19}, where $\gamma=\sigma + 2$ measures the
power of nonlinearity (see (\ref{RDNLS-EOMG})). Namely, we assume that the dynamics starts on a chaotic trajectory. Then by assumption, we will continue to
be on a chaotic trajectory, and spread. For the typical size of the wave packet at a later stage, we may recalculate the probability to keep chaoticity if we
suddenly change the disorder realization. The answer is, that for $\sigma=2$ (cubic nonlinearity), even in the limit of an infinitely spread wave packet (with
infinitesimally small densities) the probability of chaos stays finite (and can be anything between zero and one).  
For $\sigma < 2$ this chaos probability tends to one in the infinite time/spreading limit - despite the fact that the densities drop to zero. In this case chaos always wins.
Finally, for $\sigma > 2$ the chaos probability shrinks to zero. Therefore, even if our chaotic trajectory will spread forever, it will enter a phase space region
which is predominantly regular. What kind of regime is that? Is there place for Arnold diffusion? Will subdiffusive spreading slow down in that case (apparently
numerical studies do not report on anything suspicious in that case) ? 

The above studies were restricted to lattice wave equations, which introduce upper bounds for the localization length. Spatially continuous wave equations may lack these upper bounds.
Therefore an initially compact localized wave packet may overlap with normal modes whose localization length is unbounded in principle. While this may become an intricate
matter of counting overlap weights, it is instructive to see that numerical studies of such cases also indicate the appearance of the universal subdiffusive spreading as observed
for lattices \cite{iva14}.

The more models are accumulated for the above studies, the more qualitative differences are becoming visible. For instance, models can be classified
according to the number of integrals of motion (KG - one, DNLS - two). Other models differ in the connectivity in normal mode space - while 
cubic DNLS and KG equations have connectivity $K=4$ (four modes are coupled), other models discussed e.g. in \cite{mvitvlsf11},\cite{mmap12}
have connectivity $K=2$. Again the strong disorder limit of $K=4$ models yields $K=2$ in leading order, which is one of the cases where 
analytical methods are applied (see references in \cite{sfekas12}). Time might be ripe to perform comparative studies.

\begin{acknowledgement}
I thank all cited authors with whom I had the pleasure to jointly work and publish. In addition I thank
I. Aleiner,
B. L. Altshuler,
P. Anghel-Vasilescu,
D. Basko,
S. Fishman,
J. C. Garreau,
A. R. Kolovsky,
Y. Krivolapov,
Y. Lahini,
G. Modugno, 
M. Mulansky, 
R. Schilling,
D. Shepelyansky,
A. Pikovsky, 
W.-M. Wang
and 
H. Veksler
for stimulating and useful discussions.
\end{acknowledgement}
\end{document}